\documentclass[12pt]{iopart}
\usepackage{iopams}
 
\expandafter\let\csname equation*\endcsname\relax
\expandafter\let\csname endequation*\endcsname\relax
\usepackage{amsmath,amssymb,mathtools}

\usepackage{graphicx,color}
\usepackage{enumerate}

\usepackage[colorlinks,linkcolor=blue,urlcolor=blue,citecolor=blue]{hyperref}

 \newcommand{\erefss}[1]{\eqref{#1}}%
 \newcommand{\erefs}[1]{\eqref{#1}}%
 \newcommand{\frefss}[1]{figures~\ref{#1}} %
 \newcommand{\frefs}[1]{~\ref{#1}} %
 \newcommand{\ffref}[1]{Figure~\ref{#1}} %
 \newcommand{\ssref}[1]{Section~\ref{#1}}%
 \newcommand{\aref}[1]{\ref{#1}}%

\newcommand\be{\begin{equation}} 
\newcommand\ee{\end{equation}} 
\begin{document}

\title{Partial entropy production in heat transport}

\author{Deepak Gupta and Sanjib Sabhapandit}  
\address{Raman Research Institute,  Bangalore - 560080, India}

\date{\today}

\begin{abstract}
We consider a system of two Brownian particles (say A and B), coupled to each other via harmonic potential of stiffness constant $k$. Particle-A is connected to two heat baths of constant temperatures $T_1$ and $T_2$, and particle-B is connected to a single heat bath of a constant temperature $T_3$. In the steady state, the total entropy production for both particles obeys the fluctuation theorem. We compute the total entropy production due to one of the particles called as partial or apparent entropy production, in the steady state for a time segment $\tau$. When both particles are weakly interacting with each other, the fluctuation theorem for partial and apparent entropy production is studied. We find a significant deviation from the fluctuation theorem. The analytical results are also verified using numerical simulations.

\end{abstract}


\maketitle

\noindent\rule{\hsize}{2pt}
\tableofcontents
\noindent\rule{\hsize}{2pt}
\markboth{Partial entropy production in heat transport}{}

\section{Introduction}
Understanding the properties of driven small systems has challenged both theoretical and experimental techniques. Examples of small scale systems include biopolymers (such as DNA, RNA and proteins), enzymes, Brownian particles, Brownian motors, and nanoscale engines. These systems always remain in contact with a noisy environment called as heat bath, and that satisfies \emph{fluctuation-dissipation theorem} \cite{Kubo}. To drive these systems away from equilibrium, an external source of energy is needed. These sources can be an external time dependent field, temperature or chemical potential gradient, shear flow, etc. For systems driven far from equilibrium, the general framework of equilibrium statistical mechanics \cite{Huang} is not applicable.
When the system is close to equilibrium, linear response of it in the presence of field, is related to the fluctuation properties of the given system (in the absence of field) in the equilibrium. 
The \emph{fluctuation relations} go beyond the linear response theory and are valid for the system driven arbitrarily far from equilibrium. These relations are particularly useful for small scale systems where fluctuations are predominant. Within this context, Evans et al. \cite{Evans-Cohen} used numerical simulation to understand the distribution of entropy production of a shear fluid interacting with a thermal reservoir. They found an interesting symmetry relation between the probability of positive entropy production to that of negative one in a steady state, which we referred as \emph{steady state fluctuation theorem}.  Gallavotti and Cohen \cite{Gallavotti-Cohen} proved the fluctuation theorem for Hamiltonian system using chaotic assumption for the dynamics. Subsequently, Kurchan \cite{Kurchan} derived the fluctuation theorem for Langevin dynamics which was later extended by Lebowitz and Spohn \cite{lebowitz-Spohn} for general Markov process. Jarzynski's equality \cite{Jarzynski-2,Jarzynski} provides a method to estimate the difference in the free energies between two equilibrium states using nonequilibrium work done. Hatano-Sasa relation \cite{Hatano-Sasa} which is analogous to Jarzynaki's equality, deals with the transition between steady states. Evans and Searles \cite{Evans-Searles} proved the fluctuation theorem in transient regime (i.e., when the initial distribution is the equilibrium one). Crooks work fluctuation theorem \cite{Crooks-1, Crooks-2} measures the relative probabilities of work done in forward and reverse processes starting from respective equilibrium distributions.

In the realm of stochastic thermodynamics \cite{Sekimoto}, thermodynamical quantities such as heat, work, power dissipation, power injection, etc., are defined at the single trajectory level of a non-equilibrium process. These observables have probability distribution function rather than having a unique value. When such distribution functions satisfy large deviation principle \cite{Touchette}, then those are characterized by large deviation function. With the interest of obtaining these distributions and large deviation functions, several efforts have been put to test the corresponding fluctuation relations for the observables mentioned above.
In contrast to initial studies where the sole contribution to the entropy production was considered from the heat dissipation by the system in the thermal bath, Seifert \cite{Seifert-1,Seifert-2} identified entropy production along the stochastic trajectory using an example of colloidal particle and for general stochastic dynamics obeying master equation. This entropy production consists of two parts: the entropy production in the system $\Delta S_{sys}$ and the entropy production in the medium $\Delta S_{med}$. The sum of these two is called as total entropy production $\Delta S_{tot}=\Delta S_{sys}+\Delta S_{med}$. While the system entropy production remains bounded for finite potential system, the medium entropy production has major contribution towards the total entropy production in the long time limit. Thus, the corresponding fluctuation theorem holds in the long time limit. When both of these parts are considered, the total entropy production $\Delta S_{tot}$ in the steady state, satisfies the fluctuation theorem given as
\begin{eqnarray}
\dfrac{P(\Delta S_{tot}=s\tau)}{P(\Delta S_{tot}=-s\tau)}=e^{s\tau},
\label{Fl-Th}
\end{eqnarray}
for all time, where $\Delta S_{tot}$ is an extensive quantity with the observation time $\tau$. Though, total entropy production $\Delta S_{tot}$ satisfies steady state fluctuation theorem in general, this relation may not hold for other stochastic observables such as work done, heat flow, power flux, etc., \cite{VonZon-1,VonZon-2,VonZon-3,SS-1,SS-2,Apal-1,Apal-2,Verley-1,Visco,Farago}. 

Consider a Brownian particle connected to two heat reservoirs of different temperatures $T_1$ and $T_2$ (where $T_1>T_2$). According to standard thermodynamics, the heat flows from a hot to a cold reservoir. However, in this small system, due to thermal fluctuations, once in a while, the heat flows in reverse direction, although on an average, the heat current is a positive quantity which obeys second law of thermodynamics. The entropy production in the baths is $\Delta S_{med}=-(Q_1/T_1+Q_2/T_2)$ (see \sref{BP}), where $Q_i$ is the heat energy transferred by $i^{th}$ bath to the Brownian particle. This quantity may not always satisfy the fluctuation theorem \eref{Fl-Th} as shown in \sref{BP}. However, once the system entropy production $\Delta S_{sys}$ is added to medium one, resulted total entropy production $\Delta S_{tot}$ obeys the relation given in \eref{Fl-Th}. 

In all of above the examples, it is assumed that all the relevant degrees of freedom (DOFs) of the system are considered. However, there may arise a situation where the complete description of the system is not possible. The cause of incomplete information may be because the system is coarse grained or there may be hidden DOFs which affect the observed system. Recently, there has been lot of excitement in understanding the system properties when the partial information of a given system is available.   
For example, Rahav et al. \cite{Rahav-Jarzynski} considered a Markov jump process on a set of finite number of states. The jump from a given state to another state is described by transition rates. Assuming certain transition rates higher than other ones, the whole network of states is then mapped to a group of clusters. This type of coarse graining modified the entropy production and they showed that the \emph{coarse grained entropy} satisfied the fluctuation theorem provided the transition rates of the system to jump within clusters are sufficiently high. 
Similar results found in Ref. \cite{Yli}, where authors have discussed the projection of Markov process with constant transition rates to smaller number of observable aggregated states, and the resulting entropy production on the set of all aggregated state satisfied both detailed and integral fluctuation theorems.
Puglisi et al. \cite{Puglisi} showed an example where decimation of certain fast states with respect to a given threshold time does not affect the entropy production, provided it does not entirely remove the loops carrying net probability current. In all of these references, it is shown that coarse graining based on time-scale separation did not alter the underlying physics of the problem. There are some other studies which differ from the above mentioned ones where relaxation time of all relevant DOFs is much larger than that of bath DOFs.
For example, in  Ref. \cite{JMehl}, two paramagnetic colloidal particles of same size were trapped in separate non-overlapping toroidal traps. Whole system was in contact of the heat bath. Tangential forces were applied on each particle using laser field. Consequently, both particles reached in the non-equilibrium stationary state. A static magnetic field perpendicular to the plane of toroidal traps, was used to set an interaction among these particles. The total entropy production due to one of the particles was measured and the deviation from fluctuation theorem with the coupling strength was observed. In a theoretical model \cite{Borrelli}, fluctuation theorem for the entropy production of a single electron box is studied in a coupled electron box system.
In an another example \cite{Lacoste}, authors studied molecular motors which are modeled by flashing ratchet, and found that Gallavotti-Cohen symmetry is preserved only when both chemical and mechanical DOFs are considered in the theory. There are also some examples where partial information is utilized to get full system properties.
For example, in Ref. \cite{Ribezzi-Crivellari}, Ribezzi-Crivellari and Ritort have shown a method to infer full work distribution from the partial work measurement using Crooks fluctuation theorem. Amann et al. \cite{Christian} have derived a criterion to describe a non-equilibrium steady state of a Markov system of three states using the data from  sufficiently long two states trajectories. Some other studies related to the area of partial observation of a complete system  can also be seen in Refs. \cite{Chun,Patrick,Talkner,Polettini,Todd,Uhl,Sagawa,Esposito}. The main conclusion of these studies is, when a system consists of DOFs having relaxation time much larger than that of bath DOFs, then the partial system or subsystem may not behave like a complete system for large coupling strength. Recently, Gupta et al. \cite{Deepak} have discussed a fluctuation theorem for partial entropy production in weak coupling limit in a general scenario, and shown that deviation from the fluctuation theorem can be seen even in the limit of coupling strength tending to zero. In Ref. \cite{Trap}, authors have shown a technique to diminish the effect of weak coupling on the observed DOFs from the hidden variables. In both of the previous setups, the authors have chosen external stochastic Gaussian forces to drive the system into nonequilibrium state. In this paper, we consider a system of a coupled Brownian particle where one of the particles (say particle-A) is connected to two heat baths at different temperatures while the other one (say particle-B) in attached to a single heat bath. We take the interaction between these two particles to be harmonic. Here, we focus on the total entropy production in the steady state by one of the particles (say particle-A), which we call as partial or apparent entropy production $\Delta S_{tot}^A$. In the limit of vanishing coupling, deviation from the fluctuation theorem [see \eref{Fl-Th}] is studied. There are two important features of this paper: (1) We have used thermal gradient to drive the system into nonequilibrium steady state, and (2) The asymmetry function given in \eref{asymmetry-function}, has negative slope which was not observed in the earlier studies.

The paper is organized as follows. We describe the model system and give the definitions of partial entropy production in \sref{model}. \ssref{SSD} contains the joint steady state distribution $P^{full}_{ss}(U)$ of the coupled system. The generating function $Z(\lambda)\sim g(\lambda) e^{\tau\mu(\lambda)}$ is obtained in the large time limit $(\tau\to\infty)$ for both definitions of partial entropy production in \sref{mgf}, and then, we invert $Z(\lambda)$ using saddle point method which yields the probability density function for partial entropy production (\sref{PDF-sec}). In \sref{BP}, we compute the medium entropy production by a single Brownian particle connected to a thermal gradient, and show that for the entropy production to satisfy the fluctuation theorem for large but finite time, it is necessary to incorporate the system entropy production. Since we are interested in the steady state fluctuation theorem for partial entropy production in the weak coupling limit, we discuss the assumption to approximate the prefactor term $g(\lambda)\approx g_0(\lambda)$ in \sref{glambda}. The computation of the branch point singularities present in the cumulant generating function or the range of the saddle point discussed in \sref{mulambda}. In \sref{GC-sec}, we discuss the Gallavotti-Cohen symmetry of the cumulant generating function $\mu(\lambda)$. The asymmetry function which measures the deviation from the fluctuation theorem is discussed in \sref{asym-func}. \ssref{Num-sim} contains the comparison of the analytical predictions with the numerical simulations. We summarize our paper in \sref{summ}. In \aref{fast-DOF}, we show how the hidden fast DOFs effect the fluctuation theorem in the weak coupling limit for a single Brownian particle coupled to three baths of different temperatures.


\section{Model}
\label{model}
\begin{figure}
\begin{center}
    \includegraphics[width=0.5  \textwidth]{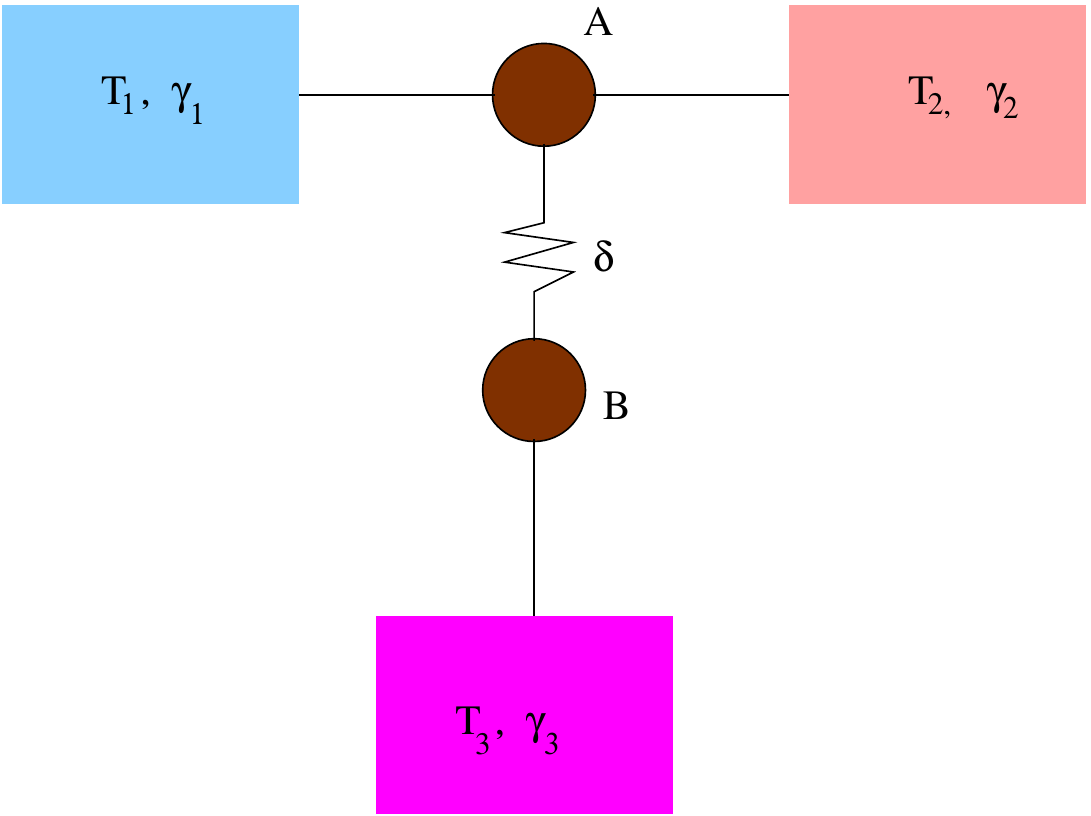}{\centering}
\end{center}
   \caption{\label{system-pics}Brownian particle-A is connected to two heat baths of temperatures (dissipation constants) $T_{1}$ ($\gamma_1$) and $T_2$ ($\gamma_2$) while particle-B is connected to a bath of temperature (dissipation constant) $T_{3}$ ($\gamma_3$). Both particles are connected by spring of coupling parameter $\delta=2 km/\gamma_1^2$ (dimensionless).}
\end{figure}

Consider a Brownian particle (say particle-A) of mass $m$, in contact with two heat baths at temperatures $T_1$ and $T_2$ ($T_1>T_2$). Let $\gamma_1$ and $\gamma_2$ are the dissipation constants of the baths with temperatures $T_1$ and $T_2$, respectively. Suppose the given Brownian particle-A is coupled harmonically with another Brownian particle (say particle-B) of mass $m$. The particle-B is in contact with a single heat bath of a constant temperature $T_3$ and the dissipation constant $\gamma_3$. The schematic diagram of the coupled Brownian particle system is shown in \fref{system-pics}. The Hamiltonian of the system is given as
\begin{equation}
\mathcal{H}(y,v_A,v_B)=\dfrac{1}{2}mv_A^2+\dfrac{1}{2}mv_B^2+\dfrac{1}{2}ky^2,
\end{equation}
where $y=x_A-x_B$ is the relative separation between particle-A and particle-B, $k$ is the stiffness constant, $v_A$ and $v_B$ are the velocities of particle-A and particle-B, respectively.
 The evolution of the given system is described by following Langevin equations
\begin{align}
\dot{y}&=v_A(t)-v_B(t)\label{dyn-1},\\
m\dot{v}_A&=-\gamma_Av_A(t)+\eta_A(t)-ky(t)\label{dyn-2},\\
m\dot{v}_B&=-\gamma_Bv_B(t)+\eta_B(t)+ky(t)\label{dyn-3},
\end{align} 
where $\eta_A(t)=\eta_1(t)+\eta_2(t)$, $\eta_B(t)=\eta_3(t)$, $\gamma_A=\gamma_1+\gamma_2$, and $\gamma_B=\gamma_3$. The thermal noises $\eta_1(t)$, $\eta_2(t)$, and $\eta_3(t)$ are from the heat baths, with mean zero and correlations $\langle\eta_{i}(t)\eta_{j}(t^\prime)\rangle=2T_{i}\gamma_i \delta_{ij}\delta(t-t^\prime)$,  where \{i,j\}=\{1,2,3\}. We set Boltzmann's constant $k_B=1$ throughout the calculations.

The goal of this paper is to understand the validity of the fluctuation theorem for the total entropy production of a subsystem or partial system of the complete system in the steady state. In the following subsections, we give two definitions of partial entropy productions.
\subsection{Partial entropy production}
\label{PEP-def}
Suppose we are interested in the total entropy production due to particle-A in a coupled Brownian particle system as shown in \fref{system-pics}. The dynamics of particle-A is given by \eref{dyn-2}.
Multiplying \eref{dyn-2} by $v_A(t)$ and integrating over time $t$ from $0$ to $\tau$ yields 
\begin{equation}
\dfrac{1}{2}m[v^2_A(\tau)-v^2_A(0)]=Q_1(t)+Q_2(t)-k\int_0^\tau dt\ y(t)v_A(t),
\label{first-law}
\end{equation}
where $Q_{i}=\int_0^\tau [\eta_i(t)-\gamma_iv_A(t)]v_A(t)\ dt$, is the heat absorbed by Brownian particle-A from the $i^{th}$ heat bath. The term on the left hand side is the change in the kinetic energy of the Brownian particle-A from time $t=0$ to $t=\tau$. The third term on the right hand side is energy change due to interaction among the particles. These baths are of infinite size and have infinite heat capacity. Therefore, these are assumed to be always in thermal equilibrium. Using standard thermodynamics, the entropy production in the baths due to Brownian particle-A can be written as
\begin{align}
\Delta S^A_{med}&=-\bigg(\dfrac{Q_1}{T_1}+\dfrac{Q_2}{T_2}\bigg)=\Delta \beta Q_1-\dfrac{k}{T_2}\int_0^\tau dt\ y(t)v_A(t)-\dfrac{m}{2T_2}[v^2_A(\tau)-v^2_A(0)]\label{eqn-2-S},
\end{align}
where $\Delta \beta=\dfrac{1}{T_2}-\dfrac{1}{T_1}$ is the difference in inverse temperatures.

Total entropy production $\Delta S_{tot}^A$ of Brownian particle-A is the sum of medium entropy production $\Delta S_{med}^A$ and system entropy production $\Delta S_{sys}^A$ of the particle-A. In the steady state, the system entropy production of the particle-A from time $t=0$ to $t=\tau$ is \cite{Seifert-1,Seifert-2}
\begin{align}
\Delta S^A_{sys}=-\ln P_{ss}(v_A(\tau))+\ln P_{ss}(v_A(0)),
\end{align}
where $P_{ss}(v_A)$ is the steady state distribution obtained after integrating the joint steady state distribution $P^{full}_{ss}(y,v_A,v_B)$ over $y$ and $v_B$
\begin{equation}
P_{ss}(v_A)=\dfrac{1}{\sqrt{2\pi H_{P}}}\exp\bigg[-\dfrac{v^2_A}{2 H_{P}}\bigg].
\end{equation}
In the above equation, $H_P$ is given by
\begin{equation}
H_P=\lim_{\tau\to\infty} \big\langle [v_A(\tau)-\langle v_A(\tau)\rangle]^2 \big\rangle=\dfrac{\gamma_3(\gamma_1+\gamma_2+\gamma_3)(\gamma_1 T_1+\gamma_2 T_2)+m k(\gamma_1 T_1+\gamma_2 T_2+\gamma_3 T_3)}{m(\gamma_1+\gamma_2+\gamma_3)(m k+\gamma_1 \gamma_3+\gamma_2 \gamma_3)}.
\end{equation}
Therefore, total entropy production due to particle-A is given as
\begin{equation}
\Delta S^A_{tot}=\Delta \beta Q_1-\dfrac{k}{T_2}\int_0^\tau dt\ y(t)v_A(t)-\frac{1}{2}\bigg[\frac{m}{T_2}-\frac{1}{H_P}\bigg][v^2_A(\tau)-v^2_A(0)].
\label{PEP}
\end{equation}

\subsection{Apparent entropy production}
Consider an experiment where we want to find the entropy production for a single Brownian particle (say particle-A) in the contact with two heat baths of temperatures (dissipation constants) $T_1$ ($\gamma_1$) and $T_2$ ($\gamma_2$) (see \fref{system-pics} with $\delta=0$). The Langevin equation for Brownian particle-A is
\begin{equation}
m\dot{v}_A=-(\gamma_1+\gamma_2)v_A(t)+\eta_1(t)+\eta_2(t).
\label{partial-dynamics}
\end{equation}
Multiplying above equation by $v_A$ and integrating over time $t$ from $0$ to $\tau$ gives \eref{first-law} with $k=0$.
Therefore, one can write the entropy production in the baths due to Brownian particle-A as
\begin{align}
\Delta S^A_{med}&=-\bigg(\dfrac{Q_1}{T_1}+\dfrac{Q_2}{T_2}\bigg)=\Delta \beta Q_1-\dfrac{m}{2T_2}[v^2_A(\tau)-v^2_A(0)].
\end{align}
The system entropy production of the Brownian particle-A in steady state is given as
\begin{align}
\Delta S^A_{sys}=-\ln\tilde{P}_{ss}(v_A(\tau))+\ln\tilde{P}_{ss}(v_A(0)).
\end{align}
In the above equation, $\tilde{P}_{ss}(v_A)$ is the steady state distribution obtained from \eref{partial-dynamics}
\begin{equation}
\tilde{P}_{ss}(v_A)=\dfrac{1}{\sqrt{2\pi H_{A}}}\exp\bigg(-\dfrac{v^2_A}{2 H_{A}}\bigg),
\label{app-ss}
\end{equation}
where 
\begin{equation*}H_{A}=\lim_{\tau\to\infty} \big\langle [v_A(\tau)-\langle v_A(\tau)\rangle]^2 \big\rangle=\dfrac{\gamma_1 T_1+\gamma_2 T_2}{m(\gamma_1+\gamma_2)}.
\end{equation*}

Therefore, total entropy production of particle-A can be written as  
\begin{equation}
\Delta S^A_{tot}=\Delta \beta Q_1-\frac{1}{2}\bigg(\dfrac{m}{T_2}-\frac{1}{H_A}\bigg)[v^2_A(\tau)-v^2_A(0)].
\label{AEP}
\end{equation}
It is important to note that equation \eref{AEP} is written by assuming that there is no other particle is coupled to the given particle-A. Therefore, an experimentalist naively uses \eref{AEP} to compute the entropy production due to a single Brownian particle-A coupled to two heat baths of distinct temperatures. If there is one more particle (say particle-B) is present and interacting harmonically with given particle-A, then the actual dynamics of particle-A will be given by \erefss{dyn-1}--\erefs{dyn-3}. To understand what an experimentalist observes without the prior knowledge of particle-B, we use \eref{AEP} for entropy production with the actual dynamics given in \erefss{dyn-1}--\erefs{dyn-3}, and then we compute the distribution of total entropy production of particle-A. This definition of entropy production we call as apparent entropy production.

In fact, we can combine both definitions of total entropy production [i.e., \erefss{PEP} and \erefs{AEP}] using following parameter 
\begin{equation}
    \Pi= 
\begin{cases}
    1,& \text{Partial entropy production},\\
    0,& \text{Apparent entropy production}.
\end{cases}
\end{equation}
Therefore, the generalized partial entropy production of particle-A reads as
\begin{equation}
\Delta S^A_{tot}=\Delta \beta Q_1-\dfrac{\Pi k}{T_2}\int_0^\tau dt\ y(t)v_A(t)-\frac{1}{2}\bigg[\frac{m}{T_2}-\frac{1}{H}\bigg][v^2_A(\tau)-v^2_A(0)],
\label{gen-EP}
\end{equation}
where $H=\Pi H_P+(1-\Pi)H_A$.

\section{Steady state distribution $P^{full}_{ss}(U)$}
\label{SSD}
We write \erefss{dyn-1}--\erefs{dyn-3} in the matrix form as
\begin{align}
\dot{y}&=A^TV(t),\label{dynamics-mat-y}\\
m\dot{V}&=-\Gamma V(t)-kAy(t)+\xi(t),
\label{dynamics-mat-V}
\end{align} 
where $A=(1,-1)^T$, $V=(v_A,v_B)^T$, $\xi=(\eta_A,\eta_B)^T$, and $\Gamma_{ij}=\delta_{ij}(\delta_{1j}\gamma_A+\delta_{2j}\gamma_{B})$ with \{i,j\}=\{1,2\}.

Consider a time dependent quantity $F(t)$. The finite time Fourier transform and its inverse are given as
\begin{align}
\tilde{F}(\omega_n)&=\dfrac{1}{\tau}\int_0^\tau dt\ F(t)e^{-i\omega_n t},\label{Fourier}\\
F(t)&=\sum_{n=-\infty}^{\infty} \tilde{F}(\omega_n)e^{i\omega_n t},
\label{inverse-Fourier}
\end{align}
where $\omega_n=2\pi n/\tau$.
Using \eref{Fourier}, one can write \erefss{dynamics-mat-y} and \erefs{dynamics-mat-V} in the frequency domain as
\begin{align}
\tilde{y}(\omega_n)&=A^TG\ \tilde{\xi}(\omega_n)-\dfrac{1}{\tau}\bigg[(\gamma_3+im\omega_n)(G_{22}-G_{12})\Delta y+mA^TG\Delta V\bigg],\label{ytilda}\\
\tilde{V}(\omega_n)&=i\omega_nG\  \tilde{\xi}(\omega_n)+\dfrac{G}{\tau}[kA\Delta y-im\omega_n\Delta V].\label{V-tilda}
\end{align}
In the above equations, $\Delta y=y(\tau)-y(0)$, $\Delta V=V(\tau)-V(0)$, and the Green's function matrix is $G(\omega_n)=[-m\omega_n^2+i\omega_n\Gamma+\Phi]^{-1}$, where $\Phi_{rl}=k(2\delta_{l,r}-1)$ with $\{l,r\}=\{1,2\}$. Therefore, we can write $\tilde{y}(\omega_n)$ and $\tilde{v}_A(\omega_n)$ as
\begin{align}
\tilde{y}(\omega_n)&=(G_{11}-G_{12})[\tilde{\eta}_1(\omega_n)+\tilde{\eta}_2(\omega_n)]+(G_{12}-G_{22})\tilde{\eta}_3-\dfrac{1}{\tau}\Delta U^Tq_1,\label{y-vA-tilde-1}\\
\tilde{v}_A(\omega_n)&=i\omega_n[G_{11}\{\tilde{\eta}_1(\omega_n)+\tilde{\eta}_2(\omega_n)\}+G_{12}\tilde{\eta}_3(\omega_n)]+\dfrac{1}{\tau}\Delta U^T q_2,
\label{y-vA-tilde-2}
\end{align}
where
\begin{align}
q_1^T&=[(\gamma_3+i\omega_n m)(G_{22}-G_{12}),m(G_{11}-G_{12}),m(G_{12}-G_{22})],\\
q_2^T&=[k(G_{11}-G_{12}),-im\omega_n G_{11},-im\omega_n G_{12}].
\end{align}
In the above equations, the matrix elements $G_{ij}=[G(\omega_n)]_{ij}$.

The row vector $U^T(\tau)=[y(\tau),V^T(\tau)]$ is given as
\begin{equation}
U^T(\tau)=\lim_{\epsilon\to 0}\sum_{n=-\infty}^{\infty} e^{-i\epsilon\omega_n}[\tilde{y}(\omega_n),\tilde{V}^T(\omega_n)].
\label{Utau}
\end{equation}
Substituting \erefss{y-vA-tilde-1} and \erefs{y-vA-tilde-2} in the above equation, we find that the terms 
\begin{align}
&\lim_{\epsilon\to 0}\sum_{n=-\infty}^{\infty}\dfrac{e^{-i\epsilon\omega_n}}{\tau}[(\gamma_3+i\omega_n m)(G_{22}-G_{12})\Delta y+mA^TG\Delta V],\nonumber\\
&\lim_{\epsilon\to 0}\sum_{n=-\infty}^{\infty}\dfrac{e^{-i\epsilon\omega_n}}{\tau}[kA^T\Delta y-im\omega_n\Delta V^T]G^T,\nonumber
\end{align}
go to zero. This is because in the limit of large $\tau$, we convert the summation into integration, and these terms have poles in the upper half of the complex $\omega$-plane. Therefore, using the calculus of residue, one can find that the contribution from these terms vanishes identically. This implies
\begin{equation}
U^T(\tau)=\lim_{\epsilon\to 0}\sum_{n=-\infty}^{\infty}e^{-i\epsilon\omega_n}[\{\tilde{\eta}_{1}(\omega_n)+\tilde{\eta}_2(\omega_n)\}q_3^T+\tilde{\eta}_3(\omega_n)q_4^T],
\label{Utau-final}
\end{equation}
where
\begin{align}
q_3^T&=(G_{11}-G_{12},i\omega_n G_{11},i\omega_n G_{12}),\\
q_4^T&=(G_{12}-G_{22},i\omega_n G_{12},i\omega_n G_{22}).
\end{align}
The mean and variance of $U(\tau)$ are
\begin{align}
\langle U(\tau)\rangle&=0,\label{mean-four}\\
\langle U(\tau)U^T(\tau)\rangle&=\dfrac{1}{\pi}\int_{-\infty}^\infty d\omega[(T_1\gamma_1+T_2\gamma_2)q_3q_3^\dagger+T_3\gamma_3q_4q_4^\dagger],
\label{var-four}
\end{align}
respectively.
In \eref{var-four}, we have used the definition of correlation function of noises in the frequency domain as given by  
\begin{equation}
\langle \tilde{\eta}_i(\omega)\tilde{\eta}_j(\omega^\prime)\rangle=\dfrac{2 T_i\gamma_i}{\tau}\delta(\omega+\omega^\prime)\delta_{ij}.
\end{equation}
From \eref{Utau-final}, it is clear that the steady state distribution of $U=(y,v_A,v_B)^T$ is Gaussian distribution whose mean and variance are given in \eref{mean-four} and \eref{var-four}, respectively:
\begin{equation}
P^{full}_{ss}(U)=\dfrac{1}{\sqrt{(2 \pi)^3 \det M}} \exp\bigg[-\dfrac{1}{2}U^TM^{-1}U\bigg],
\end{equation}
where $M_{ij}=\langle U(\tau)U^T(\tau)\rangle_{ij}$.
But from \eref{gen-EP}, we see that $\Delta S^A_{tot}$ depends on thermal noises quadratically. Therefore, the distribution of partial entropy production is non-Gaussian, and does not depend only on the mean and variance of it. Nevertheless, one can find the large but finite time distribution of it as shown in the following sections.


\section{Moment generating function}
\label{mgf}
Total entropy production for particle-A ($\Delta S_{tot}^A$) given in \eref{gen-EP}, is a stochastic quantity. Therefore, to find the probability distribution of it in the steady state, it is better to start with characteristic function or moment generating function defined as 
\begin{equation}
Z(\lambda)=\langle \exp[-\lambda \Delta S^A_{tot}]\rangle,
\label{MGF}
\end{equation}
where average is taken over set of all trajectories of noises. Rather than computing \eref{MGF} directly, we first calculate restricted characteristic function
\begin{equation}
Z(\lambda,U,\tau|U_0)=\big\langle\exp[-\lambda\Delta S^A_{tot}]\delta[U-U(\tau)]\big\rangle_{U,U_0},
\label{RMGF}
\end{equation}
where the angular brackets represent the average over trajectories starting from initial variable $U_0$ to final variable $U$. Substituting $\Delta S^A_{tot}$ from \eref{gen-EP} in \eref{RMGF}, we get
\begin{align}
Z(\lambda,U,\tau|U_0)=\exp\big[\lambda/2\big(mT^{-1}_2-H^{-1}\big)\big(U^T\Sigma U-U_0^T\Sigma U_0\big)\big]Z_W(\lambda,U,\tau|U_0),
\label{full-RMGF}
\end{align}
where $\Sigma_{ij}=\delta_{i,j}\delta_{2,j}$ with \{i,j\}=\{1,2,3\}, $W$ is 
\begin{equation}
W=\Delta \beta Q_1-\dfrac{\Pi k}{T_2}\int_0^\tau dt\ y(t)v_A(t),
\label{Wtot}
\end{equation}
and
\begin{equation}
Z_W(\lambda,U,\tau|U_0)=\big\langle e^{-\lambda W}\delta[U-U(\tau)]\big\rangle_{U,U_0}.
\end{equation}
Since the boundary terms do not contribute in the averaging process in \eref{full-RMGF}, we have taken them outside from the angular brackets.
The quantity $W$ given in \eref{Wtot}, is a functional of all trajectories. Therefore, the evolution of the restricted characteristic function $Z_{W}(\lambda,U,\tau|U_0)$ corresponding to it follows the differential equation 
\begin{equation}
\dfrac{\partial Z_W(\lambda,U,\tau|U_0)}{\partial \tau}=\mathcal{L}_\lambda Z_{W}(\lambda,U,\tau|U_0),
\label{FP-eqn}
\end{equation}
where the differential operator $\mathcal{L}_\lambda$ has the following form
\begin{align}
\mathcal{L}_\lambda&=\dfrac{1}{m}\sum_{i=A,B}\bigg[\dfrac{\partial \mathcal{H}}{\partial x_{i}}\dfrac{\partial}{\partial v_{i}}-\dfrac{\partial \mathcal{H}}{\partial v_{i}}\dfrac{\partial}{\partial x_{i}}\bigg]+\dfrac{\gamma_B v_{B}}{m}\dfrac{\partial}{\partial v_{B}}+\dfrac{\gamma_1T_1+\gamma_2T_2}{m^2}\dfrac{\partial^2}{\partial v_{A}^2}+\dfrac{\gamma_3T_3}{m^2}\dfrac{\partial^2}{\partial v_{B}^2}+\sum_{i=1}^{3}\dfrac{\gamma_i}{m}\nonumber\\
&+\lambda\bigg[\gamma_1\Delta \beta\bigg(v_A^2+\frac{T_1}{m}\bigg)+\dfrac{\Pi k y v_{A}}{T_2}\bigg]+\lambda^2\Delta \beta^2 v_{A}^2\gamma_1T_1+\dfrac{v_{_{A}}}{m}(\gamma_A+2\lambda\gamma_1T_1\Delta \beta)\dfrac{\partial}{\partial v_{A}}.
\end{align}
The differential equation \eref{FP-eqn} is subjected to the initial condition $Z_W(\lambda,U,0|U_0)=Z(\lambda,U,0|U_0)=\delta(U-U_0)$.

It is impossible to solve the differential equation given in \eref{FP-eqn}, to get full solution. Nevertheless, one can write the general solution of the differential equation as
\begin{equation}
Z_{W}(\lambda,U,\tau|U_0)=\sum_{n} e^{\tau \mu_n(\lambda)}\chi_n(U_0,\lambda)\Psi_n(U,\lambda).
\end{equation}
In the above equation, $\chi_n(U_0,\lambda)$ and $\Psi_n(U,\lambda)$ are the $n^{th}$  left and right eigenfunctions, respectively, corresponding to the eigenvalue $\mu_n(\lambda)$ of the differential operator $\mathcal{L}_\lambda$. These eigenfunctions satisfy orthonormality condition
\begin{equation}
\int \chi_n(U,\lambda)\Psi_m(U,\lambda)\ dU=\delta_{nm}.
\end{equation}

Computation of all of these eigenvalues and eigenfunctions is a non-trivial problem. In most of the physical situations, one is interested in the large time solution of the differential equation. In such cases, the solution of differential equation shown in \eref{FP-eqn}, is dominated by the largest eigenvalue of the full spectrum, i.e., $\mu(\lambda):=\max\{\mu_{n}(\lambda)\}$. Thus, for large time
\begin{equation}
Z_{W}(\lambda,U,\tau|U_0)=e^{\tau \mu(\lambda)}\chi(U_0,\lambda)\Psi(U,\lambda)+\dots,
\label{factor-Z}
\end{equation}
where $\mu(\lambda)$ is the largest eigenvalue of the differential operator $\mathcal{L}_\lambda$ and the corresponding left and right eigenfunctions are $\chi(U_0,\lambda)$  and $\Psi(U,\lambda)$, respectively. The steady state distribution one can obtain from the restricted characteristic function by substituting $\lambda=0$: $P^{full}_{ss}(U)=Z_W(0,U,\tau\to\infty|U_0)=\Psi(U,0)$. Note that $\mu(0)=0$ and $\chi(U_0,0)=1$. Even, it is impossible to evaluate the largest eigenvalues and these eigenfunctions from the differential equation. Nevertheless, there is a technique developed in \cite{kundu} with which one can find these eigenfunctions and eigenvalue exactly.

The quantity $W$ is non-linear in thermal noises. Therefore, we write it in the frequency domain using \eref{Fourier} and \eref{inverse-Fourier}, we get
\begin{align}
W=\dfrac{\tau}{2}\sum_{n=-\infty}^{\infty}\bigg[\Delta \beta I_{1n}-\dfrac{\Pi k}{T_2}I_{2n}\bigg],
\label{Wfactor}
\end{align}
where
\begin{align}
I_{1n}&=\tilde{\eta}_1(\omega_n)\tilde{v}_A(-\omega_{n})+\tilde{\eta}_1(-\omega_n)\tilde{v}_A(\omega_{n})-2 \gamma_1\tilde{v}_A(\omega_n)\tilde{v}_A(-\omega_{n}), \label{I1}\\
I_{2n}&=\tilde{y}(\omega_n)\tilde{v}_A(-\omega_{n})+\tilde{y}(-\omega_n)\tilde{v}_A(\omega_{n}).
\label{I2}
\end{align}
Substituting $\tilde{y}(\omega_n)$ and $\tilde{v}_A(\omega_n)$ from \eref{y-vA-tilde-1} and \eref{y-vA-tilde-2} in \eref{I1} and \eref{I2} yields
\begin{align}
I_{1n}&=i\omega_{n}\big[G_{11}(\tilde{\eta}_1+\tilde{\eta}_2)\tilde{\eta}^*_1+G_{12}\tilde{\eta}_3\tilde{\eta}^*_1-G^*_{11}\tilde{\eta}_1(\tilde{\eta}^*_1+\tilde{\eta}^*_2)-G^*_{12}\tilde{\eta}_1\tilde{\eta}^*_3\big]+\dfrac{\Delta U^Tq_2}{\tau}\tilde{\eta}^*_1\nonumber\\
&-2\gamma_1\omega^2_n\big[|G_{11}|^2(\tilde{\eta}_1+\tilde{\eta}_2)(\tilde{\eta}^*_1+\tilde{\eta}^*_2)+|G_{12}|^2\tilde{\eta}_3\tilde{\eta}^*_3+G_{11}G^*_{12}(\tilde{\eta}_1+\tilde{\eta}_2)\tilde{\eta}^*_3\nonumber\\
&+G_{12}G^*_{11}\tilde{\eta}_3(\tilde{\eta}^*_1+\tilde{\eta}^*_2)\big]-\dfrac{2\gamma_1}{\tau^2}\Delta U^Tq_2q^T_2\Delta U-2i\gamma_1\omega_n\dfrac{q_2^\dagger\Delta U}{\tau}[G_{11}(\tilde{\eta}_1+\tilde{\eta}_2)+G_{12}\tilde{\eta}_3]\nonumber\\
&+\dfrac{q_2^\dagger\Delta U}{\tau}\tilde{\eta}_{1}+2i\gamma_1\omega_n\dfrac{\Delta U^T q_2}{\tau}[G^*_{11}(\tilde{\eta}^*_1+\tilde{\eta}^*_2)+G^*_{12}\tilde{\eta}^*_3],
\end{align}
and
\begin{align}
I_{2n}&=i\omega_n(\tilde{\eta}_1+\tilde{\eta}_2)(\tilde{\eta}^*_1+\tilde{\eta}^*_2)(G_{12}G^*_{11}-G_{11}G^*_{12})+i\omega_n(\tilde{\eta}_1+\tilde{\eta}_2)\tilde{\eta}^*_3[G_{11}(G^*_{12}-G^*_{22})\nonumber\\
&-G^*_{12}(G_{11}-G_{12})]+i\omega_n\tilde{\eta}_3(\tilde{\eta}^*_1+\tilde{\eta}^*_2)[G_{12}(G^*_{11}-G^*_{12})-G^*_{11}(G_{12}-G_{22})]\nonumber\\
&+i\omega_n \eta_3\eta_3^*(G_{22}G^{*}_{12}-G_{12}G^*_{22})+\dfrac{q_2^\dagger \Delta U}{\tau}[(G_{11}-G_{12})(\tilde{\eta}_1+\tilde{\eta}_2)+(G_{12}-G_{22})\tilde{\eta}_3]\nonumber\\
&+\dfrac{\Delta U^Tq_2}{\tau}[(G^*_{11}-G^*_{12})(\tilde{\eta}^*_1+\tilde{\eta}^*_2)+(G^*_{12}-G^*_{22})\eta_3^*]-i\omega_n \dfrac{q_1^\dagger \Delta U}{\tau}[G_{11}(\tilde{\eta}_1+\tilde{\eta}_2)+G_{12}\tilde{\eta}_3]\nonumber\\
&+i\omega_n\dfrac{\Delta U^Tq_1}{\tau}[G^*_{11}(\tilde{\eta}^*_1+\tilde{\eta}^*_2)+G^*_{12}\tilde{\eta}^*_3]-\dfrac{\Delta U^T(q_1q_2^\dagger+q_2q_1^\dagger)\Delta U}{\tau^2}.
\end{align}
For convenience, in the above equations, we have written $G_{ij}=[G(\omega_n)]_{ij}$, $G^*_{ij}=[G(-\omega_n)]_{ij}$, $\tilde{\eta}_r=\tilde{\eta}_r(\omega_n)$, and $\tilde{\eta}_r^*=\tilde{\eta}_r(-\omega_n)$, where $r=1,2,3$.

Now, the restricted characteristic function for $W$ is given by
\begin{align}
Z_{W}(\lambda,U,\tau|U_0)&=\langle e^{-\lambda W}\delta[U-U(\tau)]\rangle_{U,U_0}=\int \dfrac{d^3\sigma}{(2\pi)^3}e^{i\sigma^T U}\langle e^{E(\tau)}\rangle_{U,U_0},
\label{ZW}
\end{align} 
where we have used the integral representation of Dirac delta function, and $E(\tau)$ is given by
\begin{equation}
E(\tau)=-\lambda W-i \sigma^T U(\tau).
\label{Etau}
\end{equation}
Using \erefs{Utau-final} and \erefss{Wfactor}, we write $E(\tau)$ as
\begin{align}
E(\tau)=\sum_{n=1}^{\infty}\big[-\lambda\tau\zeta_n^TC_n\zeta^*_n+\zeta_n^T\alpha_n+\alpha^T_{-n}\zeta_n^*+\dfrac{\lambda}{\tau}|f_n|^2\big]-\dfrac{\lambda \tau}{2}\zeta_0^TC_0\zeta_0+\zeta_0^T\alpha_0+\dfrac{\lambda}{2\tau}f_0^2,
\label{Etau-eqn}
\end{align}
where $C_n=\Delta \beta C^{\mathrm{I}}_{n}-\dfrac{\Pi k}{T_2}C^{\mathrm{II}}_n$, and the row vector containing noise components in the frequency domain is $\zeta_n^T=(\tilde{\eta}_1,\tilde{\eta}_2,\tilde{\eta}_3)$. Here the matrices $C^{\mathrm{I}}_n$ and $C_n^{\mathrm{II}}$ are
\begin{equation*}
C_n^{\mathrm{I}}=
\begin{bmatrix}
C^{\mathrm{I}}_{11}&&C^{\mathrm{I}}_{12}&&C^{\mathrm{I}}_{13}\\
C^{\mathrm{I}*}_{12}&&C^{\mathrm{I}}_{22}&&C^{\mathrm{I}}_{23}\\
C^{\mathrm{I}*}_{13}&&C^{\mathrm{I}*}_{23}&&C^{\mathrm{I}}_{33}
\end{bmatrix}
\hspace{0.5
cm}\text{and}\hspace{0.5cm}C_n^{\mathrm{II}}=
\begin{bmatrix}
C^{\mathrm{II}}_{11}&&C^{\mathrm{II}}_{12}&&C^{\mathrm{II}}_{13}\\
C^{\mathrm{II}}_{21}&&C^{\mathrm{II}}_{22}&&C^{\mathrm{II}}_{23}\\
C^{\mathrm{II}*}_{13}&&C^{\mathrm{II}*}_{23}&&C^{\mathrm{II}}_{33}
\end{bmatrix},
\end{equation*}
whose the matrix elements are
\begin{align*}
C^{\mathrm{I}}_{11}&=i\omega_n(G_{11}-G^*_{11})-2\gamma_1\omega_n^2|G_{11}|^2,\\
C^{\mathrm{I}}_{12}&=-i\omega_n G^*_{11}-2\gamma_1\omega_n^2|G_{11}|^2,\\
C^{\mathrm{I}}_{13}&=-i\omega_n G^*_{12}-2\gamma_1\omega_n^2G_{11}G^*_{12},\\
C^{\mathrm{I}}_{22}&=-2\gamma_1\omega_n^2|G_{11}|^2,\\
C^{\mathrm{I}}_{23}&=-2\gamma_1\omega_n^2G_{11}G^*_{12},\\
C^{\mathrm{I}}_{33}&=-2\gamma_1\omega_n^2|G_{12}|^2,\\
C^{\mathrm{II}}_{11}&=C^{\mathrm{II}}_{12}=C^{\mathrm{II}}_{21}=C^{\mathrm{II}}_{22}=i\omega_n[G_{12}G^*_{11}-G_{11}G^*_{12}],\\
C^{\mathrm{II}}_{13}&=C^{\mathrm{II}}_{23}=i\omega[G_{11}(G^*_{12}-G^*_{22})-G^*_{12}(G_{11}-G_{12})],\\
C^{\mathrm{II}}_{33}&=i\omega_n[G_{22}G^{*}_{12}-G_{12}G^*_{22}].
\end{align*}

The column vector $\alpha_n$ is
\begin{align}
\alpha_n=-\lambda\begin{bmatrix}
a_{11}^T\Delta U\\
a_{21}^T\Delta U\\
a_{31}^T\Delta U\\
\end{bmatrix}
-ie^{-i\epsilon\omega_n}\begin{bmatrix}
q_3^T\sigma\\
q_3^T\sigma\\
q_4^T\sigma
\end{bmatrix},
\end{align}
in which
\begin{align}
a_{11}^T&=\bigg[\Delta \beta(1-2i\gamma_1\omega_n G_{11})-\dfrac{\Pi k}{T_2}(G_{11}-G_{12})\bigg]q_{2}^\dagger+\dfrac{i\omega_n \Pi k}{T_2}G_{11}q_1^\dagger,\nonumber\\
a_{21}^T&=\bigg[-2i\gamma_1\omega_n \Delta \beta G_{11}-\dfrac{\Pi k}{T_2}(G_{11}-G_{12})\bigg]q_{2}^\dagger+\dfrac{i\omega_n \Pi k}{T_2}G_{11}q_1^\dagger,\nonumber\\
a_{31}^T&=\bigg[-2i\gamma_1\omega_n \Delta \beta G_{12}-\dfrac{\Pi k}{T_2}(G_{12}-G_{22})\bigg]q_{2}^\dagger+\dfrac{i\omega_n \Pi k}{T_2}G_{12}q_1^\dagger.\nonumber
\end{align}
In \eref{Etau-eqn}, $|f_n|^2$ is given by
\begin{equation*}
|f_n|^2=\Delta U^T\bigg[2\Delta \beta\gamma_1q_2q_2^\dagger-\dfrac{\Pi k}{T_2}(q_1q_2^\dagger+q_2q_1^\dagger)\bigg]\Delta U.
\end{equation*}
Therefore,
\begin{align}
\langle e^{E(\tau)}\rangle_{U,U_0}=\prod_{n=1}^{\infty}\bigg\langle \exp\bigg[-\lambda \tau\zeta_n^TC_n\zeta^*_n+\zeta_n^T\alpha_n+\alpha^T_{-n}\zeta^*_n+\dfrac{\lambda}{\tau}|f_n|^2\bigg]\bigg\rangle\nonumber\\
\times\bigg\langle \exp\bigg[-\dfrac{\lambda \tau}{2}\zeta_0^TC_0\zeta_0+\zeta_0^T\alpha_0+\dfrac{\lambda}{2\tau}f_0^2\bigg]\bigg\rangle,
\label{avg-E}
\end{align}
where the angular brackets represent the average over the noise distribution. For $n=0$, the average is over the distribution $P(\zeta_0)=(2\pi)^{-3/2}(\det{\Lambda})^{-1/2}\exp[-\frac{1}{2} \zeta_0^T\Lambda^{-1}\zeta_0]$ whereas for each $n\geq1$, average is over distribution $P(\zeta_n)=\pi^{-3}(\det{\Lambda})^{-1}\exp[-\zeta_n^T\Lambda^{-1}\zeta^*_n]$. The diagonal matrix $\Lambda=\mathrm{diag}\bigg(\dfrac{2T_1\gamma_1}{\tau},\dfrac{2T_2\gamma_2}{\tau},\dfrac{2T_3\gamma_3}{\tau}\bigg)$.
After some simplification, \eref{avg-E} becomes
\begin{align}
\langle e^{E(\tau)}\rangle_{U,U_0}=e^{\tau \mu(\lambda)}\exp\bigg[\dfrac{1}{2}\sum_{n=-\infty}^{\infty}\bigg(\alpha^T_{-n}\Omega^{-1}_n\alpha_n+\dfrac{\lambda |f_n|^2}{\tau}\bigg)\bigg].
\end{align}
In the large time limit ($\tau\to\infty$), we convert the summation in the above equation into integration. Therefore, we get
\begin{align}
\langle e^{E(\tau)}\rangle_{U,U_0}=e^{\tau \mu(\lambda)}\exp\bigg[-\dfrac{1}{2}\sigma^TH_{1}(\lambda)\sigma+i\Delta U^TH_2(\lambda)\sigma+\dfrac{1}{2}\Delta U^TH_3(\lambda)\Delta U\bigg],
\end{align}
where one can identify $\mu(\lambda)$, $H_1(\lambda)$, $H_2(\lambda)$, and $H_3(\lambda)$ as
\begin{align}
\mu(\lambda)&=-\dfrac{1}{4\pi}\int_{-\infty}^{\infty}d\omega\ \ln[\det(\Lambda\Omega)],\label{mu-lambda}\\
H_1(\lambda)&=\dfrac{\tau}{2\pi}\int_{-\infty}^{\infty}d\omega\ \rho^T\Omega^{-1}\phi,\label{H1-lambda}\\
H_2(\lambda)&=-\dfrac{\tau}{2\pi}\lim_{\epsilon\to0}\int_{-\infty}^{\infty}d\omega\ e^{-i\epsilon\omega}\ b_1^T\Omega^{-1}\phi,\label{H2-lambda}\\
H_{3}(\lambda)&=\dfrac{\tau}{2\pi}\int_{\infty}^{\infty}d\omega\bigg[b_1^T\Omega^{-1}b_2+\dfrac{\lambda}{\tau}\big\{2\Delta \beta\gamma_1q_2q_2^\dagger-\dfrac{\Pi k}{T_2}(q_1q_2^\dagger+q_2q_1^\dagger)\big\}\bigg],\label{H3-lambda}
\end{align}
with $\Omega=[\Lambda^{-1}+\lambda\tau C_n]$, $\rho^T=(q_3^*,q_3^*,q_4^*)$,
$b_1^T=-\lambda(b_{11},b_{12},b_{13})$, $b_2=-\lambda(a^T_{11},a^T_{21},a_{31}^T)^T$, and $\phi=(q_3,q_3,q_4)^T$. The column vectors $b_{1j}$ are given as
\begin{align}
b_{11}&=q_2\bigg[\Delta \beta(1+2i\gamma_1\omega G^*_{11})-\dfrac{\Pi k}{T_2}(G^*_{11}-G^*_{12})\bigg]-\dfrac{i\omega \Pi k}{T_2} q_1G^*_{11},\nonumber\\
b_{12}&=q_2\bigg[2i\gamma_1\omega \Delta \beta G^*_{11}-\dfrac{\Pi k}{T_2}(G^*_{11}-G^*_{12})\bigg]-\dfrac{i\omega \Pi k}{T_2} q_1G^*_{11},\nonumber\\
b_{13}&=q_2\bigg[2i\gamma_1\omega \Delta \beta G^*_{12}-\dfrac{\Pi k}{T_2}(G^*_{12}-G^*_{22})\bigg]-\dfrac{i\omega \Pi k}{T_2} G^*_{12}q_1.\nonumber
\end{align}
Therefore, the characteristic function $Z_{W}(\lambda,U,\tau|U_0)$ can be rewritten as
\begin{align}
Z_{W}(\lambda,U,\tau|U_0)&=\int \dfrac{d^3\sigma}{(2\pi)^3}e^{i\sigma^T U}\langle e^{E(\tau)}\rangle_{U,U_0}\nonumber\\
&=\dfrac{e^{\tau \mu(\lambda)}e^{\frac{1}{2}\Delta U^TH_3\Delta U}e^{-\frac{1}{2}(U^T+\Delta U^TH_2)H_1^{-1}(U+H_2^T\Delta U)}}{\sqrt{(2\pi)^3\det H_1(\lambda)}}.
\end{align}
Following \eref{factor-Z}, we can factorize the restricted characteristic function  $Z_{W}(\lambda,U,\tau|U_0)$ into left and right eigenfunctions. Consequently, the matrices $H_1(\lambda)$, $H_2(\lambda)$, $H_3(\lambda)$ satisfy the condition $H_3-H_2H_1^{-1}H^T_2-H^{-1}_1H^T_2=0$. Therefore, we write 
\begin{align}
Z_{W}(\lambda,U,\tau|U_0)=\dfrac{e^{\tau \mu(\lambda)}e^{-\frac{1}{2}U^T L_1(\lambda)U}e^{-\frac{1}{2}U_0^T L_2(\lambda)U_0}}{\sqrt{(2\pi)^3\det H_1(\lambda)}},
\end{align}
where the matrices $L_{1}(\lambda)=H_1^{-1}+H_1^{-1}H_2^{T}$ and $L_2=-H_1^{-1}H_2^{T}$.

Using \eref{full-RMGF}, we can write the characteristic function for total entropy production for particle-A
\begin{align}
Z(\lambda,U,\tau|U_0)=\dfrac{e^{\tau \mu(\lambda)}e^{-\frac{1}{2}U^T \tilde{L}_1(\lambda)U}e^{-\frac{1}{2}U_0^T \tilde{L}_2(\lambda)U_0}}{\sqrt{(2\pi)^3\det H_1(\lambda)}},
\end{align}
where the matrices $L_1(\lambda)$ and $L_2(\lambda)$ modify as
\begin{align}
\tilde{L}_{1}(\lambda)&=L_1(\lambda)-\lambda(mT^{-1}_2-H^{-1})\Sigma, \nonumber\\
\tilde{L}_{2}(\lambda)&=L_2(\lambda)+\lambda(mT^{-1}_2-H^{-1})\Sigma. \nonumber
\end{align}
Therefore, the characteristic function $Z(\lambda)$ for partial and apparent entropy production $\Delta S_{tot}^A$ can be obtained by integrating the restricted characteristic function $Z(\lambda,U,\tau|U_0)$ over the initial steady state distribution $P^{full}_{ss}(U_0)$ and the final variable $U$, we get
\begin{align}
Z(\lambda)=\int dU\int dU_0 P^{full}_{ss}(U_0)Z(\lambda,U,\tau|U_0)\approx g(\lambda)e^{\tau \mu(\lambda)}+\dots,
\label{char-function}
\end{align}
where the prefactor is given by
\begin{equation}
g(\lambda)=[\det H_1(0)\det H_{1}(\lambda)\det \tilde{L}_1(\lambda)\det[\tilde{L}_2(\lambda)+H^{-1}_1(0)]]^{-1/2}.
\label{g-lambda-exp}
\end{equation}

\subsection{Cumulant generating function}
The cumulant generating function $\mu(\lambda)$ given in \eref{mu-lambda}, has the following form 
\begin{equation}
\mu(\lambda)=-\dfrac{1}{4 \pi \tau_\gamma}\int_{-\infty}^\infty du\ \ln\bigg[1+\dfrac{h(u,\lambda)}{q(u)}\bigg],
\label{mu-integral}
\end{equation}
where 
\begin{align}
h(u,\lambda)&=4 \lambda(1-\lambda)\beta_{12}\big[\alpha_{12}(1-\beta_{12})^2\{u^4+u^2(\alpha_{13}^2-\delta)+\delta^2/4\}+\delta^2\alpha_{13}\beta_{13}(\beta_{12}+\Pi-1)/4\big]\nonumber\\
&-\lambda \delta^2 \alpha_{13}\big[(\beta_{12}+\beta_{13}\lambda \Pi-\beta_{13} \lambda)
(\beta_{12}+\Pi-1)+\alpha_{12}\beta_{12}\Pi(\beta_{12}-\beta_{13}+\beta_{13}\lambda\Pi)\big],\\
q(u)&=\beta_{12}^2\big[u^6+u^4\{(1+\alpha_{12})^2+\alpha_{13}^2-2\delta\}+u^2\{(1+\alpha_{12})^2-\delta\}(\alpha^2_{13}-\delta)\nonumber\\
&+\delta^2(1+\alpha_{12}+\alpha_{13})^2/4\big].
\end{align}
Here $\beta_{1j}=\dfrac{T_j}{T_1}$ and $\alpha_{1j}=\dfrac{\gamma_j}{\gamma_1}$ with $j=2,3$, the coupling parameter $\delta=\dfrac{2km}{\gamma^2_1}$, and $\tau_\gamma=m/\gamma_1$ is the viscous relaxation time. For simplicity, we assume $\alpha_{12}=\alpha_{13}=1$, i.e., $\gamma_1=\gamma_2=\gamma_3=\gamma$. Therefore, $h(u,\lambda)$ and $q(u)$ become
\begin{align}
h(u,\lambda)&=4\lambda(1-\lambda)\beta_{12}\big[(1-\beta_{12})^2\{u^4+u^2(1-\delta)+\delta^2/4\}+\delta^2\beta_{13}(\beta_{12}+\Pi-1)/4\big]\nonumber\\
&-\lambda \delta^2\big[(\beta_{12}+\beta_{13}\lambda\Pi-\beta_{13} \lambda)(\beta_{12}+\Pi-1)+\beta_{12}\Pi\{\beta_{12}-\beta_{13}+\beta_{13}\lambda\Pi\}\big],\\
q(u)&=\beta_{12}^2[u^6+u^4(5-2 \delta)+u^2(4-\delta)(1-\delta)+9\delta^2/4].
\end{align}
\section{Probability distribution function}
\label{PDF-sec}
The probability distribution function for the total entropy production $\Delta S_{tot}^A$ or any observable whose characteristic function $Z(\lambda)$ is of the form given by \eref{char-function}, is obtained by inverting it using inverse transformation
 \begin{align}
 P(\Delta S^A_{tot}=s\tau/\tau_{\gamma})=\int_{-i\infty}^{+i\infty}\dfrac{d\lambda}{2\pi i}Z(\lambda) e^{\lambda s\tau/\tau_\gamma}\approx\int_{-i\infty}^{+i\infty}\dfrac{d\lambda}{2\pi i}g(\lambda) e^{(\tau/\tau_\gamma) [\tilde{\mu}(\lambda)+\lambda s]},
 \label{PDF-integral}
 \end{align}
  where $\tilde{\mu}(\lambda)=\tau_\gamma \mu(\lambda)$ is the scaled cumulant generating function, and the contour of integration is taken along the direction of imaginary axis passing through the origin of the complex $\lambda$-plane.
 If both $\tilde{\mu}(\lambda)$ and $g(\lambda)$ are analytic functions of $\lambda$, then, for large time ($\tau\gg\tau_\gamma$), we can approximate the above integral using saddle-point method. Therefore, we get
 \begin{equation} 
 P(\Delta S^A_{tot}=s\tau/\tau_{\gamma})\approx\dfrac{g(\lambda^*)e^{(\tau/\tau_\gamma) h(\lambda^*)}}{\sqrt{2\pi(\tau/\tau_\gamma) |h^{\prime\prime}(\lambda^*)|}},
 \label{PDF}
 \end{equation}
 where the function $h(\lambda^*)=\tilde{\mu}(\lambda^*)+\lambda^* s$. In the above equation,
 \begin{equation}
h^{\prime\prime}(\lambda^*)=\dfrac{\partial^2 h(\lambda)}{\partial \lambda^2}\bigg|_{\lambda=\lambda^*(s)},
 \end{equation}
and the saddle point $\lambda^*(s)$ is calculated by solving the following equation
 \begin{equation}
 \dfrac{\partial\tilde{\mu}(\lambda)}{\partial \lambda}\bigg|_{\lambda=\lambda^*(s)}=-s.
 \label{saddle-point}
 \end{equation}
Now, assume that both $g(\lambda)$ and $\tilde{\mu}(\lambda)$ satisfy  Gallavotti-Cohen symmetry, i.e., $g(\lambda)=g(1-\lambda)$ and $\tilde{\mu}(\lambda)=\tilde{\mu}(1-\lambda)$ (condition I). Therefore, we can write the probability distribution function for negative entropy production as
\begin{align}
 P(\Delta S^A_{tot}=-s\tau/\tau_{\gamma})&\approx\int_{-i\infty}^{+i\infty}\dfrac{d\lambda}{2\pi i}g(1-\lambda) e^{(\tau/\tau_\gamma)[\tilde{\mu}(1-\lambda)+(1-\lambda)s-s]}\nonumber\\
 &\approx e^{-s \tau /\tau_\gamma}\int_{1-i\infty}^{1+i\infty}\dfrac{d\lambda}{2\pi i}g(\lambda) e^{(\tau/\tau_\gamma)[\tilde{\mu}(\lambda)+\lambda s]}.
 \label{neg-pdf}
\end{align}
If both $g(\lambda)$ and $\tilde{\mu}(\lambda)$ do not have singularities between $\lambda\in[0,1]$ (condition II), then, we can easily shift the contour of integration from $(1-i\infty,1+i\infty)$ to $(-i\infty,+i\infty)$ easily. Therefore, from \eref{PDF-integral} and \eref{neg-pdf}, we get
\begin{equation}
\dfrac{P(\Delta S^A_{tot}=s\tau/\tau_{\gamma})}{P(\Delta S^A_{tot}=-s\tau/\tau_{\gamma})}\approx e^{ s\tau/\tau_\gamma}.
\label{cal-FT}
\end{equation}
The equation \eref{cal-FT} is the fluctuation theorem for total entropy production $\Delta S^A_{tot}$ in steady state. If $g(\lambda)$ and $\tilde{\mu}(\lambda)$ satisfy both conditions I and II, then the corresponding observable (for instance, in our case, the observable is the total entropy production $\Delta S^A_{tot}$) will satisfy fluctuation theorem. Note that we have proved this result for large but finite time. 

In the following section, we discuss the result for the entropy production for a single Brownian particle connected to a thermal gradient.

\section{Single Brownian particle in contact with two heat baths ($\delta=0$)}
\label{BP}
Consider a single Brownian particle-A is in contact with two heat baths at temperatures $T_1$ and $T_2$ (see \fref{system-pics} with $\delta=0$). The Langevin equation is given as
\begin{equation}
m\dot{v}_A=-(\gamma_1+\gamma_2)v_A(t)+\eta_1(t)+\eta_{2}(t).
\end{equation}
Entropy production in both baths due to particle-A is given as
\begin{equation}
\Delta S_{med}=-\bigg[\dfrac{Q_1}{T_1}+\dfrac{Q_2}{T_2}\bigg]=\Delta \beta Q_1-\dfrac{m}{2T_2}[v_{A}^2(\tau)-v_{A}^2(0)],
\end{equation}
where $Q_1=\int_0^\tau dt\ [\eta_1-\gamma_1v_A]v_A(t)$, is the heat energy given by the bath of the temperature $T_1$ and dissipation constant $\gamma_1$ to the Brownian particle-A. Therefore, the restricted characteristic function for medium entropy production is given as
\begin{align}
Z(\lambda,v_{A},\tau|v_A(0))&=\exp\bigg[\frac{\lambda m}{2T_2}\{v^2_A-v^2_A(0)\}\bigg]\big\langle e^{-\lambda \Delta \beta Q_1}\delta[v_{A}-v_{A}(\tau)]\big\rangle_{v_A,v_A(0)}\nonumber\\
&=\exp\bigg[\frac{\lambda m}{2T_2}\{v^2_A-v^2_A(0)\}\bigg]\tilde{Z}(\lambda,v_{A},\tau|v_A(0)),
\end{align} 
where $\tilde{Z}(\lambda,v_{A},\tau|v_A(0))$ satisfies the following differential equation 
\begin{align}
\dfrac{\partial \tilde{Z}(\lambda,v_{A},\tau|v_{A}(0))}{\partial \tau}=\mathcal{L}_\lambda^A \tilde{Z}(\lambda,v_{A},\tau|v_{A}(0)).
\label{FP-BP}
\end{align}
In the above equation, the differential operator $\mathcal{L}_\lambda^A$ is given as
\begin{align}
\mathcal{L}_\lambda^A=\dfrac{\gamma_1T_1+\gamma_2T_2}{m^2}\dfrac{\partial^2}{\partial v_A^2}+\dfrac{v_A}{m}(\gamma_1+\gamma_2+2\lambda\Delta\beta\gamma_1T_1)\dfrac{\partial}{\partial v_{A}}
+\bigg[\dfrac{\gamma_1+\gamma_2}{m}\nonumber\\
+\lambda \Delta \beta\bigg(\gamma_1v_A^2+\dfrac{\gamma_1 T_1}{m}\bigg)+\lambda^2\Delta\beta^2v_A^2\gamma_1T_1\bigg].
\end{align}
The differential equation \eref{FP-BP} is subjected to initial condition $\tilde{Z}(\lambda,v_{A},\tau|v_{A}(0))=\delta[v_{A}-v_{A}(0)]$. One can solve this differential equation exactly \cite{Visco,Efficiency}. In the limit of large $\tau$, we get
\begin{align}
Z(\lambda,v_{A},\tau|v_A(0))&\approx e^{(\tau/t_\gamma) \mu_0(\lambda)}\sqrt{\dfrac{m(\gamma_1+\gamma_2)\nu(\lambda)}{2\pi(\gamma_1T_1+\gamma_2T_2)}}\nonumber\\
&\times\exp\bigg[-\dfrac{m(\gamma_1+\gamma_2)v_A^2}{4(\gamma_1T_1+\gamma_2T_2)}\{\nu(\lambda)-2\lambda+1\}\bigg]\nonumber\\
&\times\exp\bigg[-\dfrac{m(\gamma_1+\gamma_2)v_A^2(0)}{4(\gamma_1T_1+\gamma_2T_2)}\{\nu(\lambda)+2\lambda-1\}\bigg],
\end{align}
where $t_\gamma=2m/(\gamma_1+\gamma_2)$. Integrating the above equation over the initial steady state distribution $Z(0,v_A,\tau\to\infty|v_A(0))=\tilde{P}_{ss}(v_A(0))$ given in \eref{app-ss}, and final variables $v_A$, we get
\begin{align}
Z(\lambda)\approx e^{(\tau/t_\gamma) \mu_0(\lambda)}g_0(\lambda),
\label{single-Z}
\end{align}
where
\begin{align}
\mu_0(\lambda)&=1-\nu(\lambda)\label{mu0-SP},\\
g_0(\lambda)&=\dfrac{2\sqrt{\nu(\lambda)}}{\sqrt{1+\nu(\lambda)-2\lambda}\sqrt{1+\nu(\lambda)+2\lambda}},
\label{mu-g-BP}
\end{align}
with
 \begin{equation}
\nu(\lambda)=\sqrt{1+4\alpha\lambda(1-\lambda)}=\sqrt{4\alpha(\lambda_+-\lambda)(\lambda-\lambda_-)}.
 \end{equation}
In the above equation, $\lambda_\pm=1/2[1\pm\sqrt{1+1/\alpha}]$ in which $\alpha=\gamma_1\gamma_2 T_1 T_2 \Delta \beta^2/(\gamma_1+\gamma_2)^2$. 
In \eref{mu-g-BP}, the first factor in the denominator comes from the integration over final variable $v_A$ while the second one comes from the integration over the initial steady state distribution $\tilde{P}_{ss}(v_A(0))$. Here, $\mu_0(\lambda)$ is analytic function when $\lambda\in(\lambda_-,\lambda_+)$. First denominator in $g_0(\lambda)$ has one branch point at $\lambda=\lambda_a=1$ for all $\alpha \in(0,\infty)$ where $\lambda_a \in(\lambda_-,\lambda_+)$ when $1-2\lambda_+<0$ while second denominator has a branch point at $\lambda=\lambda_b=(\alpha-1)/(\alpha+1)$ for $\alpha<1/3$ where $\lambda_b\in(\lambda_-,\lambda_+)$ when $1+2\lambda_-<0$. 

The probability distribution function of the medium entropy production $\Delta S_{med}$ can be obtained by inverting $Z(\lambda)$ given in \eref{single-Z}, using inverse transform [see \eref{PDF}]. Here, the saddle point $\lambda_0^*(s)$ is given by solving the following equation
\begin{equation}
\dfrac{\partial\mu_0(\lambda)}{\partial \lambda}\bigg|_{\lambda=\lambda^*_0(s)}=-s.
\end{equation}
 When $\alpha>1/3$, $g_0(\lambda)$ has only one singularity, i.e., at $\lambda=\lambda_a$. Therefore, the saddle point $\lambda_0^*(s)$ moves from $\lambda_-$ to $\lambda_a$ as $s$ decreases from $s=+\infty$ to $s_a$, and gets stuck at $\lambda=\lambda_a$, where $s_a$ is the solution of $\lambda^*_0(s_a)=\lambda_a$. Therefore, $P(\Delta S_{med}=s \tau/t_\gamma)\sim e^{(\tau/t_\gamma) [\mu_0(\lambda_-)+\lambda_- s] }$ as $s\to +\infty$.
Around $s=s_a$, i.e., $\lambda^*_0=\lambda_a-\epsilon_a$, where $0<\epsilon_a\ll1$, $P(\Delta S_{med}=s \tau/t_\gamma)\sim e^{(\tau/t_\gamma) [\mu_0(\lambda_a)+\lambda_as] }$. This implies for large s, we find \cite{SS-1,SS-2,Apal-1,Apal-2,Visco,Deepak,Trap}
\begin{align}
\lim_{(\tau/t_\gamma)\to\infty}\dfrac{t_\gamma}{\tau}\ln\dfrac{P(\Delta S_{med}=s\tau/t_\gamma)}{P(\Delta S_{med}=-s\tau/t_\gamma)}=\mu_0(\lambda_-)-\mu_0(\lambda_a)+[\lambda_-+\lambda_a]s.
\label{medium-EP}
\end{align}
Similarly, for $\alpha<1/3$, $g(\lambda)$ has both singularities, i.e., $\lambda_a$ and $\lambda_b$. In this case, \eref{medium-EP} modifies as
\begin{align}
\lim_{(\tau/t_\gamma)\to\infty}\dfrac{t_\gamma}{\tau}\ln\dfrac{P(\Delta S_{med}=s\tau/t_\gamma)}{P(\Delta S_{med}=-s\tau/t_\gamma)}=\mu_0(\lambda_b)-\mu_0(\lambda_a)+[\lambda_b+\lambda_a]s.
\end{align}
Therefore, we find that entropy production in the medium will not satisfy fluctuation theorem for large scaled parameter $s$ for any $\alpha\in(0,\infty)$.
But it is interesting to note that once we incorporate the entropy production of the system into entropy production in the medium, i.e., $\Delta S_{tot}=\Delta S_{med}+\Delta S_{sys}$, where $\Delta S_{sys}=-\ln \tilde{P}(v_{A}(\tau))+\ln \tilde{P}(v_{A}(0))$, in which $\tilde{P}_{ss}(v_A(\tau))$ is given in \eref{app-ss}, the prefactor term $g_0(\lambda)$ corresponding to $\Delta S_{tot}$ modifies to
\begin{equation}
g_0(\lambda)=\dfrac{2\sqrt{\nu(\lambda)}}{1+\nu(\lambda)},
\label{g0}
\end{equation}
and it is analytic function within the same domain as that of $\mu_0(\lambda)$. Therefore, both $\mu_0(\lambda)$ and $g_0(\lambda)$ satisfy condition I and II. Consequently, in this case, total entropy production satisfies fluctuation theorem for all $\alpha$.

\section{Analysis for prefactor $g(\lambda)$ for $\delta\neq 0$}
\label{glambda}
In our problem, it is not possible to obtain the analytical expression for the prefactor term $g(\lambda)$ given in \eref{g-lambda-exp}, as it requires the computation of matrices $H_1(\lambda)$, $H_2(\lambda)$ and $H_3(\lambda)$ [see \erefss{H1-lambda}--\eref{H3-lambda}] which is a non-trivial problem.  Since we are interested in the weak coupling limit ($\delta\to 0$), we can write
\begin{equation} 
g(\lambda)\approx g_0(\lambda)+\delta^b g_{1}(\lambda), \quad \quad\text{where}\quad\quad b>0.
\end{equation}
The term $g_0(\lambda)$ is the same prefactor term as given in \eref{g0}. The term $g_1(\lambda)$ may have singularities, but in the limit $\delta \to 0$, we can approximate the correction term as \cite{Deepak,Trap}
\begin{equation}
g(\lambda)\approx g_{0}(\lambda).
\end{equation}
\section{Analysis for cumulant generating function $\mu(\lambda)$ for $\delta\neq 0$}
\label{mulambda}
Computation of the integral given in \eref{mu-integral}, is quite involved and not very illuminating to us. Therefore, we compute $\mu(\lambda)$  numerically for fix parameters $\delta$, $\beta_{12}$, and $\beta_{13}$ as a function of $\lambda$. The range of $\lambda$ is not arbitrary, it is given by the saddle point $\lambda^*(s)$ which is the solution of \eref{saddle-point}. As one varies the scaled parameter $s$, the saddle point $\lambda^*(s)$ moves on the real line between two end points $\lambda_\pm^\delta$, i.e., $\lambda^*(s\to\mp \infty)\to\lambda_\pm^\delta$. In the following, we identify these branch points singularities $\lambda_\pm^\delta$ present in $\mu(\lambda)$. We see that arguments of the logarithm of the integrand in \eref{mu-integral} are $q(u)$ and $[h(u,\lambda)+q(u)]$. The function 
\begin{equation}
q(u)=\beta_{12}^2\big[u^6+u^4(5-2 \delta)+u^2(4-\delta)(1-\delta)+9\delta^2/4\big]
\end{equation}
is positive for all real $u\in(-\infty,\infty)$. We write $[h(u,\lambda)+q(u)]$ as
\begin{equation}
h(u,\lambda)+q(u)=-p_1(u)\lambda^2+p_2(u)\lambda+q(u)=0,
\label{h+q}
\end{equation}  
where 
\begin{align}
p_1(u)=4\beta_{12}\big[(1-\beta_{12})^2\{u^4+u^2(1-\delta)+\delta^2/4\}+\delta^2\beta_{13}(\beta_{12}+\Pi-1)/4\big]\nonumber\\+\delta^2\beta_{13}[(\Pi-1)(\beta_{12}+\Pi-1)+\beta_{12}\Pi^2],
\end{align}
and 
\begin{align}
p_2(u)=4\beta_{12}\big[(1-\beta_{12})^2\{u^4+u^2(1-\delta)+\delta^2/4\}+\delta^2\beta_{13}(\beta_{12}+\Pi-1)/4\big]\nonumber\\
-\delta^2\beta_{12}[\beta_{12}+\Pi-1+\Pi(\beta_{12}-\beta_{13})].
\end{align}
The roots of the quadratic equation \eref{h+q} are given as
\begin{equation}
\lambda^\delta_{\pm}(u)=\dfrac{p_2(u)\pm\sqrt{p_2(u)^2+4p_1(u)q(u)}}{2p_1(u)}.
\label{roots}
\end{equation}
\begin{figure}
\begin{center}
    \includegraphics[width=0.4  \textwidth]{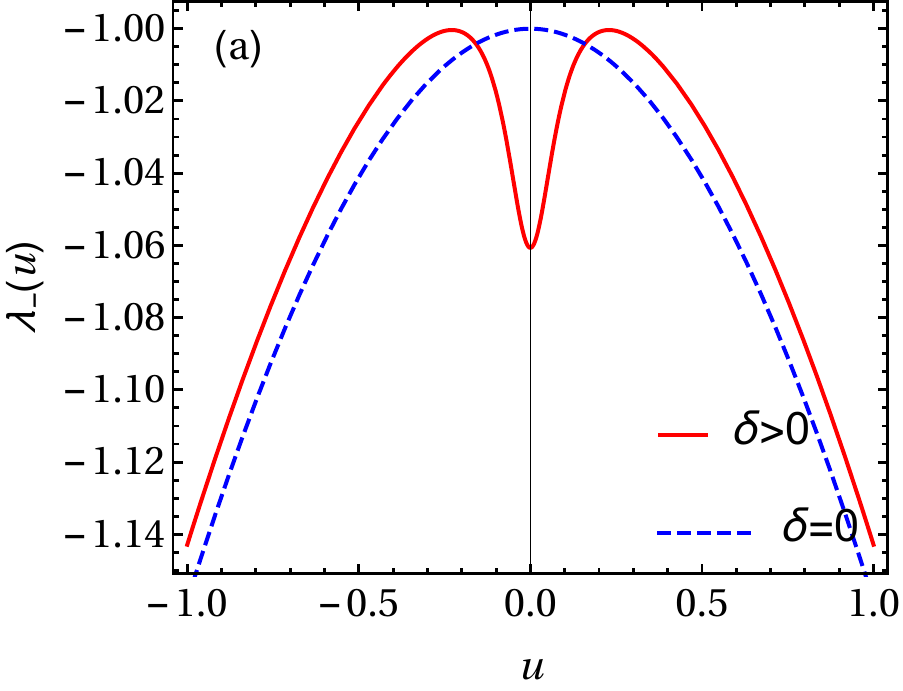}
    \includegraphics[width=0.4 \textwidth]{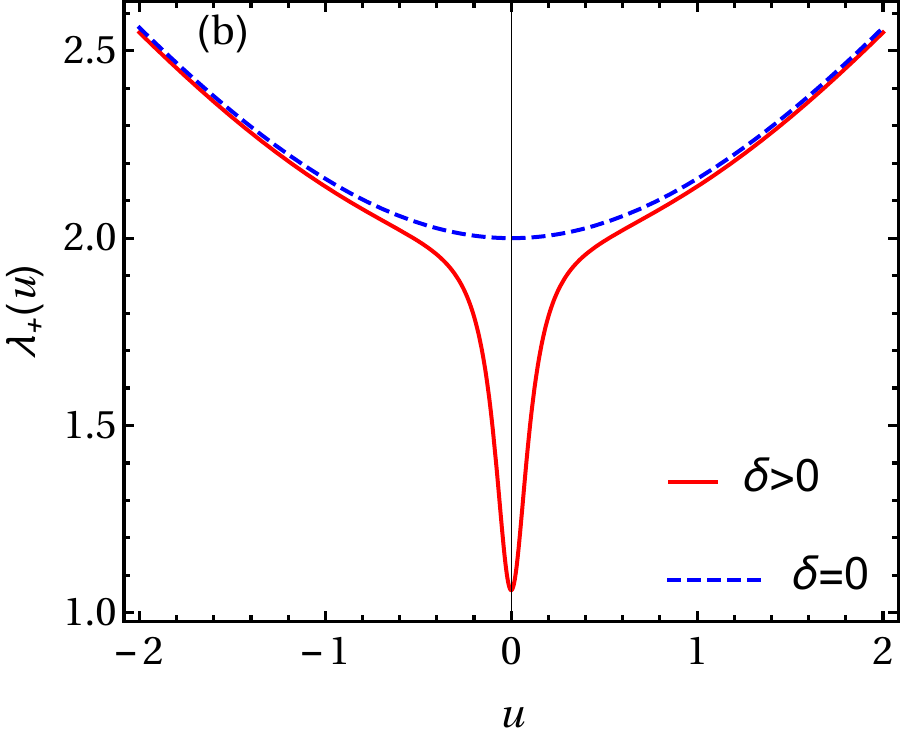}\\
     \includegraphics[width=0.4  \textwidth]{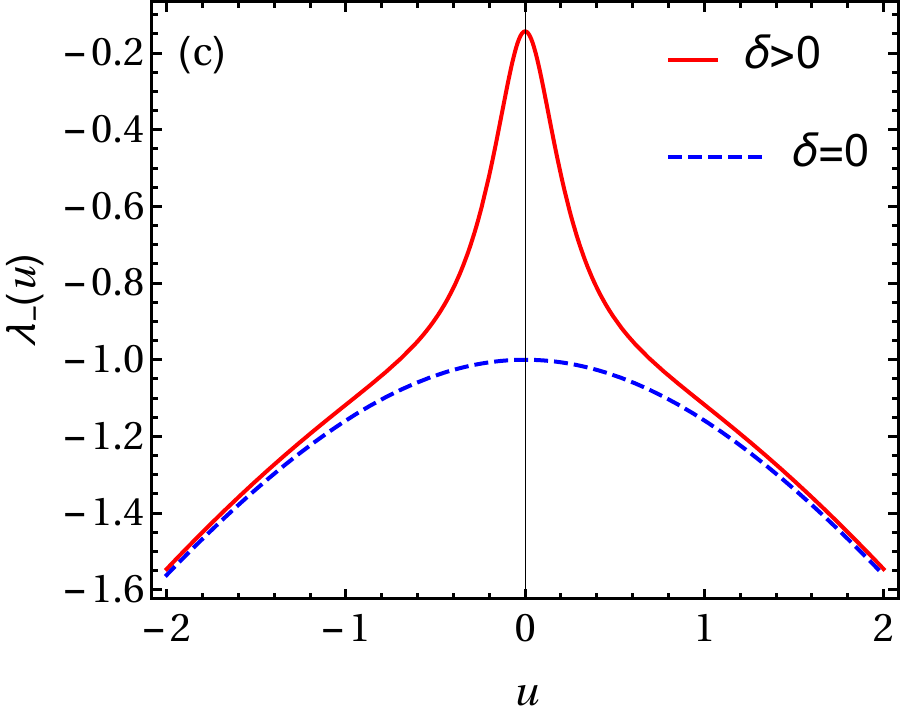}
    \includegraphics[width=0.4\textwidth]{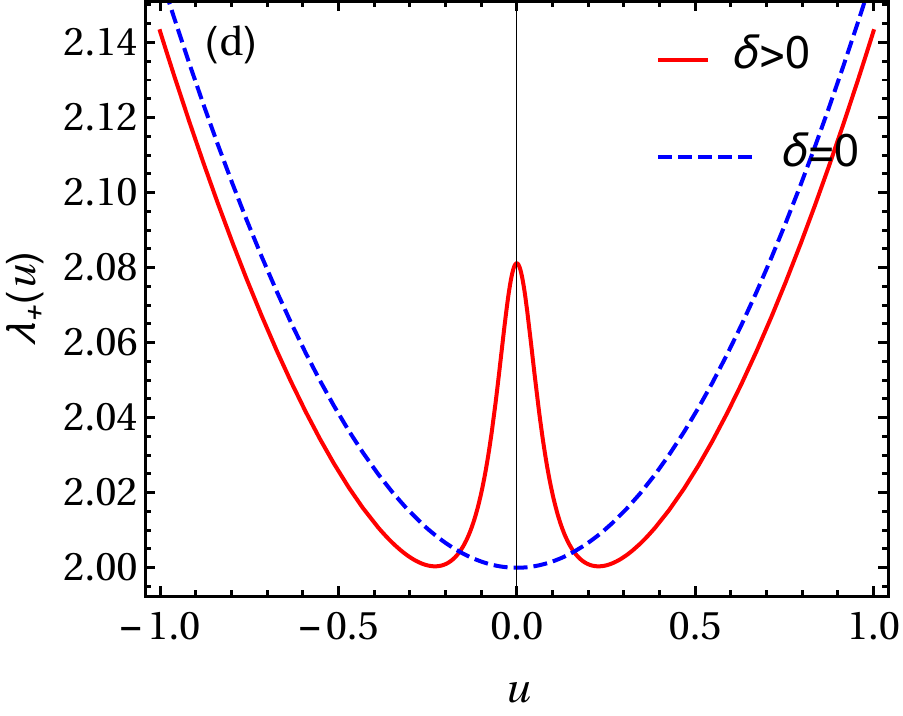}
 \caption{\label{lambda+-} The figures show the variations of $\lambda^\delta_{-}(u)$ and $\lambda^\delta_{+}(u)$ with respect to $u$ in which blue dashed lines correspond to $\delta$=0 (coupling is absent) whereas red solid lines correspond to $\delta>0$ case for fixed $\beta_{12}$, $\beta_{13}$, and $\Pi$.}
\end{center}
\end{figure}  
\ffref{lambda+-} shows the variation of $\lambda^\delta_\pm(u)$ with respect to $u$ for fixed $\beta_{12}$, $\beta_{13}$, $\delta$, and $\Pi$, in which red solid and blue dashed lines correspond to $\delta>0$ and $\delta=0$ case, respectively. It is clear from \fref{lambda+-}, both $\lambda^\delta_{\pm}(u)$ have either one extremum or two extrema depending upon the choice of parameters $\beta_{12}$ and $\beta_{13}$ for given $\Pi$ and $\delta$. We see that the curvature of the functions $\lambda^\delta_\pm(u)$ changes around $u=0$. Therefore, the equation
\begin{equation}
\lambda^{\prime\prime}_\pm(0)=\dfrac{\partial^2\lambda^\delta_\pm(u)}{\partial u^2}\bigg|_{u=0}
\label{phase-boundaries}
\end{equation}
decides the extrema of the function $\lambda^\delta_\pm(u)$ for given $\Pi$ and $\delta$.

For $\Pi=1$, i.e., partial entropy production
\begin{equation}
\dfrac{\partial^2\lambda^\delta_\pm(u)}{\partial u^2}\bigg|_{u=0}=\dfrac{2\beta_{12}(1-\delta)r_P^\pm(\beta_{12},\beta_{13},\delta)}{\delta^2x_1[(1-\beta_{12})^2+\beta_{13}(1+\beta_{12})]^2},
\label{curve-1}
\end{equation}
where 
\begin{align*}
r_P^\pm(\beta_{12},\beta_{13},\delta)=(1-\beta_{12})^2[4x_1\mp(1+6\beta_{12}+\beta_{12}^2)]
\pm\beta_{13}(1+\beta_{12})[3(1-\beta_{12})^2\nonumber\\+4\beta_{13}(1+\beta_{12})]\mp[(1-\beta_{12})^2+\beta_{13}(1+\beta_{12})]^2\delta.
\end{align*}
In equation \eref{curve-1}, we take $\delta<1$, and function $x_1$ is given by
\begin{align*}
x_1=\sqrt{(1+\beta_{12})(1+\beta_{12}+\beta_{13})[(\beta_{12}-1/2)^2+\beta_{13}(1+\beta_{12})+3/4]},
\end{align*}
where $x_1$ is a positive functions of $\beta_{12}$ and $\beta_{13}$.

\begin{figure}
\begin{center}
    \includegraphics[width=0.4  \textwidth]{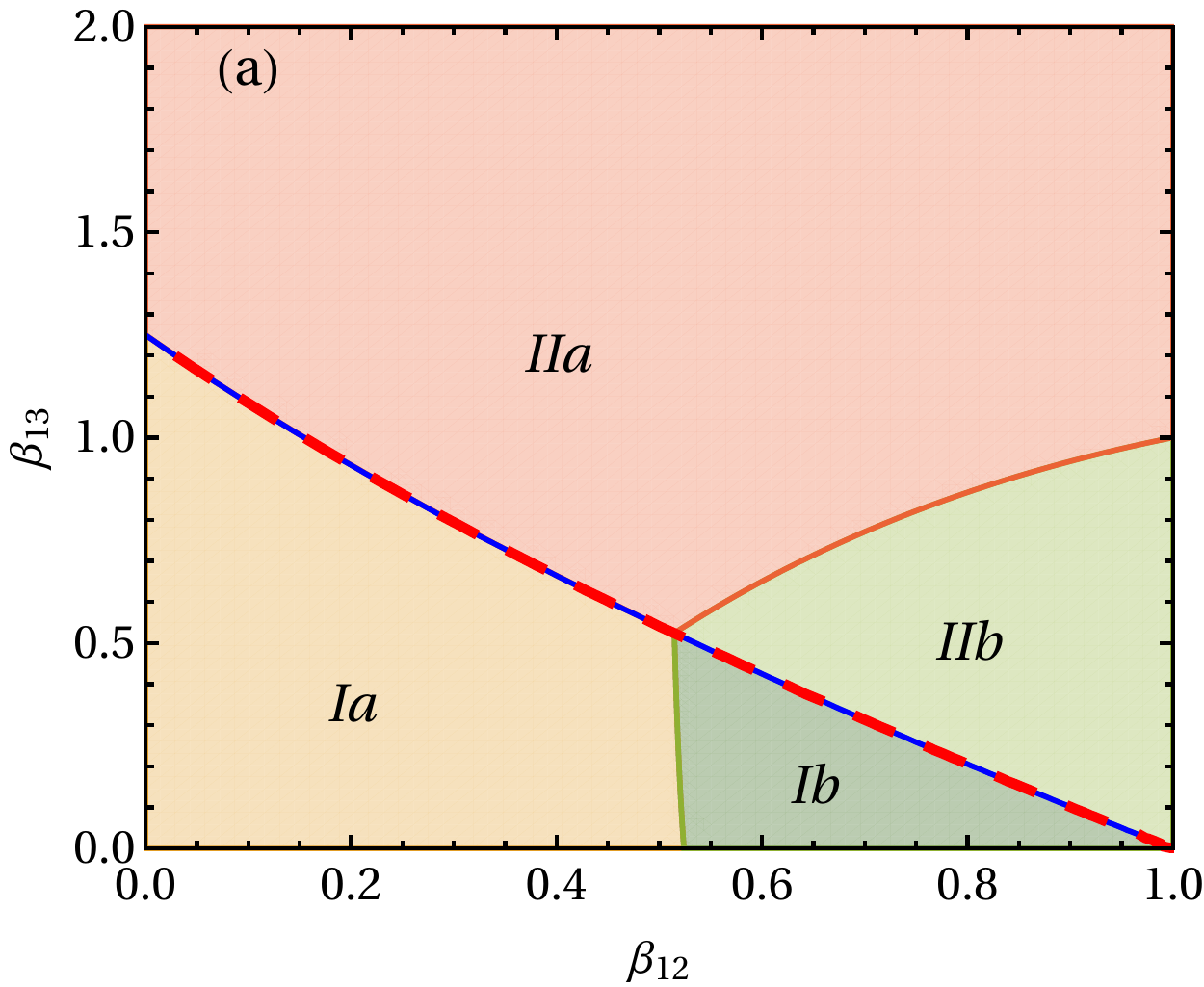}
    \includegraphics[width=0.39 \textwidth]{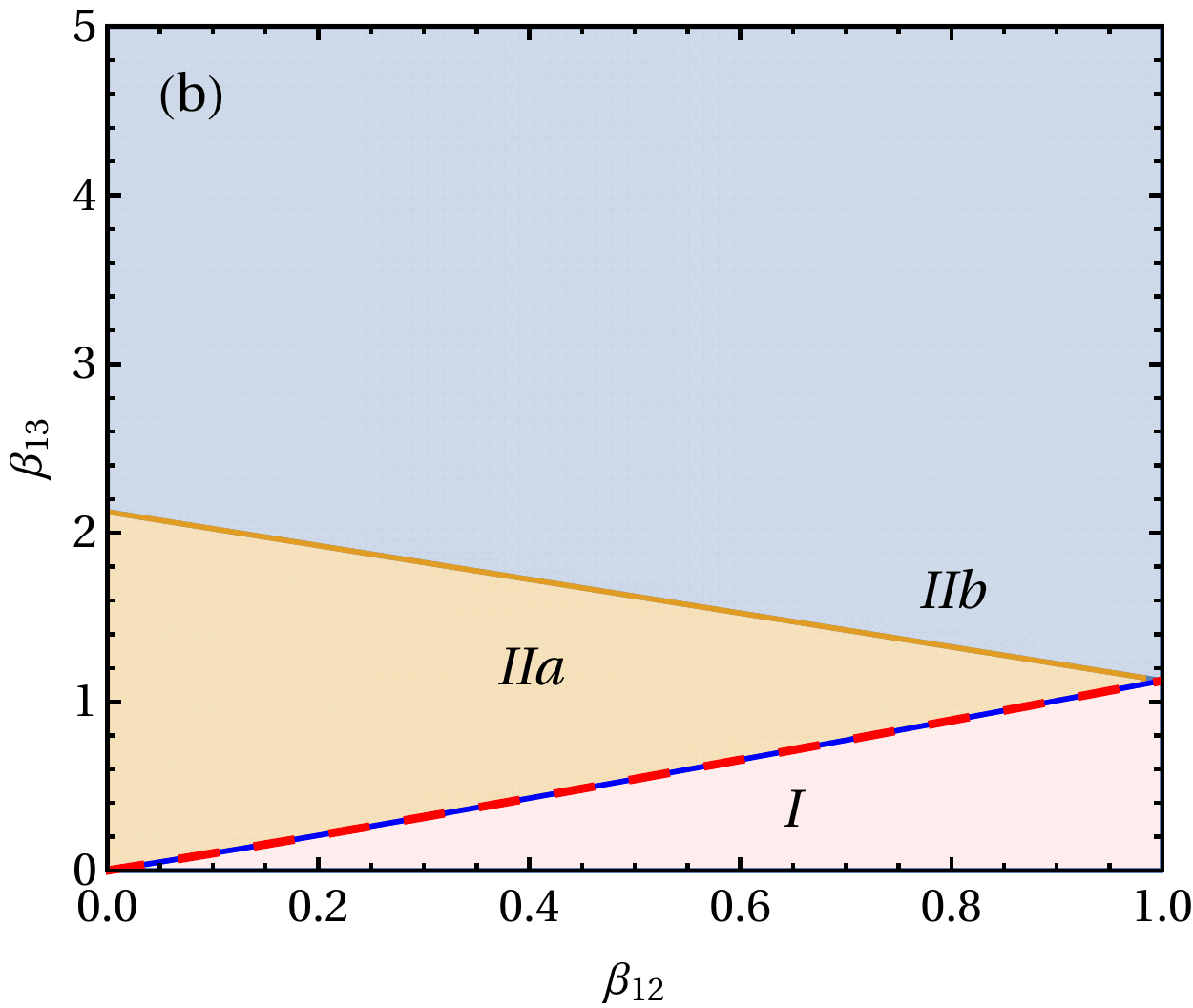}
 \bigskip
 \caption{\label{phase-diagram}
Phase diagrams for partial (a) and apparent (b) entropy production are shown. The red dashed contours correspond to \erefss{c-1} and \erefs{c-2} in figures (a) and (b), respectively, in the limit $\delta\to 0$. There are four subregions in figure (a). In subregion Ia, the function $f_1(\beta_{12},\beta_{13})>0$ whereas $f_1(\beta_{12},\beta_{13})<0$ in the subregion Ib. The function $f_2(\beta_{12},\beta_{13})>0$ in subregion IIa and $f_2(\beta_{12},\beta_{13})<0$ in subregion IIb. There are two subregions of region II in figure (b). Subregion IIa corresponds to $f_3(\beta_{12},\beta_{13})>0$ whereas $f_3(\beta_{12},\beta_{13})<0$ in subregion IIb.}
\end{center}
\end{figure}

While the function $r_P^+(\beta_{12},\beta_{13},\delta)>0$ for all $\beta_{12}$, $\beta_{13}$ at $\delta<1$ which indicates that the function $\lambda^\delta_+(u)$ has similar behavior as shown in \fref{lambda+-}(b),  the function $r_P^-(\beta_{12},\beta_{13},\delta)$ changes sign depending upon the choice of parameters $\beta_{12}$, $\beta_{13}$ at $\delta<1$. Therefore, the contour separating these two regions is given by
\begin{equation}
r_P^-(\beta_{12},\beta_{13},\delta)=0.
\label{c-1}
\end{equation}
We plot the phase diagram as shown in \fref{phase-diagram}(a) in the limit $\delta\to 0$, in which red dashed contour corresponds to above equation. Thus, $\lambda^\delta_-(u)$ has similar behaviours as shown in \frefss{lambda+-}(a) and \frefs{lambda+-}(c) for region I and II, respectively, of \fref{phase-diagram}(a).

For $\Pi=0$, i.e., apparent entropy production
\begin{equation}
\dfrac{\partial^2\lambda^\delta_\pm(u)}{\partial u^2}\bigg|_{u=0}=\dfrac{2\beta_{12}(1-\delta)r_A^\pm(\beta_{12},\beta_{13},\delta)}{\delta^2x_2(1-\beta_{12})(\beta_{12}+\beta_{13})^2},
\label{curve-2}
\end{equation}
where 
\begin{align*}
r_A^\pm(\beta_{12},\beta_{13},\delta)=\mp[(\beta_{12}-\beta_{13})(4\pm 2x_2)+\beta_{12}\beta_{13}(1+2\delta)+\beta^2_{12}(3+\delta)-\beta^2_{13}(2-\delta))],
\end{align*}
In \eref{curve-2}, we take $\delta<1$, and function $x_2$ is given by
\begin{equation}
x_2=\sqrt{(1+\beta_{12}+\beta_{13})(4+\beta_{12}+\beta_{13})},
\end{equation}
where $x_2$ is positive function of $\beta_{12}$ and $\beta_{13}$.

The function $r_A^-(\beta_{12},\beta_{13},\delta)>0$ for all $\beta_{12},\ \beta_{13}$ and $\delta<1$ which suggests that the function $\lambda^\delta_-(u)$ has variation with respect to $u$ as shown in \fref{lambda+-}(a). The function $r_A^+(\beta_{12},\beta_{13},\delta)$ can be either positive or negative depending upon the choice of $\beta_{12}$, $\beta_{13}$ at $\delta<1$. Therefore, the equation of contour separating these two regions is given as
\begin{equation}
r^+_{_A}(\beta_{12},\beta_{13},\delta)=0
\label{c-2}
\end{equation}
In the limit $\delta\to 0$, the phase diagram is shown in \fref{phase-diagram}(b) where red dashed contour is given by above equation. Thus, the root $\lambda_+^\delta(u)$ has the variations as shown in \frefss{lambda+-}(d) and \frefs{lambda+-}(b) in region I and II, respectively, of \fref{phase-diagram}(b). Note that in both phase diagram, the axis (not shown) corresponds to $\delta$ is perpendicular to the plane of paper.

The extrema of $\lambda^\delta_\pm(u)$ shown in \fref{lambda+-}, give the cut-off on the real line of the complex $\lambda$-plane within which the cumulant generating function $\mu(\lambda)$ is a real function. Note that the saddle point $\lambda^*(s)$ also lies within this domain. 
 The equation for extremum is given by
\begin{equation}
\dfrac{\partial \lambda^\delta_\pm (u)}{\partial u}\bigg|_{u=u^*_\pm}=0,
\label{extremum}
\end{equation}
where $u^*_\pm$ is the solution of the following equation
\begin{align}
[\sqrt{p_2^2(u^*_\pm)+4 p_1(u^*_\pm)q(u^*_\pm)}\pm p_2(u^*_\pm)][p_1(u^*_\pm)p^\prime_2(u^*_\pm)-p_2(u^*_\pm)p^\prime_1(u^*_\pm)]\nonumber\\
\pm 2p_1(u^*_\pm)[p_1(u^*_\pm)q^\prime(u^*_\pm)-q(u^*_\pm)p^\prime_1(u^*_\pm)]=0.
\label{u-critical}
\end{align}
In the above equation, $\prime$ represents the derivative with respect to $u$. In the weak coupling limit ($\delta \to 0$), we compute $u^*_\pm$ (upto leading order in $\delta$) using perturbation theory for both definitions of entropy production. 
In the case of partial entropy  production ($\Pi=1$)
\begin{equation*}
u^*_{-}=
\begin{cases}
\pm\sqrt{\delta}\dfrac{[5-3\beta_{12}^2-4\beta_{13}-2\beta_{12}(1+2\beta_{13})]^{1/4}}{\sqrt{2(1-\beta_{12}})}+o(\sqrt{\delta})\quad\text{for region I of \fref{phase-diagram}(a)},\\
0 \quad\quad \quad\quad \quad\quad \quad\quad \quad\quad \quad\quad \quad\quad \quad\quad \quad\quad \quad\quad\quad\hspace{0.2cm}\quad\text{for region II of \fref{phase-diagram}(a)}, 
\end{cases}
\end{equation*}
and $u^*_+=0$ for region I and II of \fref{phase-diagram}(a). Thus, $\lambda^\delta_-(u_-^*)\to \lambda_-$ in the region I whereas $\lambda^\delta_-(u_-^*) \to \tilde{\lambda}_-$ in the region II of \fref{phase-diagram}(a) in the limit of $\delta \to 0$. Similarly, $\lambda_+^\delta(u^*_+)\to\tilde{\lambda}_+$ for both regions I and II of \fref{phase-diagram}(a) in the limit $\delta \to 0.$

In the case of apparent entropy production ($\Pi=0$)
\begin{equation*}
u^*_{+}=
\begin{cases}
\pm\sqrt{\delta}\dfrac{[5\beta_{12}^2+4\beta_{12}(1-\beta_{13})-4\beta_{13}]^{1/4}}{\sqrt{2\beta_{12}}}+o(\sqrt{\delta})\quad\hspace{1.0cm}\text{for region I of \fref{phase-diagram}(b)},\\
0 \hspace{8.5cm} \quad \quad\text{for region II of \fref{phase-diagram}(b)},
\end{cases}
\end{equation*}
whereas $u^*_-=\sqrt{\delta/2}+o(\sqrt{\delta})$ for region I and II of \fref{phase-diagram}(b). 
In the weak coupling limit, $\lambda^\delta_+(u_+^*)\to \lambda_+$ in the region-I and $\lambda^\delta_+(u_+^*)\to \tilde{\lambda}_+$ in the region II of \fref{phase-diagram}(b). Similarly, $\lambda_-^\delta(u^*_-)\to\lambda_-$ for both regions I and II of \fref{phase-diagram}(b) in the limit $\delta \to 0.$

$\tilde{\lambda}_\pm$ for $\Pi=1$ are given by
\begin{equation}
\tilde{\lambda}_\pm=\dfrac{1+\beta_{13}+\beta_{12}(-4+\beta_{12}+\beta_{13})\pm x_1}{2[1+\beta_{13}+\beta_{12}(-2+\beta_{12}+\beta_{13})]},
\label{lambda-pmt-Pi=1}
\end{equation}
whereas $\tilde{\lambda}_+$ for $\Pi=0$ is
\begin{equation}
\tilde{\lambda}_+=\dfrac{\beta_{12}(2-\beta_{12}-\beta_{13}+x_2)}{2(1-\beta_{12})(\beta_{12}+\beta_{13})}.
\label{lambda-pmt-Pi=0}
\end{equation}
 One can find $\lambda_\pm$ in \sref{BP}.

In the \fref{phase-diagram}(a), region I has two subregions (Ia and Ib) which are decided by the function 
\begin{equation}
f_{1}(\beta_{12},\beta_{13})=\lambda_{-}+\tilde{\lambda}_{+},
\end{equation} 
where $\lambda_-$ and $\tilde{\lambda}_{+}$ are given in \sref{BP} and \eref{lambda-pmt-Pi=1}, respectively. In subregion Ia, the function $f_{1}(\beta_{12},\beta_{13})>0$ whereas in subregion Ib, it is less than zero. Similarly in \fref{phase-diagram}(a), region II has also two subregions (IIa and IIb) which are decided by the function 
\begin{equation}
f_{2}(\beta_{12},\beta_{13})=\tilde{\lambda}_{-}+\tilde{\lambda}_{+},
\end{equation}
where $\tilde{\lambda}_\pm$ are given in \eref{lambda-pmt-Pi=1}. The function $f_{2}(\beta_{12},\beta_{13})>0$ in the subregion IIa whereas in subregion IIb, it is negative. 

Similarly in the \fref{phase-diagram}(b), region II has two subregions (IIa and IIb) which are decided by the function
\begin{equation}
f_{3}(\beta_{12},\beta_{13})=\lambda_{-}+\tilde{\lambda}_{+}
\end{equation}
where $\lambda_-$ and $\tilde{\lambda}_{+}$ are given in \sref{BP} and \eref{lambda-pmt-Pi=0}, respectively. While the function $f_{3}(\beta_{12},\beta_{13})$ is positive in subregion IIa, it is negative in subregion IIb of \fref{phase-diagram}(b)  .
\subsection{Gallavotti-Cohen symmetry for cumulant generating function $\mu(\lambda)$}
\label{GC-sec}
\begin{figure}
\begin{center}
    \includegraphics[width=0.38  \textwidth]{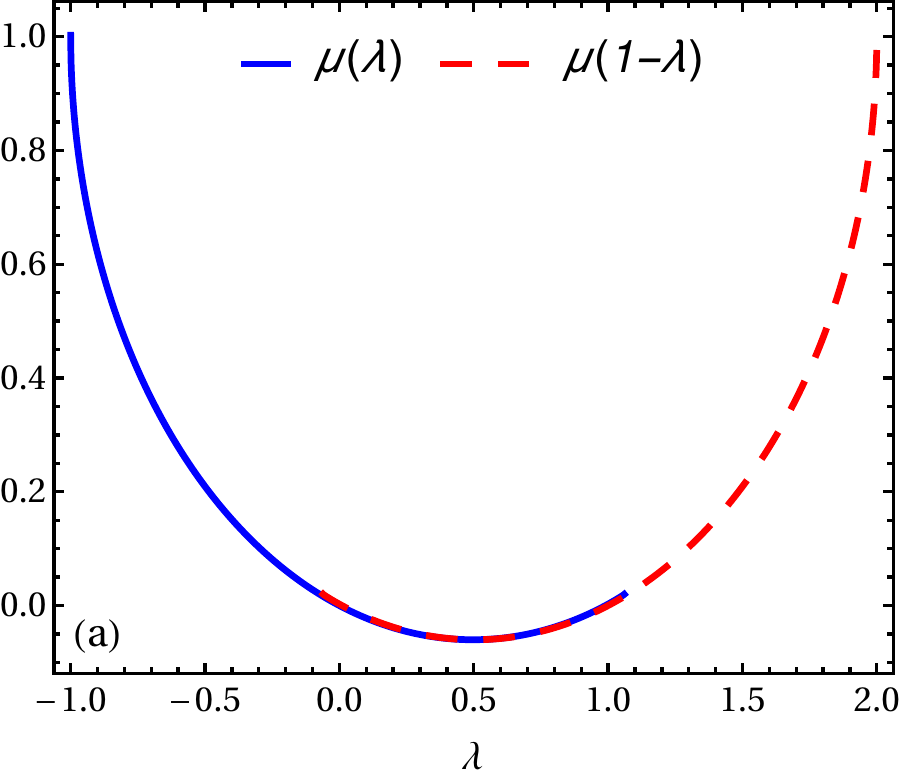}
    \includegraphics[width=0.4 \textwidth]{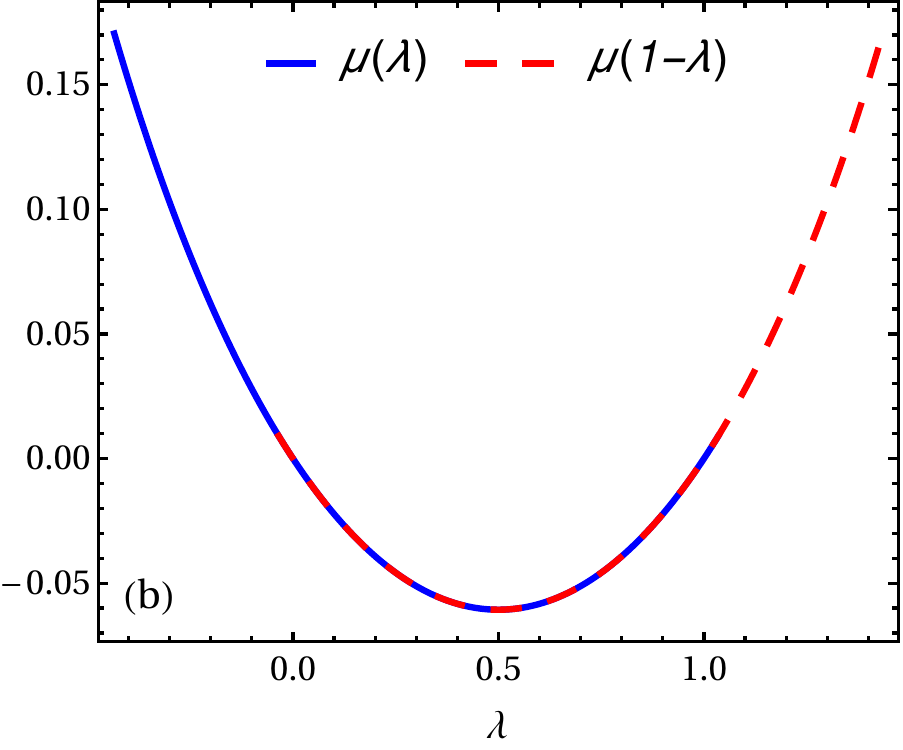}\\
     \includegraphics[width=0.4  \textwidth]{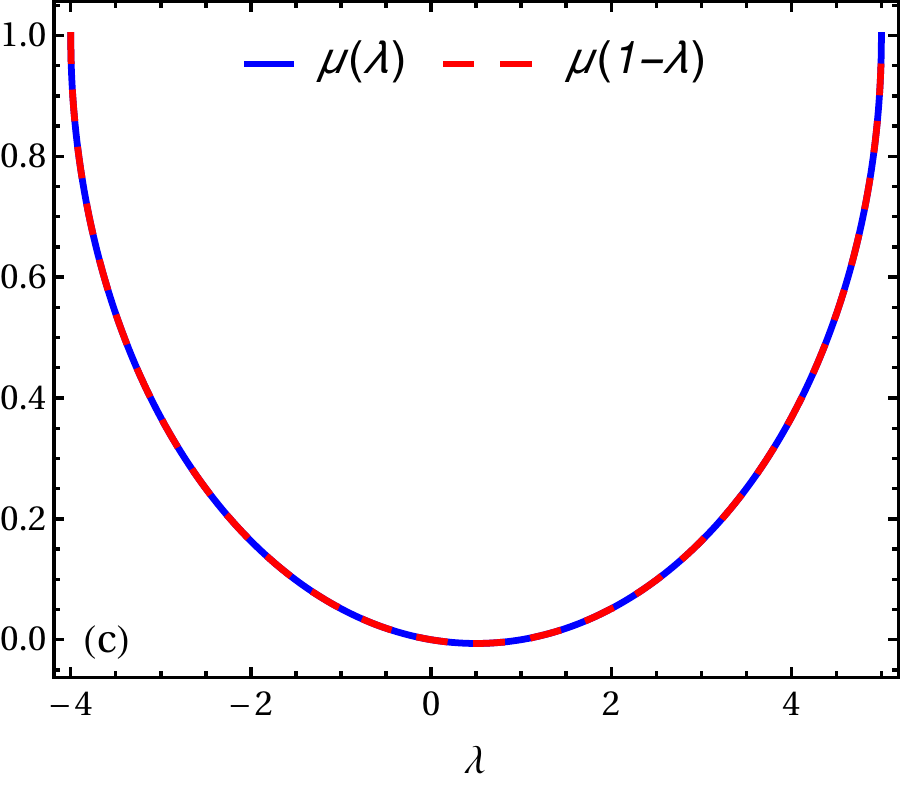}
    \includegraphics[width=0.4\textwidth]{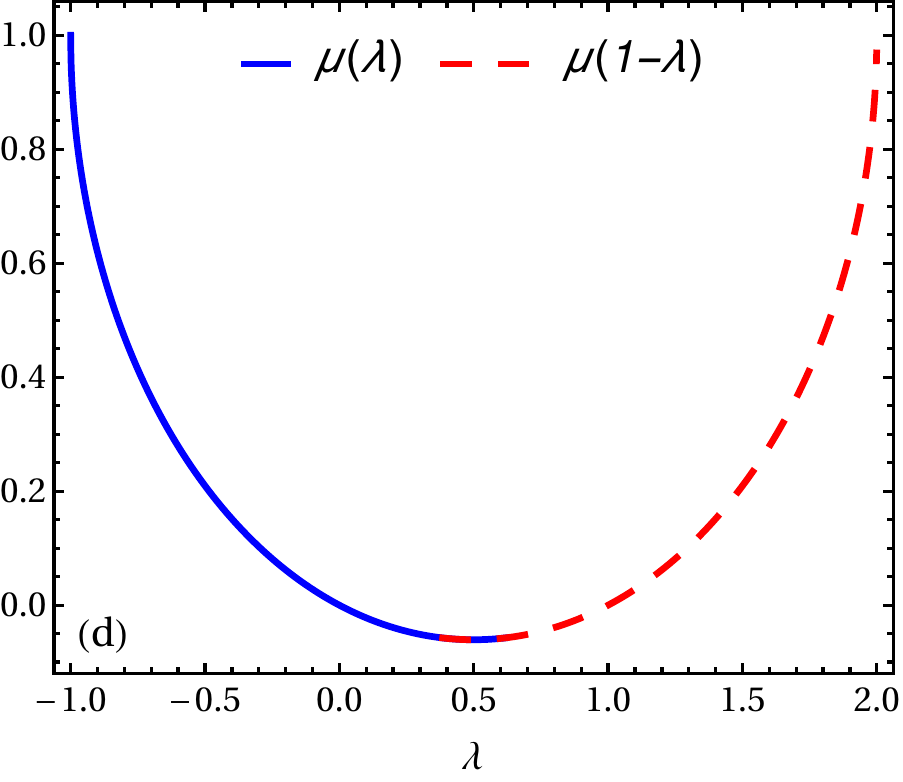}
 \caption{\label{GC-sym} The cumulant generating function $\mu(\lambda)$ (blue solid line) and $\mu(1-\lambda)$ (red dashed line) are plotted against $\lambda$ for: (a) region I of \fref{phase-diagram}(a), (b) region II of \fref{phase-diagram}(a), (c) region I of \fref{phase-diagram}(b), and (d) region II of \fref{phase-diagram}(b). All of the above figures are plotted for fixed coupling parameter $\delta=10^{-10}$. Except for region I of \fref{phase-diagram}(b), the cumulant generating function $\mu(\lambda)$ does not satisfy the Gallavotti-Cohen symmetry even in the limit $\delta\to0$. }
\end{center}
\end{figure}

It can be seen from \eref{mu-integral} that $\mu(\lambda)$ does not satisfy Gallavotti-Cohen symmetry, i.e., $\mu(\lambda)\neq \mu(1-\lambda)$ for large $\delta$, which is also a signature of partial measurement. In \fref{GC-sym}, we have plotted the cumulant generating function $\mu(\lambda)$ (blue solid line) and $\mu(1-\lambda)$ (red dashed line) against $\lambda$ for: (a) region I of \fref{phase-diagram}(a), (b) region II of \fref{phase-diagram}(a), (c) region I of \fref{phase-diagram}(b), and (d) region II of \fref{phase-diagram}(b). All of the these figures are plotted for fixed coupling parameter $\delta=10^{-10}$. Except for region I of \fref{phase-diagram}(b), the cumulant generating function does not satisfy the Gallavotti-Cohen symmetry for all $\lambda$. Therefore, we expect that the fluctuation theorem for apparent entropy production in the steady state, may hold in the region I of \fref{phase-diagram}(b) in the limit $\delta\to0$. 
%
%
\begin{figure*}
 \begin{center}
    \includegraphics[width=0.4  \textwidth]{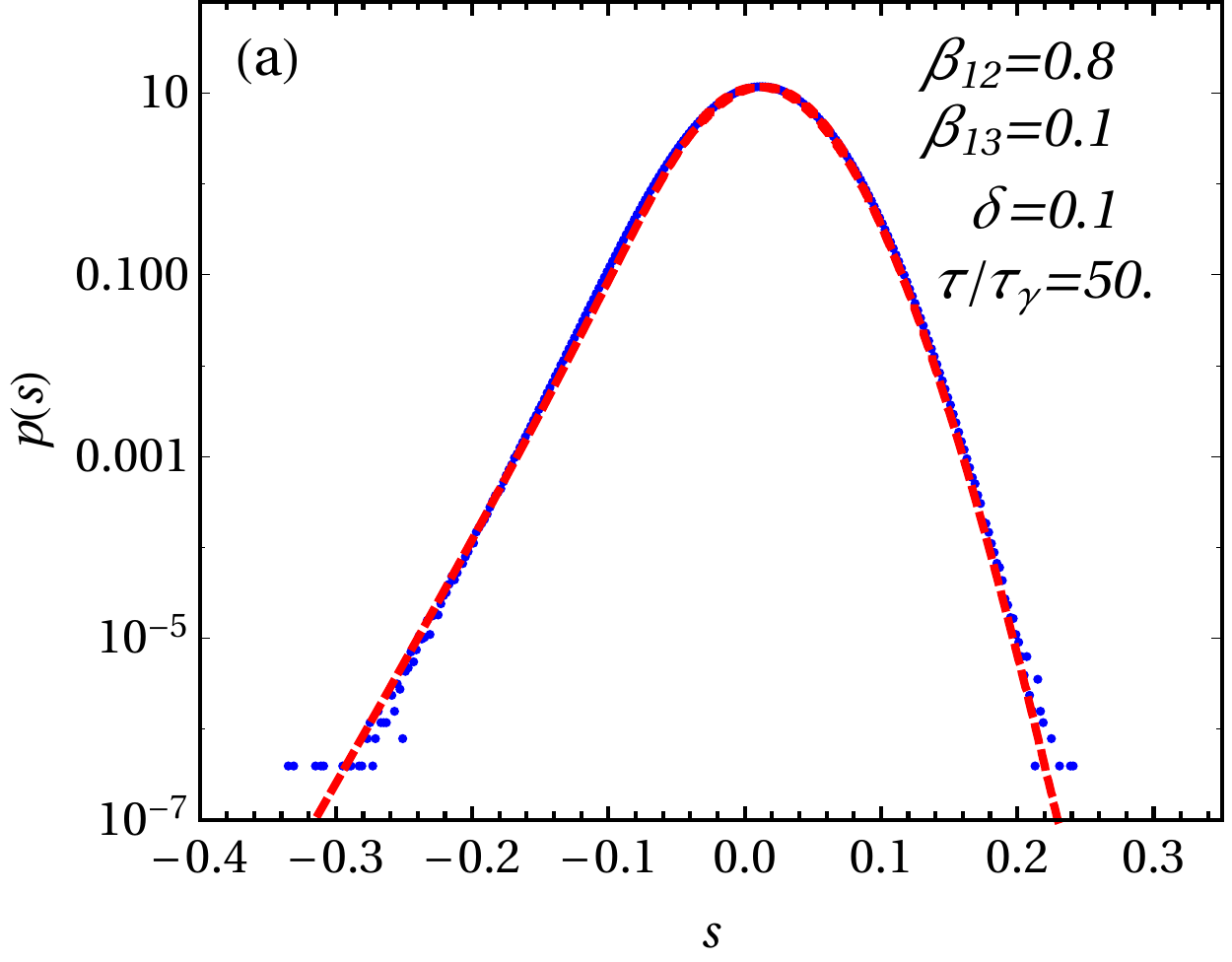}
    \includegraphics[width=0.4 \textwidth]{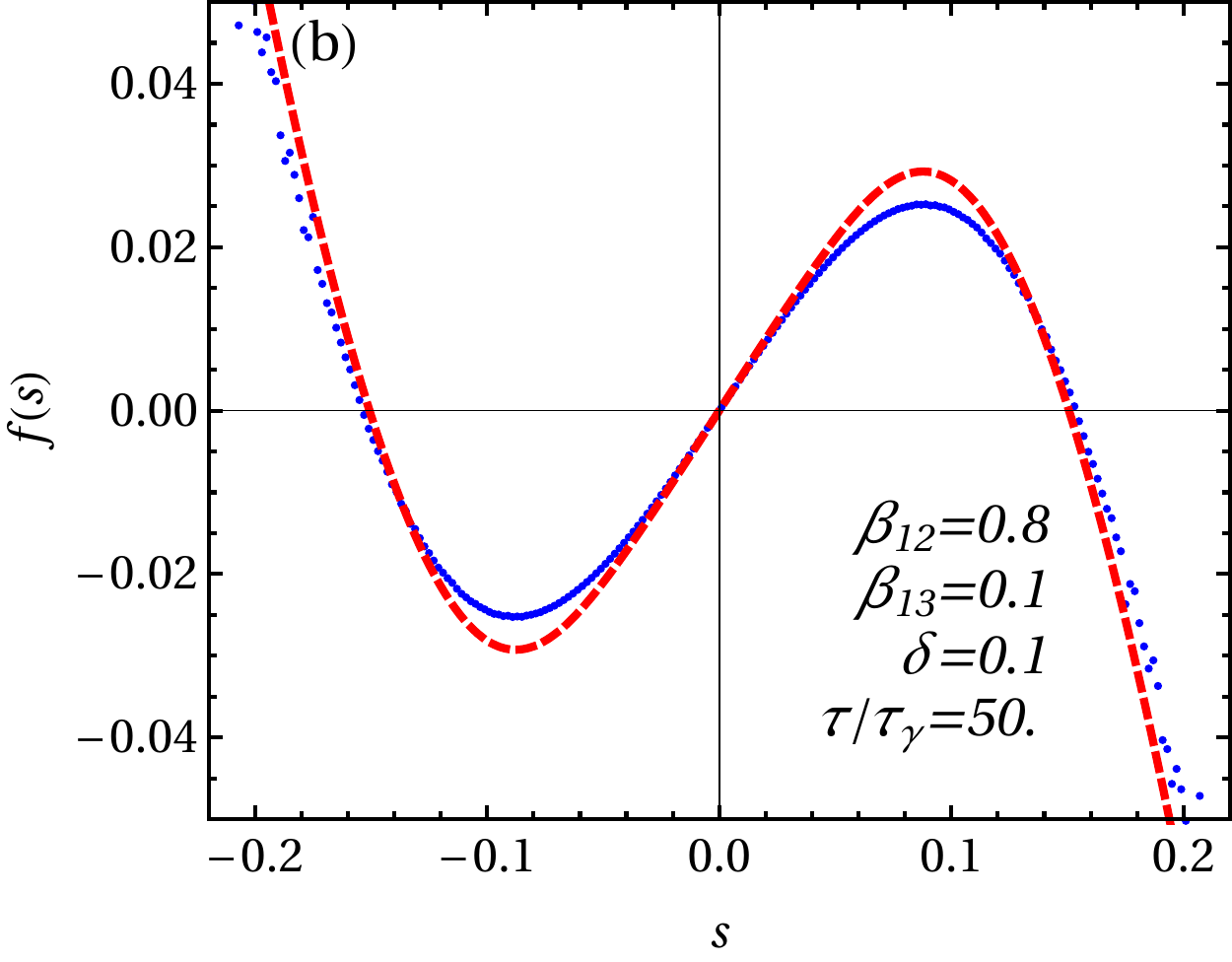}\\
    \includegraphics[width=0.4 \textwidth]{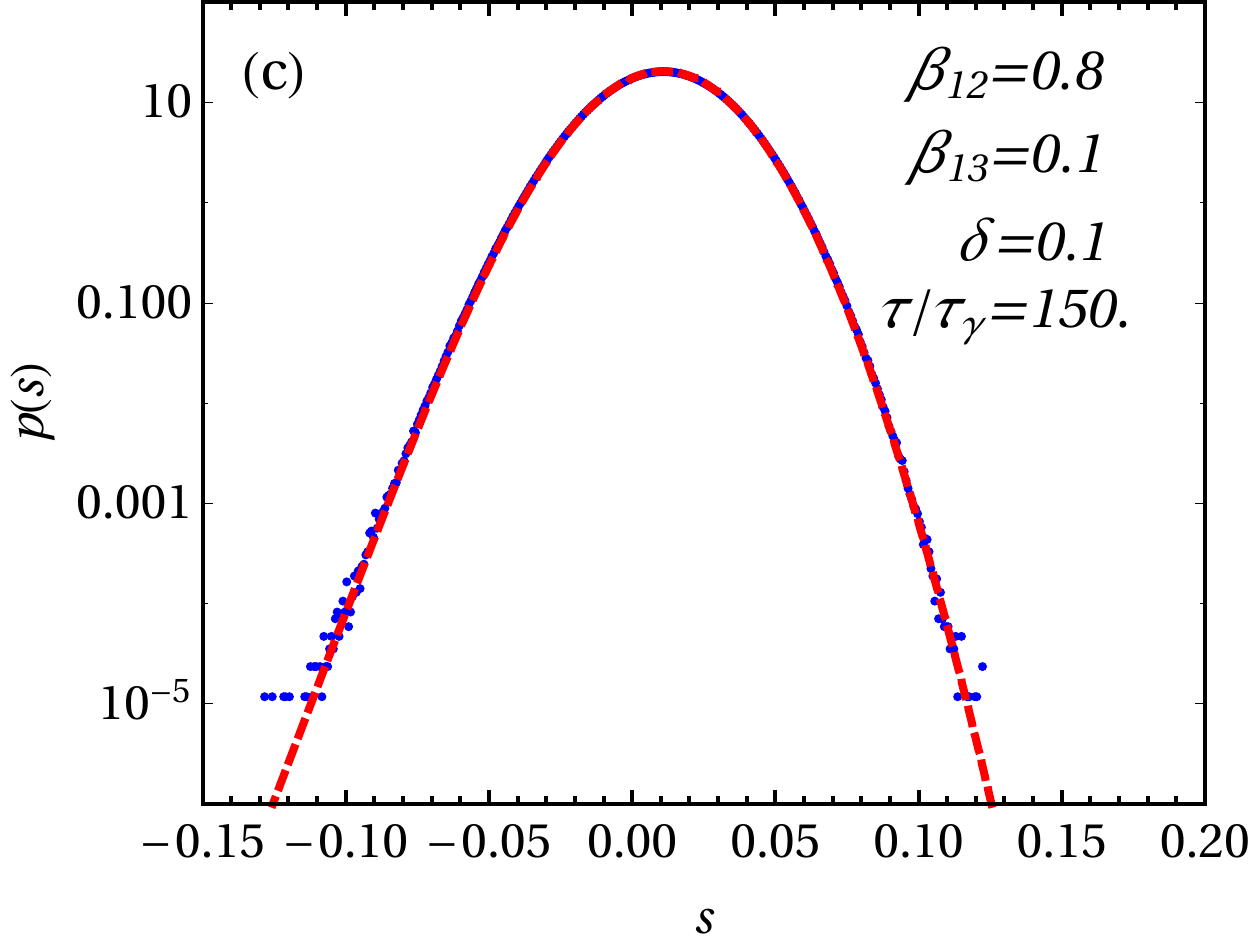}
    \includegraphics[width=0.4  \textwidth]{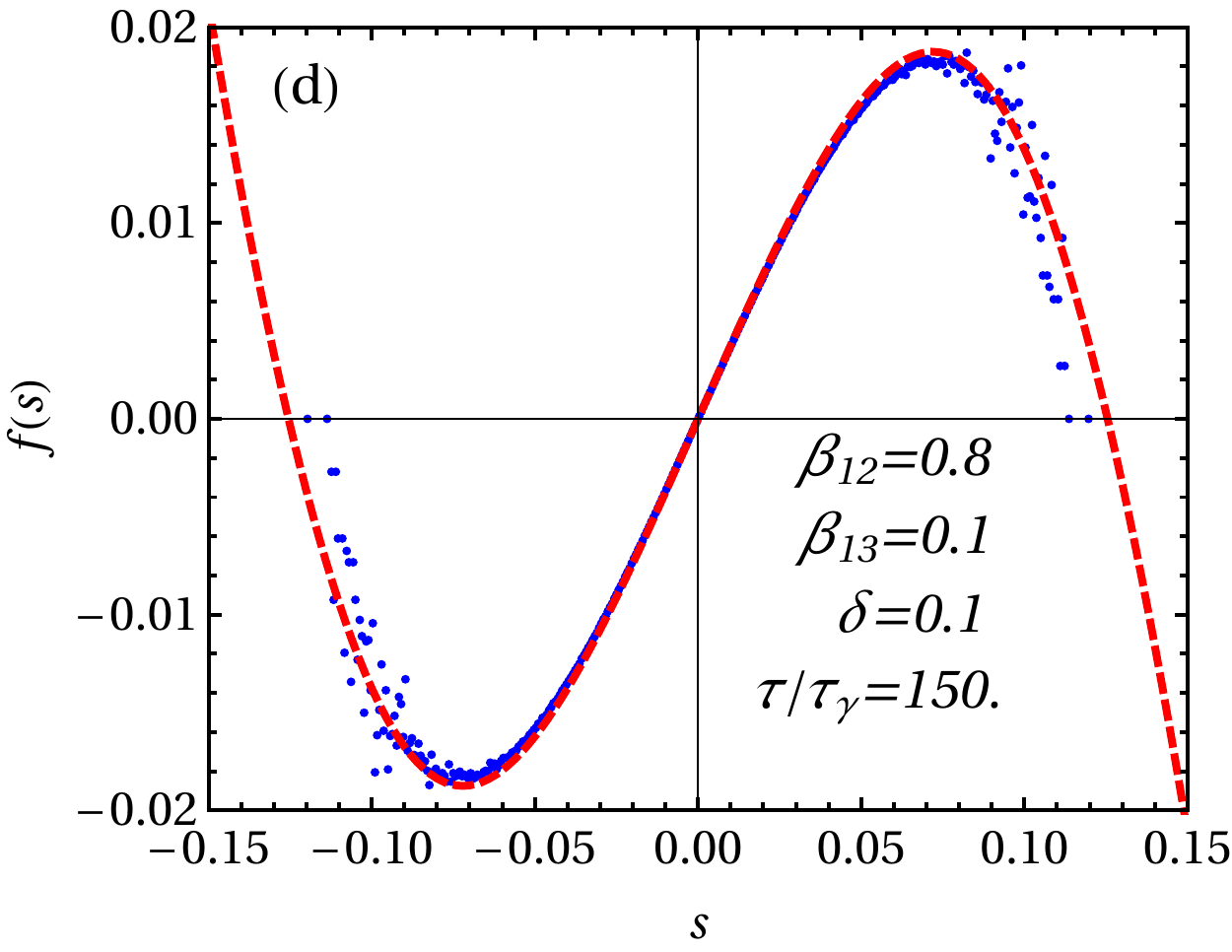}
 \caption{\label{time}A comparison of analytically obtained results (red dashed lines) of probability density function $p(s)$ and asymmetry function $f(s)$ given in \eref{pdf} and \eref{finite-time}, respectively, with the numerical simulations (blue dots) is shown for partial entropy production with time $\tau/\tau_\gamma=50.0$ [figures (a) and (b)] and $\tau/\tau_\gamma=150.0$ [figures (c) and (d)]. All of the above figures are shown for coupling strength $\delta=0.1$.}
 \end{center}
\end{figure*}  

\begin{figure*}
 \begin{tabular}{ccc}
    \includegraphics[width=0.33  \textwidth]{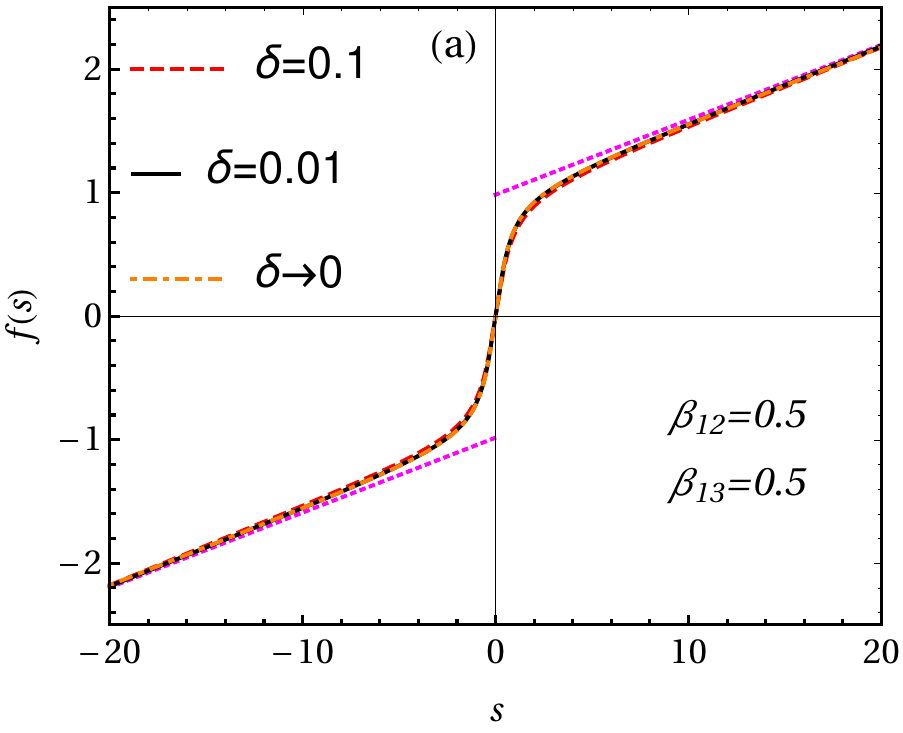}
    \includegraphics[width=0.34 \textwidth]{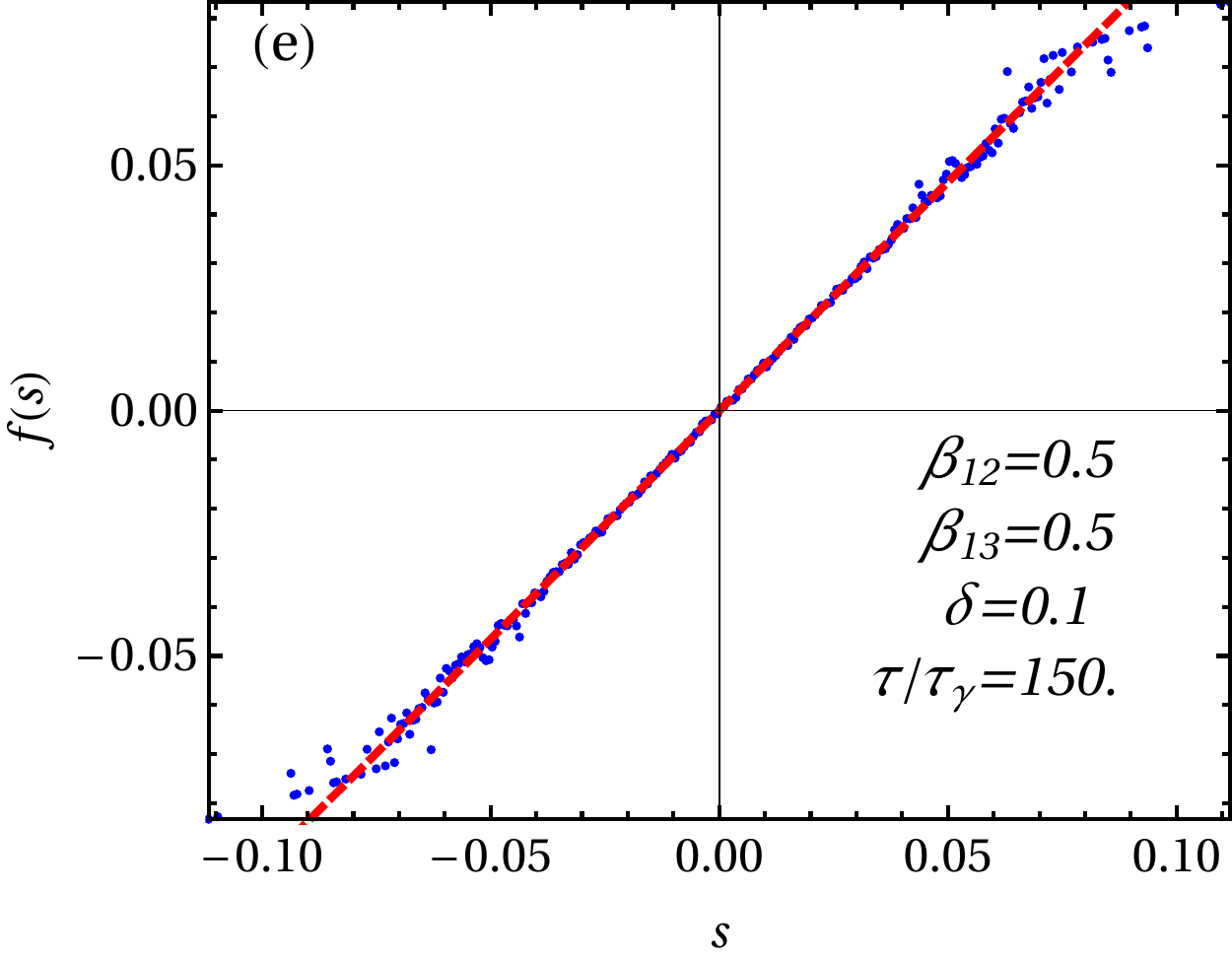}
    \includegraphics[width=0.33 \textwidth]{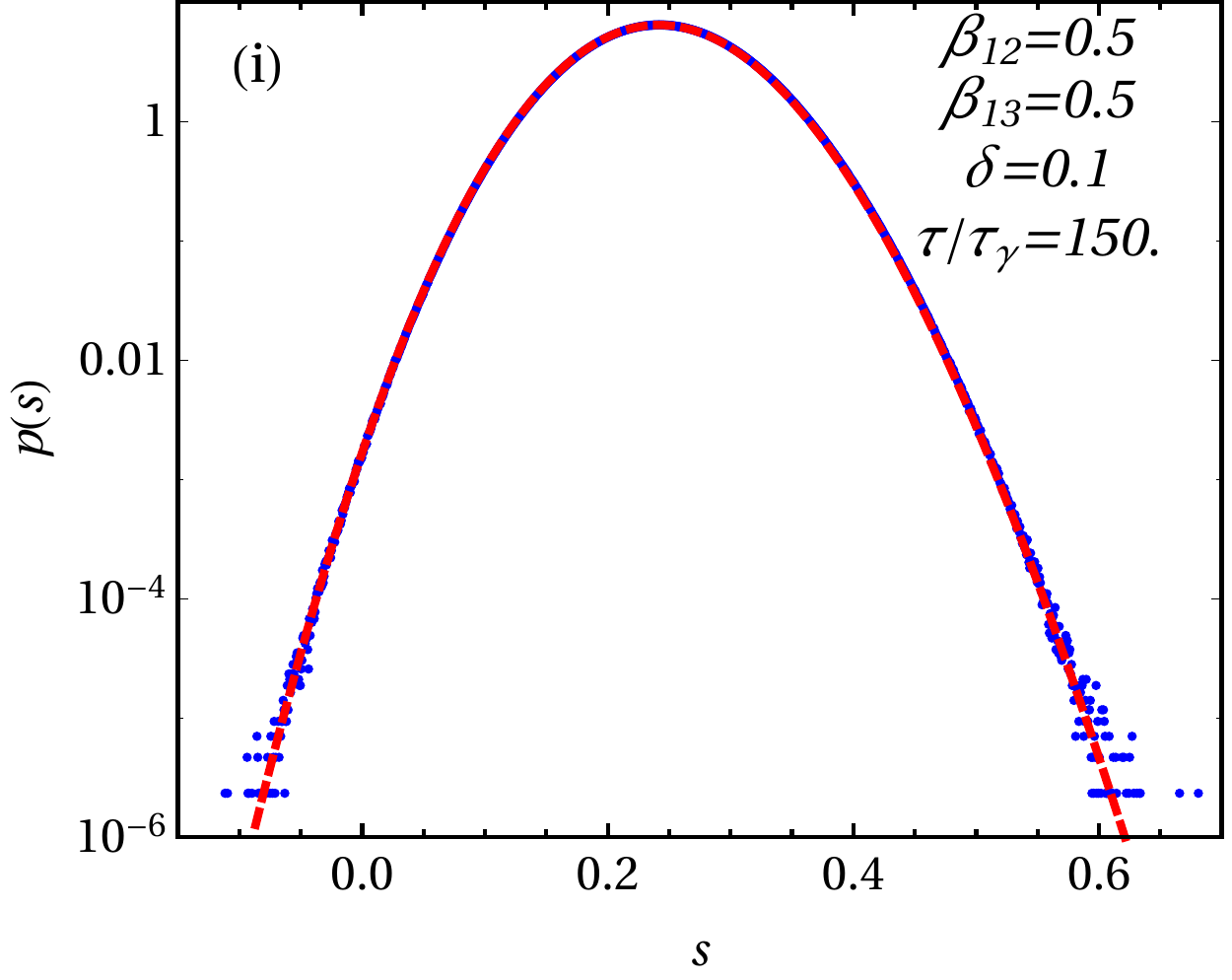}\\
    \includegraphics[width=0.33  \textwidth]{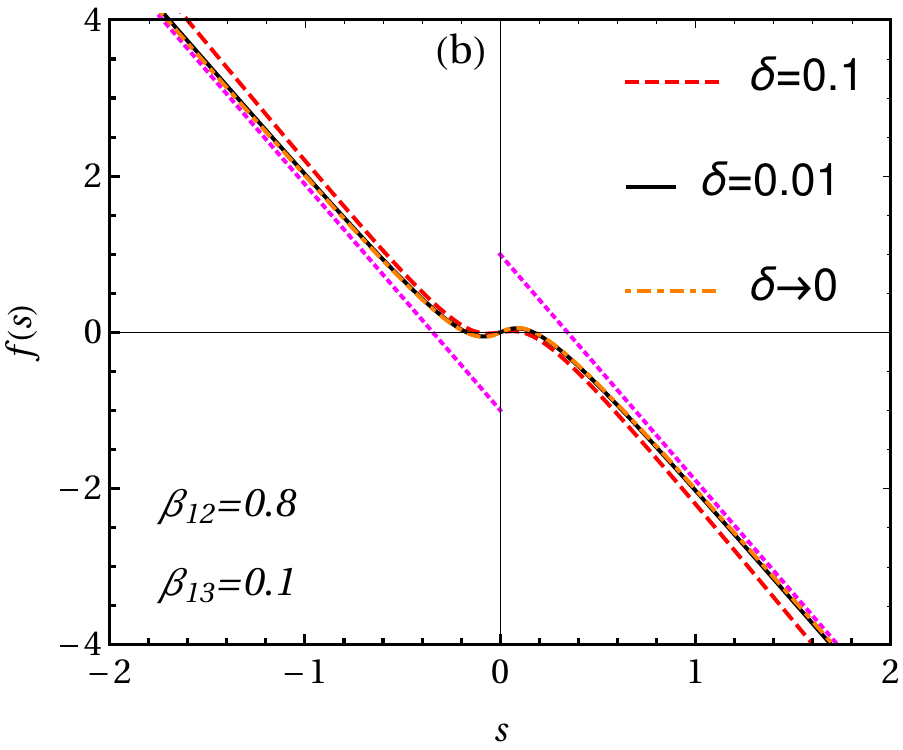}
    \includegraphics[width=0.34 \textwidth]{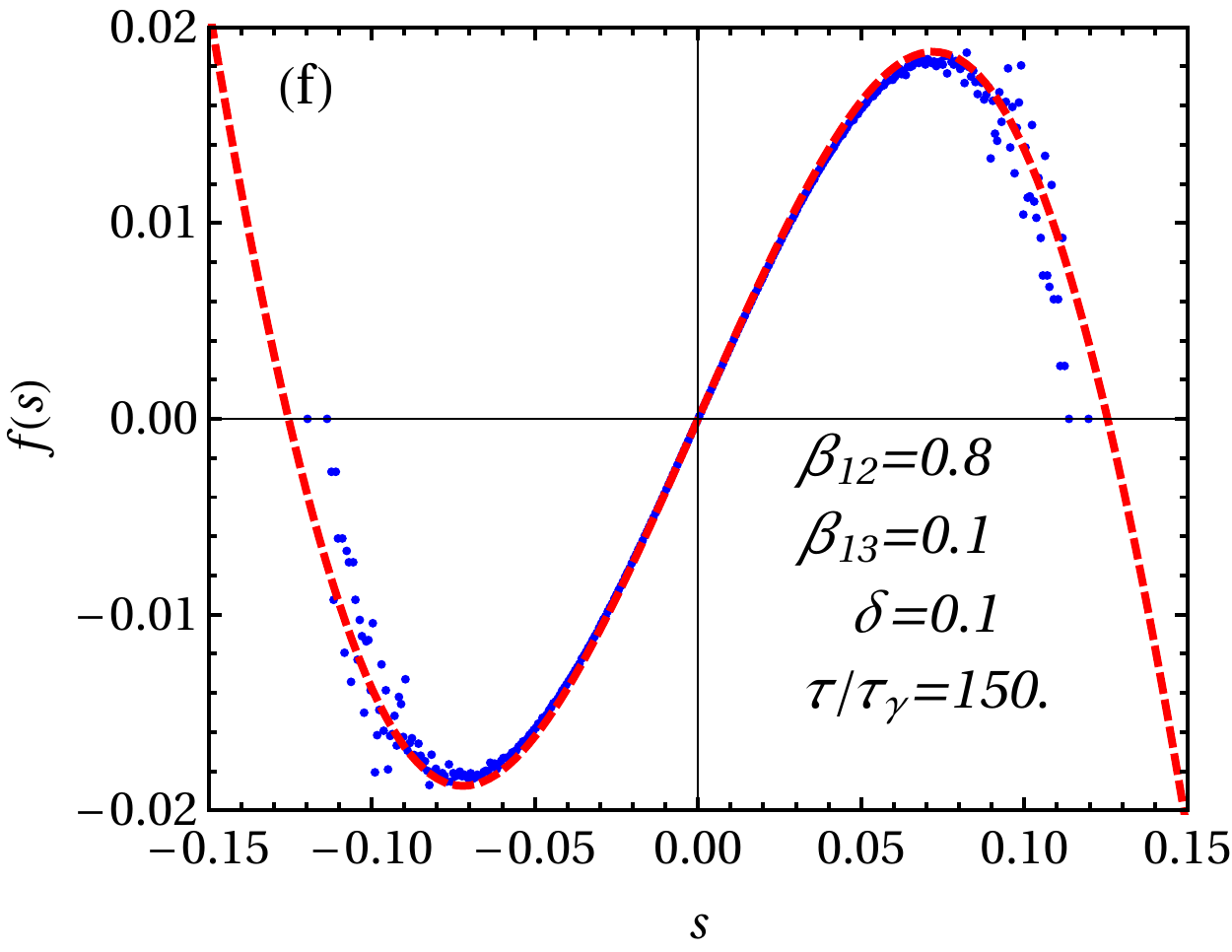}
    \includegraphics[width=0.33 \textwidth]{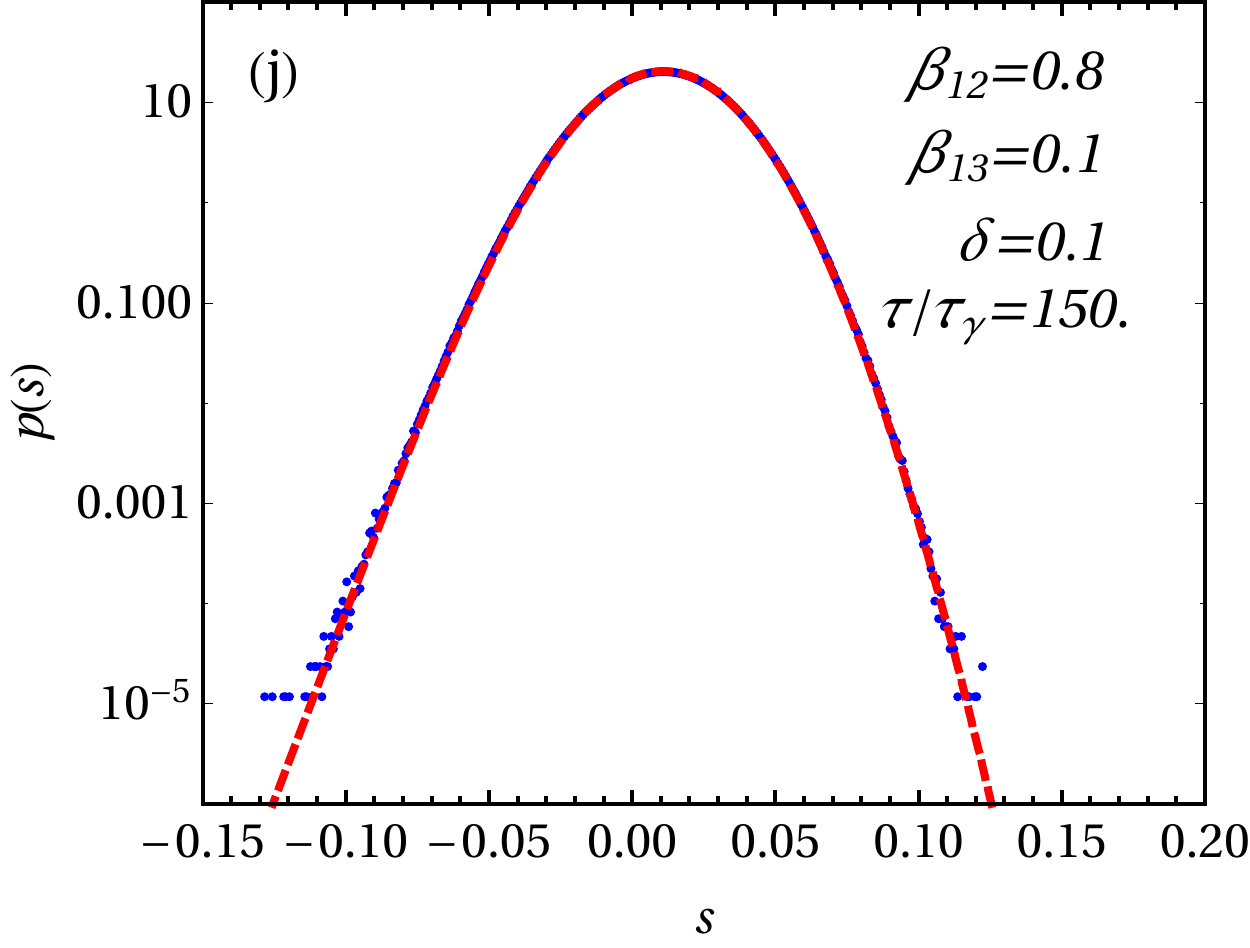}\\
    \includegraphics[width=0.33  \textwidth]{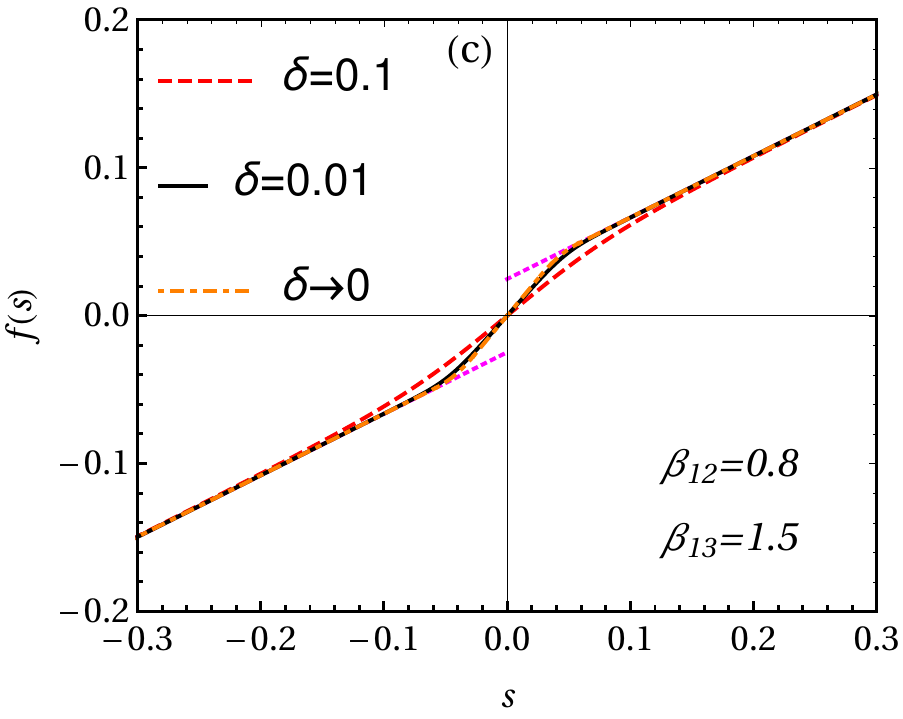}
    \includegraphics[width=0.34 \textwidth]{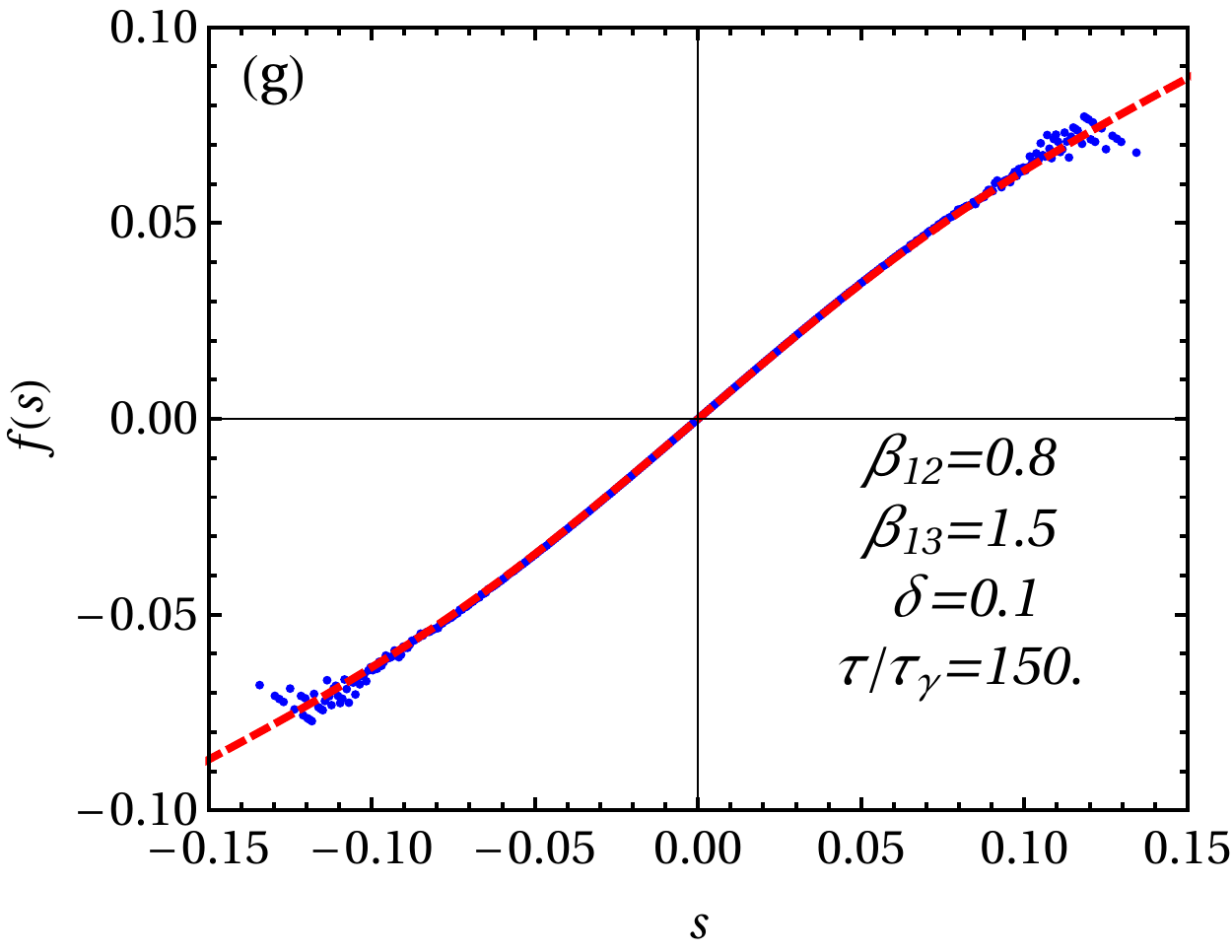}
    \includegraphics[width=0.33 \textwidth]{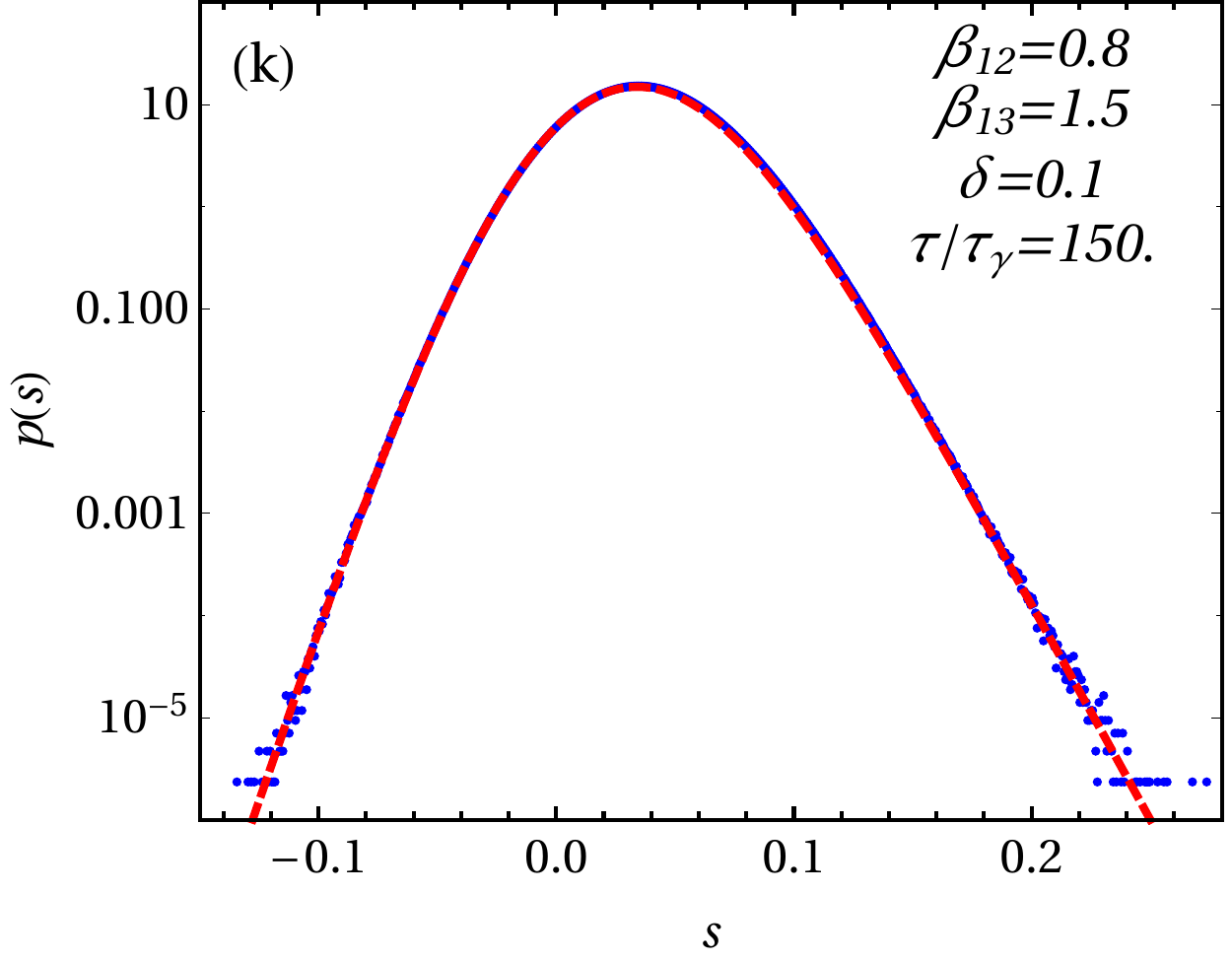}\\
    \includegraphics[width=0.33  \textwidth]{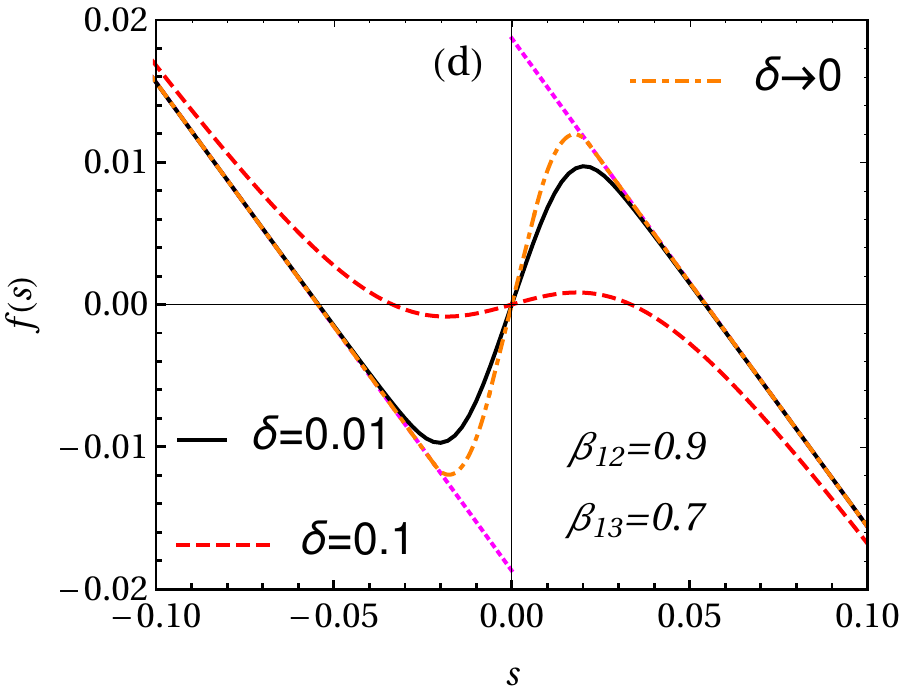}
    \includegraphics[width=0.34 \textwidth]{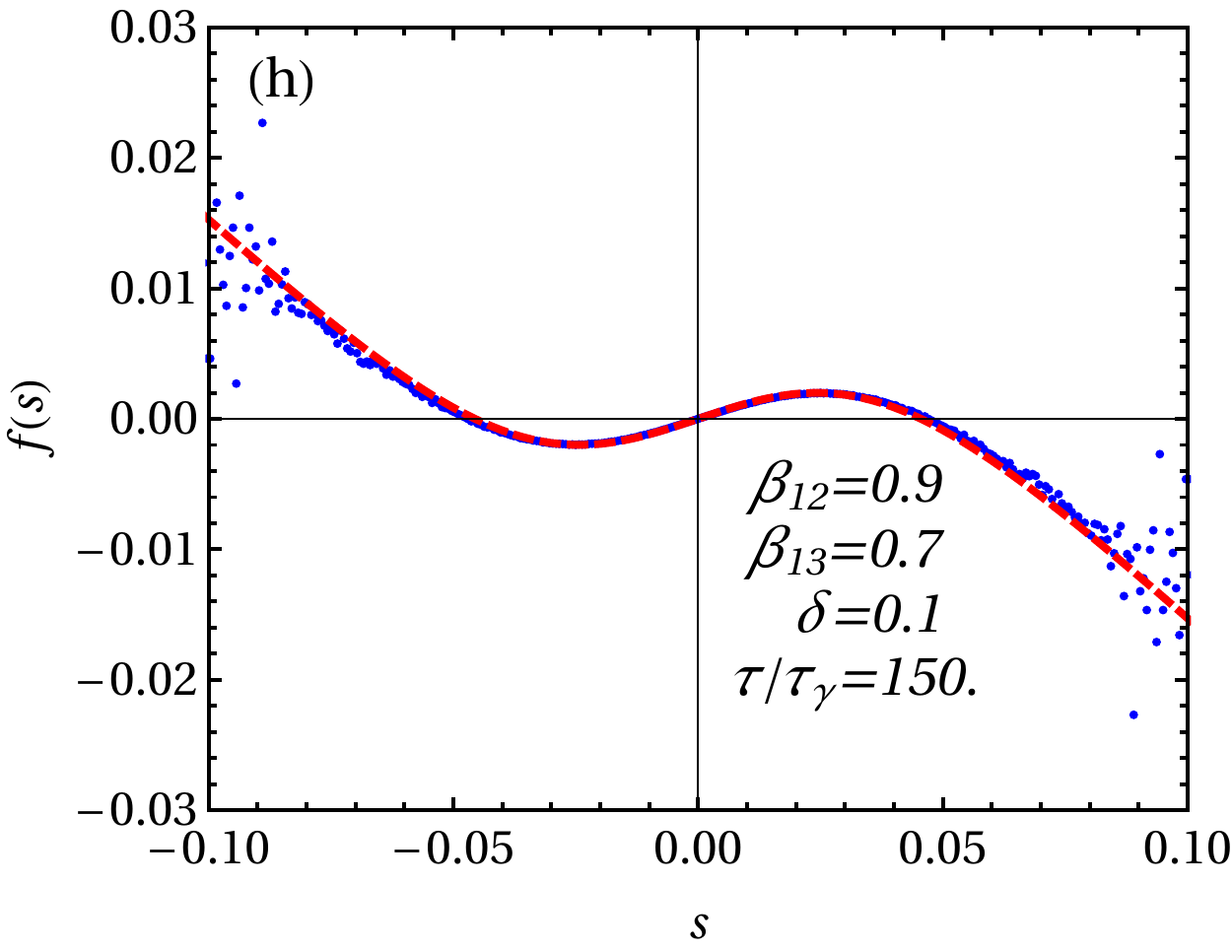}
    \includegraphics[width=0.33 \textwidth]{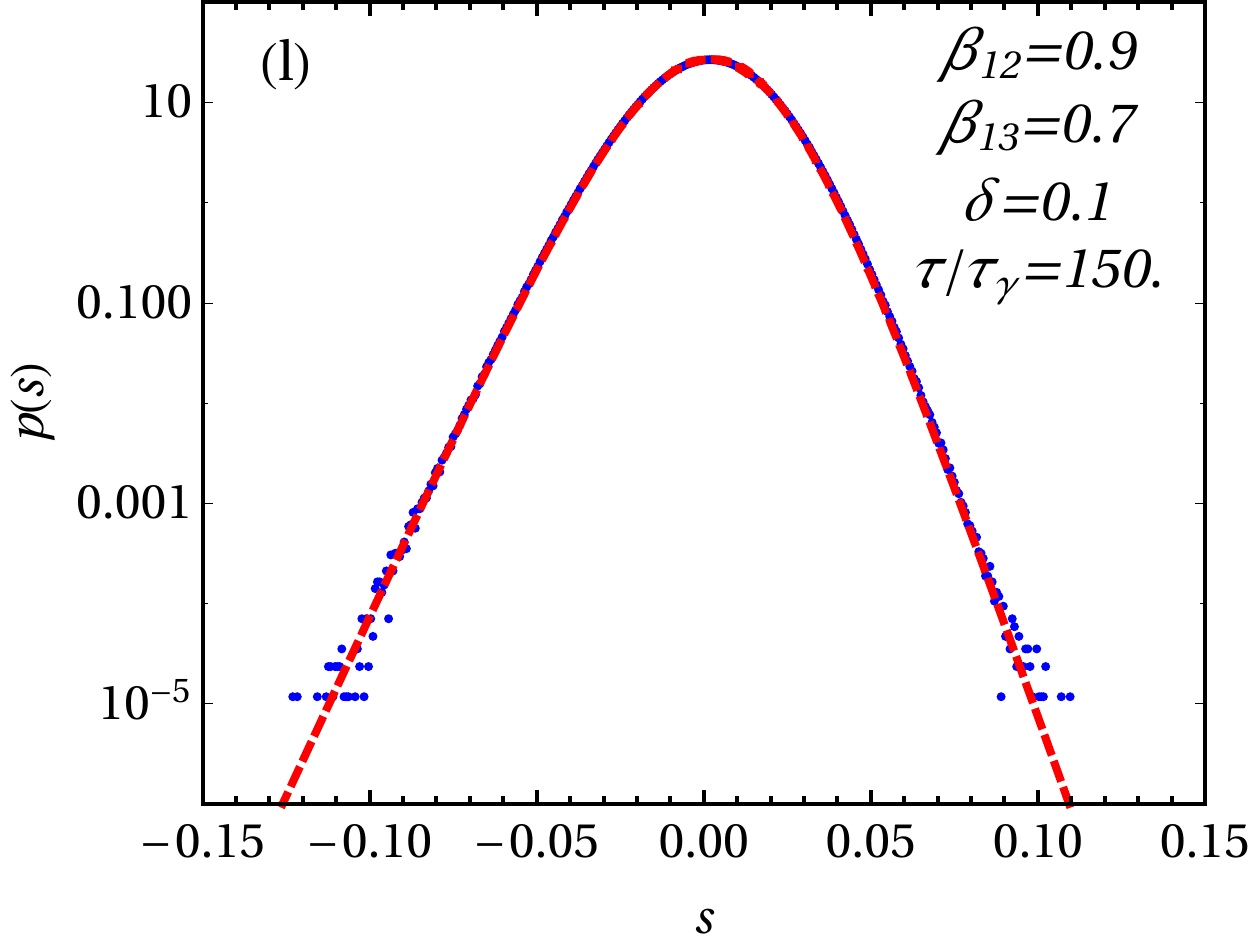}
 \end{tabular}
 \bigskip
 \caption{\label{PEP-fig} The analytically evaluated asymmetry function $f(s)$ given in \eref{asymmetry-function} for partial entropy production is plotted against the scaled variable $s=\Delta S^A_{tot}\tau_\gamma/\tau$ in figures (a)--(d) for respective $\beta_{12}$ and $\beta_{13}$ of phase diagram shown in \fref{phase-diagram}(a). These plots are obtained for $\delta=0.1$ (red dashed line) and $\delta=0.01$ (black solid line). The asymmetry functions in the limit $\delta \to 0$ (orange dotdashed line) are also plotted for respective cases \cite{Deepak}. The asymptotic behaviours for asymmetry function $f(s)$ for partial entropy production given in \erefs{asym-fs-PEP} are shown by magenta tiny dashed lines. The comparison of analytical results (red dashed line)  for asymmetry function $f(s)$ given by \eref{finite-time} and probability density function $p(s)$ given by \eref{pdf} with the numerical simulations (blue dots) of partial entropy production are shown in figures (e)--(h) and figures (i)--(l), respectively. These comparison are shown for fixed $\delta=0.1$ and $\tau/\tau_\gamma=150.0$.}
\end{figure*}  
\begin{figure*}
 \begin{tabular}{ccc}
    \includegraphics[width=0.33  \textwidth]{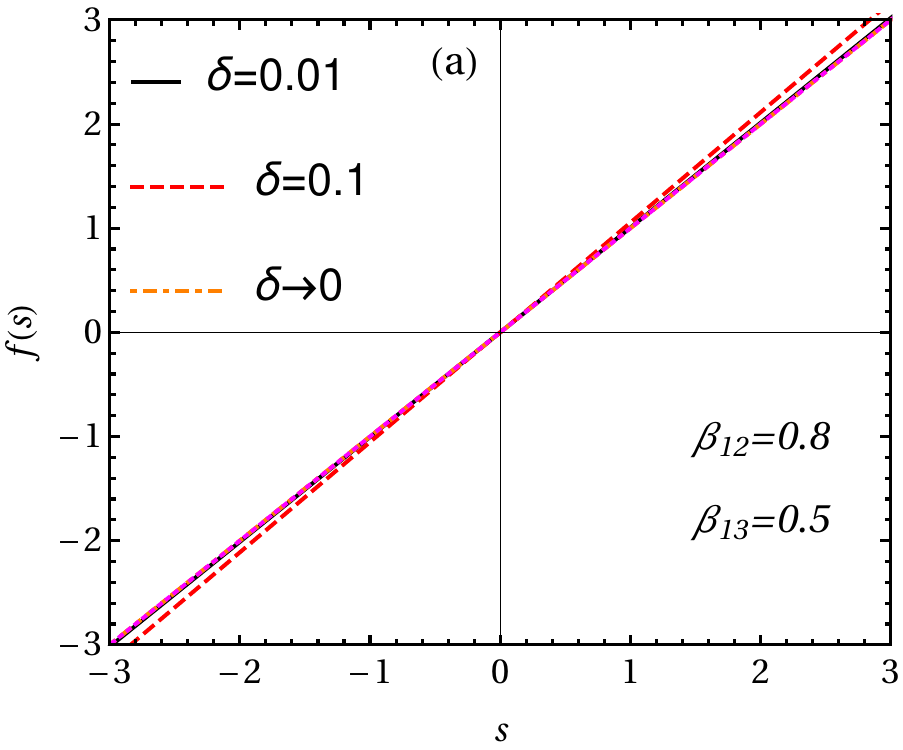}
    \includegraphics[width=0.34 \textwidth]{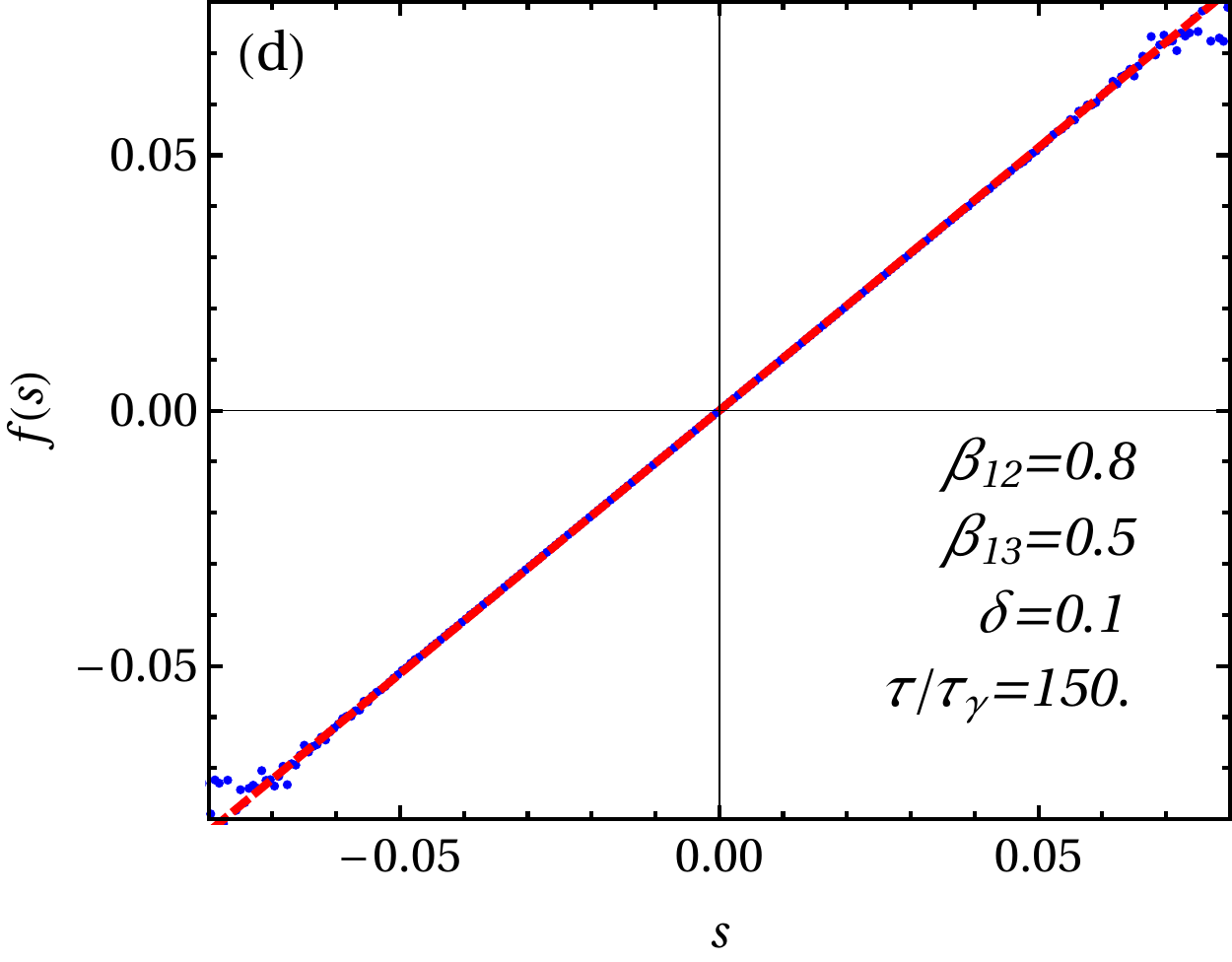}
    \includegraphics[width=0.33 \textwidth]{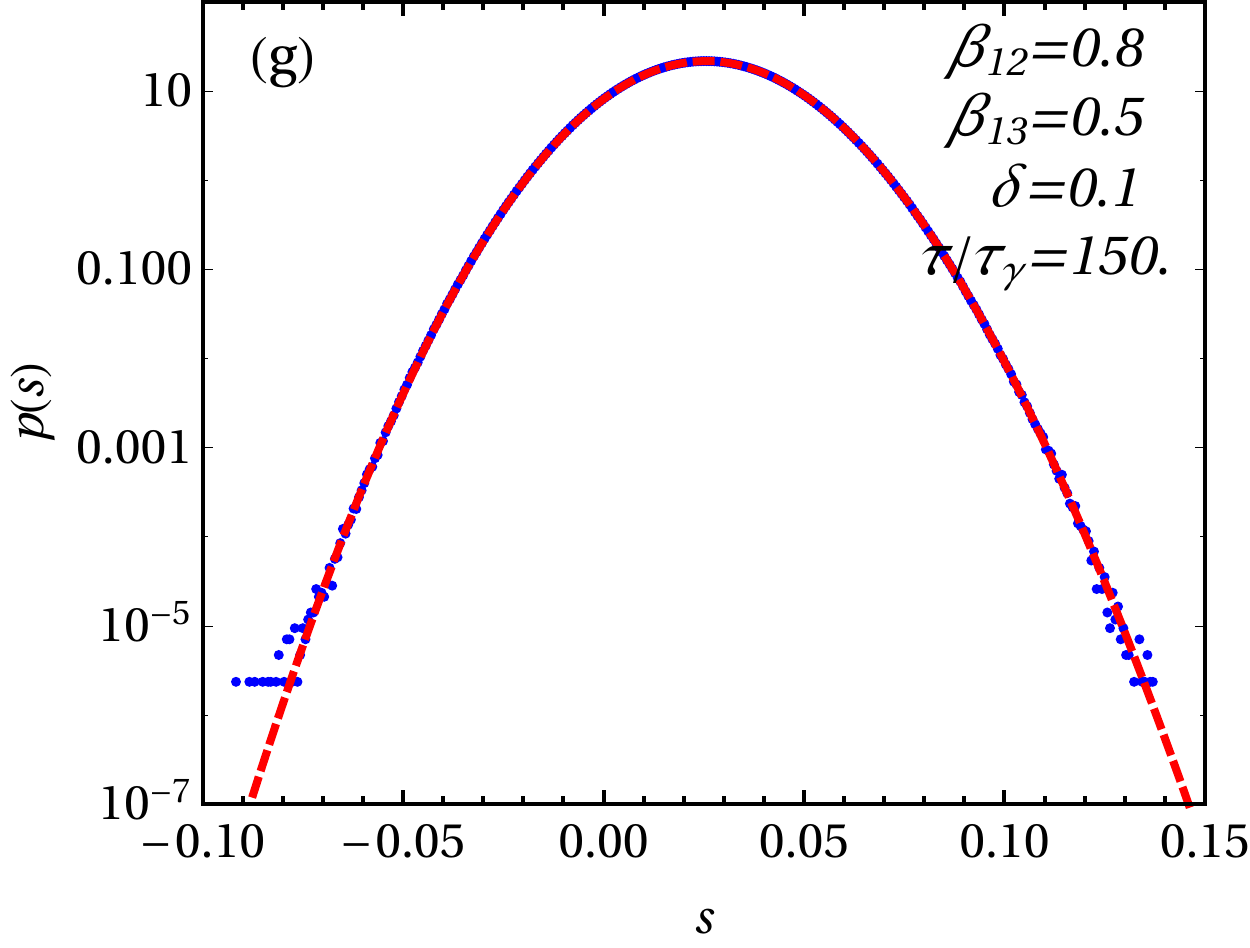}\\
    \includegraphics[width=0.33  \textwidth]{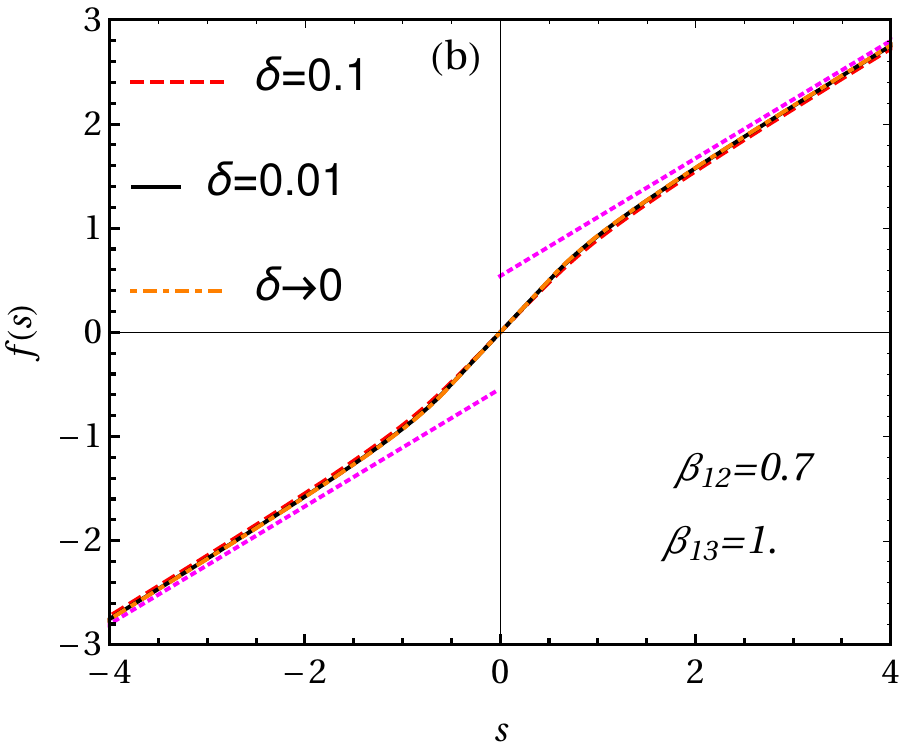}
    \includegraphics[width=0.34 \textwidth]{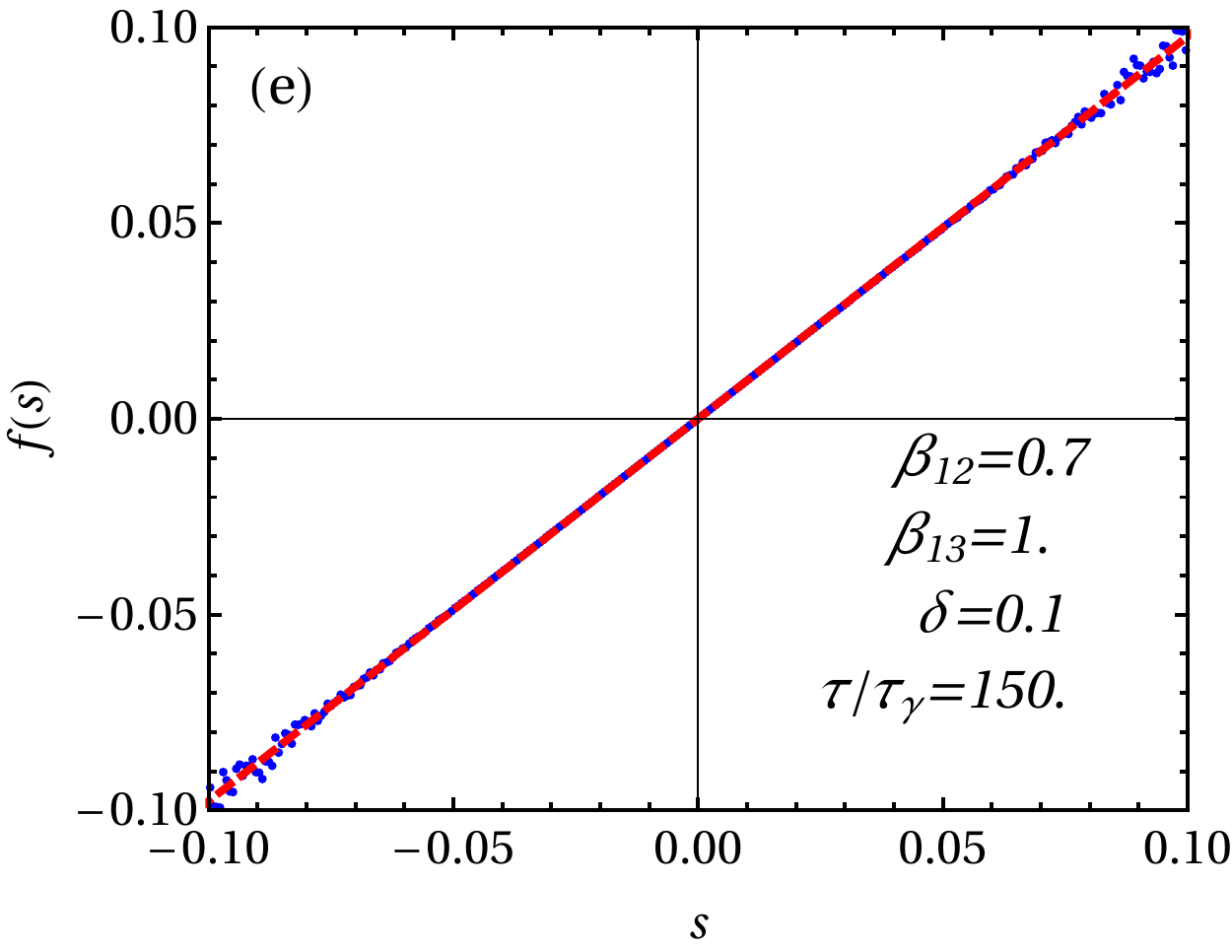}
    \includegraphics[width=0.33 \textwidth]{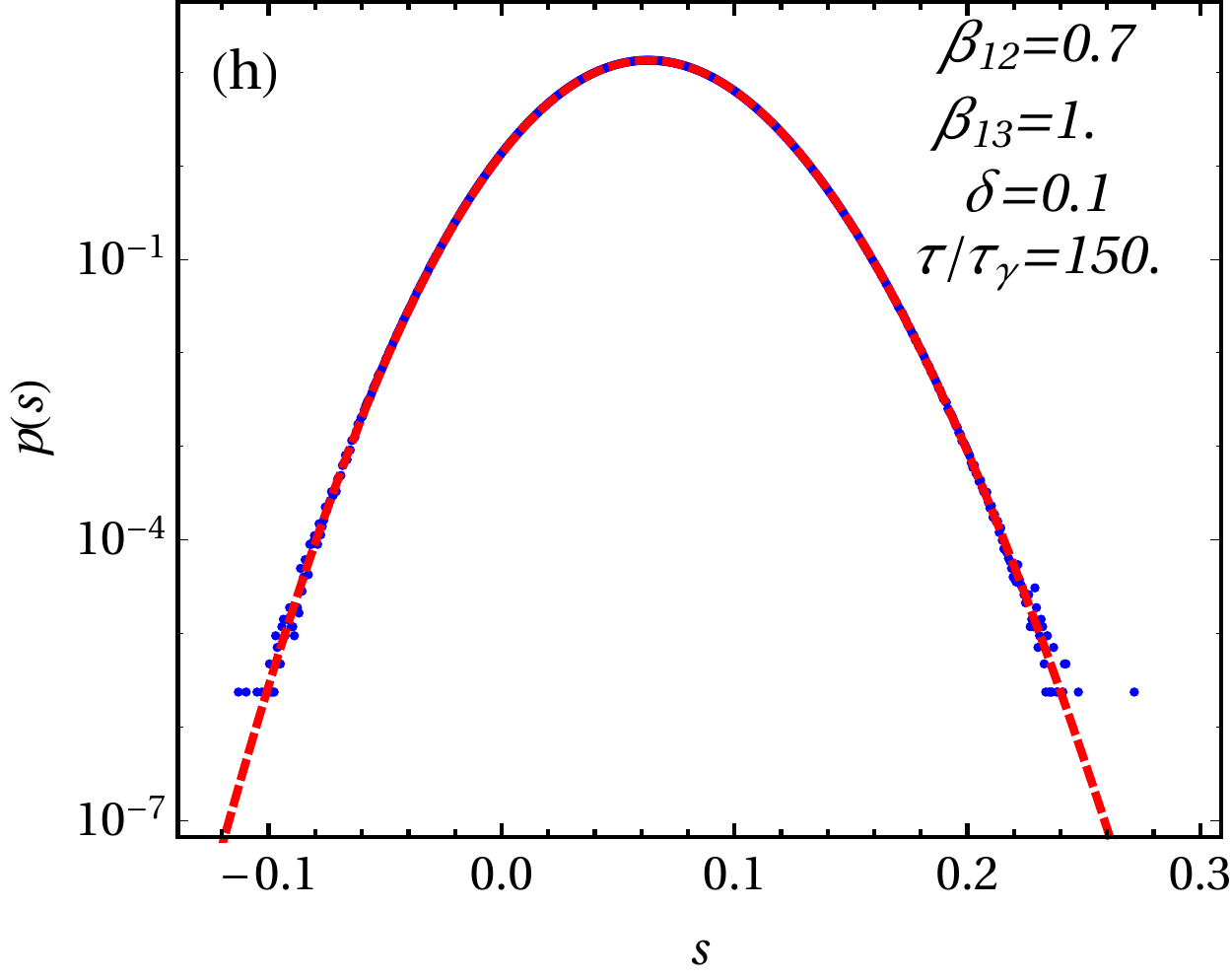}\\
    \includegraphics[width=0.33  \textwidth]{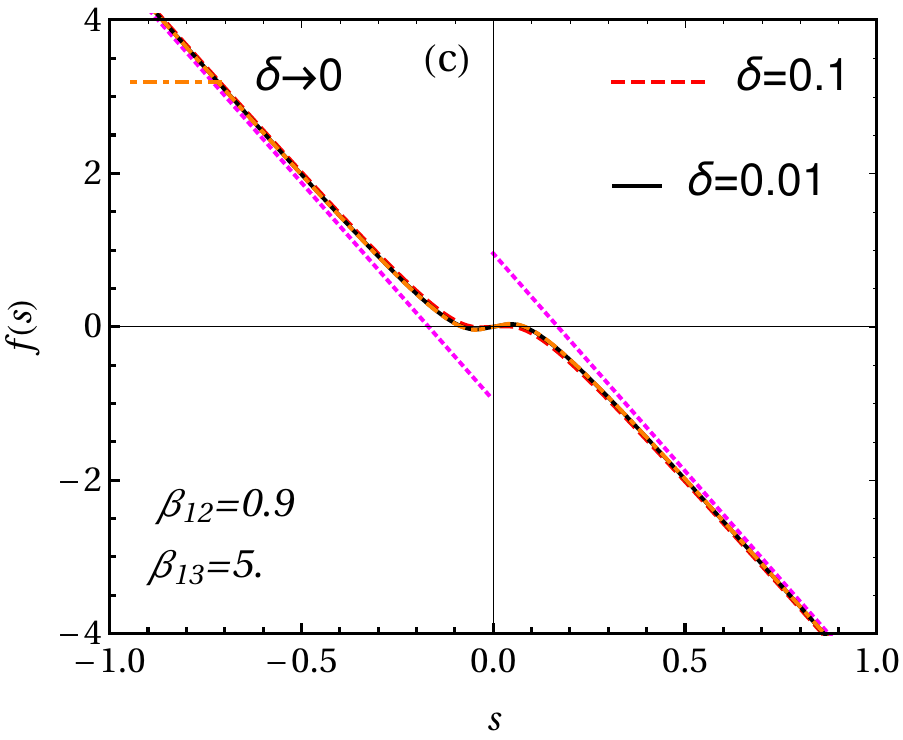}
    \includegraphics[width=0.34 \textwidth]{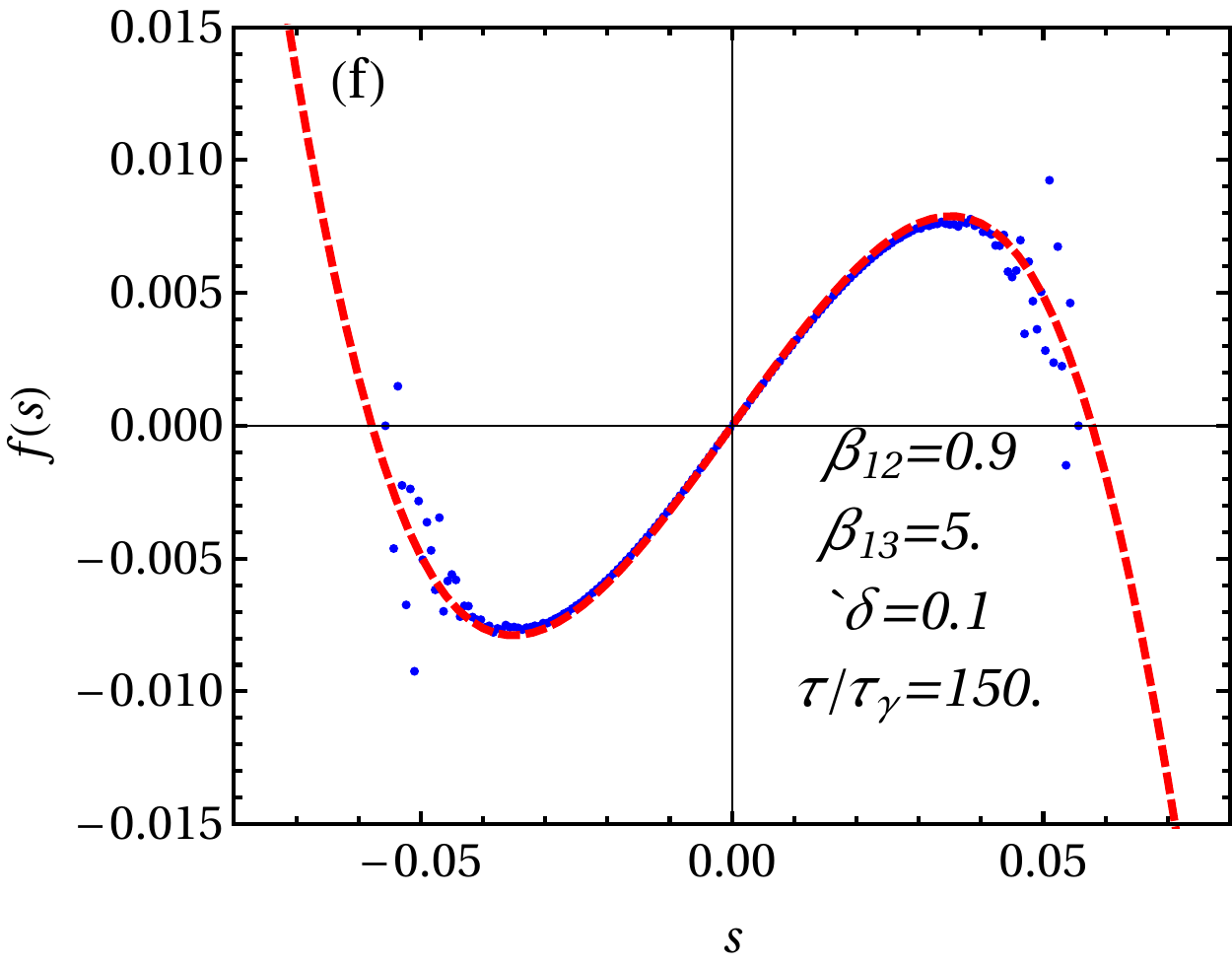}
    \includegraphics[width=0.33 \textwidth]{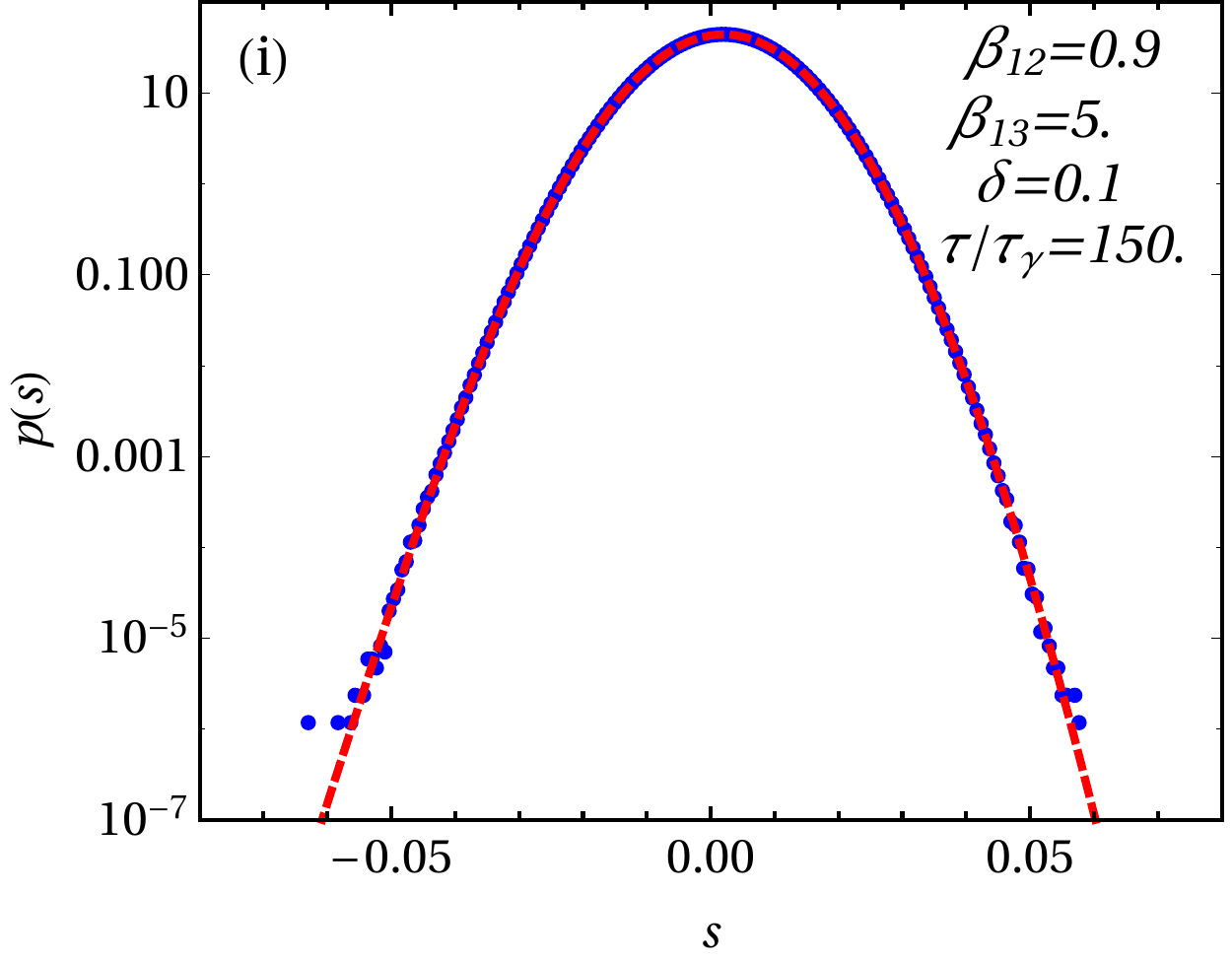}
 \end{tabular}
 \bigskip
 \caption{\label{AEP-fig} The analytical expression for $f(s)$ given in \eref{asymmetry-function} of apparent entropy production is plotted against the scaled variable $s=\Delta S^A_{tot}\tau_\gamma/\tau$ in figures (a)--(c) for respective $\beta_{12}$ and $\beta_{13}$ of phase diagram shown in \fref{phase-diagram}(b) for $\delta=0.1$ (red dashed line) and $\delta=0.01$ (black solid line). The asymmetry functions in the limit $\delta \to 0$ (orange dotdashed line) are also plotted for respective cases \cite{Deepak}.  Magenta tiny dashed lines represent the asymptotic expressions of $f(s)$ given in \eref{asym-fs-AEP}. In figures (d)--(f) and figures (g)--(i), we compared the asymmetry function $f(s)$ given in \eref{finite-time} and probability density function given in \eref{pdf} with the numerical simulation results, respectively, for apparent entropy production for given $\delta=0.1$ and $\tau/\tau_\gamma=150.$}
\end{figure*}  
%
%
\section{Large deviation function, fluctuation theorem and asymmetry function}
\label{asym-func}
Both $\mu(\lambda)$ and $g_0(\lambda)$ are real functions within any pair of singularities, i.e., $\lambda^\delta_-(u_-^*)$ and $\lambda^\delta_+(u_+^*)$ depending upon the choice of $\beta_{12}$ and $\beta_{13}$ in the weak coupling limit. The saddle point $\lambda^*(s)$ moves from $\lambda^\delta_-(u_-^*)$ to $\lambda^\delta_+(u_+^*)$ as $s$ decreases from $+\infty$ to $-\infty$. Therefore, the probability distribution function $P(\Delta S_{tot}^A)$ is given by \eref{PDF} and the corresponding probability density function $p(s)$ for large $s$ has the following form
\begin{equation}
p(s)\approx 
\begin{cases}
e^{\tau/\tau_\gamma I(s)}, \hspace{2cm} \text{s} \to +\infty,\\
e^{\tau/\tau_\gamma I(-s)}, \hspace{1.81cm} \text{s} \to -\infty,
\end{cases}
\label{result-1}
\end{equation}
where the probability density function is
\begin{equation} 
p(s)= (\tau/\tau_\gamma) P(\Delta S_{tot}^A=s\tau/\tau_\gamma),
\label{pdf}
\end{equation} and $I(s):=h(\lambda^*(s))$ is the large deviation function \cite{Touchette}.
We define an asymmetry function $f(s)$ as
\begin{equation}
f(s)=\dfrac{\tau_\gamma}{\tau}\ln\bigg[\dfrac{p(s)}{p(-s)}\bigg].
\label{finite-time}
\end{equation}
Thus, in the large time limit $\tau/\tau_\gamma\to\infty$, the asymmetry function is given by
\begin{equation}
f(s)=\lim_{\tau/\tau_\gamma\to\infty}\bigg[\dfrac{p(s)}{p(-s)}\bigg]=I(s)-I(-s).
\label{asymmetry-function}
\end{equation}

The quantity of interest is variation of the asymmetry function $f(s)$ with the scaled parameter $s=\Delta S_{tot}^A \tau_\gamma/\tau$, and deviation from $f(s)=s$ will indicate the violation of steady state fluctuation theorem. In the limit $\delta\to 0$, the asymptotic behavior of $f(s)$ in the case of partial entropy production for $s\to\infty$ is given by \cite{Deepak}
\begin{equation}
f(s)=\begin{cases}
\mu_0(\lambda_-)-\mu_0(\tilde{\lambda}_+)+(\lambda_-+\tilde{\lambda}_+)s,\quad\quad\quad\quad \text{for region I of \fref{phase-diagram}(a)},\\
\mu_0(\tilde{\lambda}_-)-\mu_0(\tilde{\lambda}_+)+(\tilde{\lambda}_-+\tilde{\lambda}_+)s,\quad\quad\quad\quad \text{for region II of \fref{phase-diagram}(a)},
\end{cases} 
\label{asym-fs-PEP}
\end{equation}
whereas for apparent entropy production
\begin{equation}
f(s)=\begin{cases}
s,\quad\quad\quad\quad\quad\quad\quad\quad\quad\quad\quad\quad\quad\quad\quad\quad\quad \text{for region I of \fref{phase-diagram}(b)},\\
\mu_0(\lambda_-)-\mu_0(\tilde{\lambda}_+)+(\lambda_-+\tilde{\lambda}_+)s,\quad\quad\quad\quad \text{for region II of \fref{phase-diagram}(b)},
\end{cases} 
\label{asym-fs-AEP}
\end{equation}
and $f(-s)=-f(s)$. In \eref{asym-fs-PEP} and \eref{asym-fs-AEP}, the cumulant generating function $\mu_0(\lambda)$ corresponds to the case $\delta=0$, is given in \eref{mu0-SP}.

We plot analytical asymmetry function $f(s)$ given in \eref{asymmetry-function} against $s$ in \frefss{PEP-fig}(a)--\frefs{PEP-fig}(d) and \frefss{AEP-fig}(a)--\frefs{AEP-fig}(c)  for partial and apparent entropy production, respectively, for $\delta=0.1$ (red dashed line) and $\delta=0.01$ (black solid line). The asymmetry function $f(s)$ in the limit $\delta\to 0$ (orange dotdashed line) is also plotted for each case \cite{Deepak}. In these figures, magenta tiny dashed lines correspond to the asymptotic expression of asymmetry function $f(s)$ given by \eref{asym-fs-PEP} and \eref{asym-fs-AEP} for partial and apparent entropy production, respectively. One can see, as the coupling parameter $\delta$ reduces, the asymmetry function $f(s)$ converges to that of $\delta \to 0$ case. 

\section{Numerical simulation}
\label{Num-sim}
In \frefss{time}(a)--\frefs{time}(d), we show the comparison of theoretical predictions (red dashed line) of probability density function $p(s)$ given by \eref{pdf} and the asymmetry function $f(s)$ given by \eref{finite-time} for partial entropy production for coupling parameter $\delta=0.1$ with the numerical simulation results (blue dots) with time $\tau=50.0$ [\frefss{time}(a) and \frefs{time}(b)] and $\tau=150.0$ [\frefss{time}(c) and \frefs{time}(d)]. This comparison indicates that as the observation time $\tau/\tau_\gamma$ increases, the agreement between theoretical predictions and numerical simulation results gets better.

We compare the analytical asymmetry functions $f(s)$ given in \eref{finite-time} (red dashed line) with the numerical simulation results (blue dots) for $\delta=0.1$ in \frefss{PEP-fig}(e)--\frefs{PEP-fig}(h) and \frefss{AEP-fig}(d)--\frefs{AEP-fig}(f) for partial and apparent entropy production, respectively. The probability density function $p(s)$ (red dashed line) given in \eref{pdf} is also compared with numerical simulation results (blue dots) for $\delta=0.1$ in \frefss{PEP-fig}(i)--\frefs{PEP-fig}(l) and \frefss{AEP-fig}(g)--\frefs{AEP-fig}(i) for partial and apparent entropy production, respectively. All of these results are compared at time $\tau/\tau_\gamma=150$, and show that there is nice agreement between theoretical predictions and numerical simulations.


\section{Summary}
\label{summ}
We have considered a coupled Brownian particle system. Both particles are connected by harmonic spring of stiffness $k$. One of the particles is connected to a thermal gradient and the other one is connected to a single bath of a constant temperature. The main goal of this paper is to understand the deviation of fluctuation theorem for total entropy production of one of the the particles (say particle-A) in a steady state when the interaction between the particle is weak. We have given two definition of total entropy production of partial system: partial and apparent entropy production. For convenience, we defined five dimensionless parameters: (1) coupling constant $\delta=2km/\gamma_1^2$, (2) $\beta_{12}=T_2/T_1$, (3) $\beta_{13}=T_3/T_1$, (4) $\alpha_{12}=\gamma_2/\gamma_1$, and (5) $\alpha_{13}=\gamma_3/\gamma_1$. When $\alpha_{12}=\alpha_{13}=1$, we plotted phase diagrams in $(\beta_{12},\beta_{13})$ plane, for both definitions of entropy production, in the limit $\delta\to0$. In the weak coupling limit ($\delta\to 0$), we have found that fluctuation theorem for total entropy production for one of the particles in the steady state is satisfied only in the region I of the phase diagram shown in \fref{phase-diagram}(b). The results given above are also supported by the numerical simulations, and they have very nice agreement. 

In a different model system where a single Brownian particle is connected with three heat baths of distinct temperatures (see \aref{fast-DOF}). The coupling $\delta$ of one of the baths with the particle is assumed to be very weak. We compute the total entropy production from the Brownian particle due to two heat baths [see \eref{tot-entropy}]. This model is exactly solvable. In the limit of $\delta \to 0$, we see the fluctuation theorem for partial entropy production is restored in the steady state. 
The results given so far in the paper sound contradictory with the result obtained from this model.  This is because, in the paper, we considered the coupling between slow variables (particle A and B) whereas in this example, we considered the coupling between slow DOF (Brownian particle)  and fast DOFs (heat bath). Therefore, in the weak coupling limit, one may see the violation of fluctuation theorem for the partial entropy production when slow DOFs are being coupled.

\makeatletter%
\def\@sect#1#2#3#4#5#6[#7]#8{\ifnum #2>\c@secnumdepth
  \let\@svsec\@empty\else
  \refstepcounter{#1}\edef\@svsec{\csname the#1\endcsname. }\fi
  \@tempskipa #5\relax
  \ifdim \@tempskipa>\z@
  \begingroup #6\relax
  \noindent{\hskip #3\relax\@svsec}{\interlinepenalty \@M #8\par}%
  \endgroup
  \csname #1mark\endcsname{#7}\addcontentsline
          {toc}{#1}{\ifnum #2>\c@secnumdepth \else
            \protect\numberline{}{\csname the#1\endcsname}\fi
            . #7}\else
          \def\@svsechd{#6\hskip #3\relax  
            \@svsec #8\csname #1mark\endcsname
                    {#7}\addcontentsline
                    {toc}{#1}{\ifnum #2>\c@secnumdepth \else
                      \protect\numberline{}{\csname the#1\endcsname}\fi
                      . #7}}\fi
          \@xsect{#5}}
\makeatother%

\appendix
\section{Single Brownian particle in connect with three heat baths}
\label{fast-DOF}
Consider a single Brownian particle of mass $m$, in the contact with three heat reservoirs of temperatures $T_1, T_2$, and $T_3$. Let $\gamma_1, \gamma_2$, and $\gamma_3$ are the dissipation constants of the heat baths with temperatures $T_1, T_2$, and $T_3$, respectively. Suppose the Brownian particle is connected to one of the baths ($T_3, \gamma_3$) with a coupling parameter $\delta$. The velocity $v(t)$ of the Brownian particle evolves according to underdamped Langevin equation
\begin{equation}
m\dot{v}=[-\gamma_1 v(t)+\eta_1(t)]+[-\gamma_2 v(t)+\eta_2(t)]+\delta [-\gamma_3 v(t)+\eta_3(t)],
\label{dynamics}
\end{equation}
The thermal noises $\eta_1$, $\eta_2$ and $\eta_3$ acting on the Brownian particle, have mean zero and correlations $\langle \eta_i(t)\eta_j(t^\prime) \rangle=2 \gamma_i T_i \delta_{i,j}\delta(t-t^\prime)$. Since the velocity is linear in thermal noises, the steady state distribution is given by
\begin{equation}
P_{ss}(v(\tau))=\dfrac{1}{\sqrt{2 \pi \sigma^2_v}} \exp\bigg[{-\dfrac{v^2(\tau)}{2 \sigma_v^2}}\bigg], 
\label{steady-state}
\end{equation}
where 
$$\sigma_v^2=\dfrac{\gamma_1T_1+\gamma_2 T_2+\delta^2\gamma_3 T_3}{m(\gamma_1+\gamma_2+\delta \gamma_3)}.$$
 
Here the observable is the total entropy production $\Delta S^{12}_{tot}$ of the Brownian particle due to two heat baths with temperatures (dissipation constants) $T_1$ ($\gamma_1$) and $T_2$ ($\gamma_2$) in the steady state for a duration $\tau$
\begin{align}
\Delta S^{12}_{tot}&=-\bigg(\dfrac{Q_1}{T_1}+\dfrac{Q_2}{T_2}\bigg)-\ln P_{ss}(v(\tau))+\ln P_{ss}(v(0))=\mathcal{Q}+\dfrac{1}{2 \sigma_v^2}[v^2(\tau)-v^2(0)],
\label{tot-entropy}
\end{align}
where
\begin{equation}
\mathcal{Q}=a_0 Q_1+b_0 Q_2,
\label{W}
\end{equation}
in which $a_0=-1/T_1$ and $b_0=-1/T_2$.

The Fokker-Planck equation for the restricted characteristic function of $\mathcal{Q}$ is given by
\begin{align}
\dfrac{\partial Z_{\mathcal{Q}}(\lambda,v,\tau|v_0)}{\partial \tau}&=\bigg[\dfrac{\gamma_1T_1+\gamma_2T_2+\delta^2\gamma_3T_3}{m^2}\dfrac{\partial^2}{\partial v^2}+\dfrac{\gamma_1+\gamma_2+\delta\gamma_3+2\lambda(a_0\gamma_1T_1+b_0\gamma_2T_2)}{m}\nonumber\\
&+\bigg\{\dfrac{\gamma_1+\gamma_2+\delta\gamma_3+2\lambda(a_0\gamma_1T_1+b_0\gamma_2 T_2)}{m}\bigg\}v\dfrac{\partial}{\partial v}+\lambda[(a_0\gamma_1+b_0\gamma_2)v^2\nonumber\\&-(a_0\gamma_1T_1+b_0\gamma_2T_2)/m]
+(a_0^2\gamma_1T_1+b_0^2 \gamma_2T_2)\lambda^2v^2\bigg]Z_\mathcal{Q}(\lambda,v,\tau|v_0).
\label{FP-Z}
\end{align}
The differential equation given above is subjected to initial condition $Z_\mathcal{Q}(\lambda,v,\tau=0|v_0)=\delta(v-v_0)$. 

For simplicity, we choose $\gamma_1=\gamma_2=\gamma_3=\gamma$. In the large time limit $(\tau\to\infty)$, the solution of \eref{FP-Z} is given by \cite{Visco,Efficiency}
\begin{align}
Z_\mathcal{Q}(\lambda,v,\tau|v_0)= e^{(\tau/\tau_\gamma) \mu(\lambda)} \sqrt{\dfrac{m\nu(\lambda)}{2 \pi \Theta}}\exp \bigg[-\dfrac{m v^2}{4\Theta}[2+\delta-4\lambda+\nu(\lambda)]\bigg]\nonumber\\
\times\exp \bigg[-\dfrac{m v_0^2}{4\Theta}[-2-\delta+4\lambda+\nu(\lambda)]\bigg]+\dots.
\end{align}
In the above equation, the characteristic time scale $\tau_\gamma=m/\gamma$, $\Theta=T_1+T_2+\delta^2 T_3$, and
\begin{align}
\mu(\lambda)=\dfrac{1}{2}[2+\delta-\nu(\lambda)],
\end{align}
with 
\begin{equation}
\nu(\lambda)=\sqrt{(2+\delta-4\lambda)^2-4\Theta\lambda(1-\lambda)(a_0+b_0)}=\sqrt{(\alpha_0-16)(\lambda^\delta_+-\lambda)(\lambda-\lambda_-^\delta)}.
\end{equation}
In the above equation, the branch points $\lambda_\pm^\delta$ are
\begin{equation}
\lambda^\delta_\pm=\dfrac{1}{2(\alpha_0-16)}[\alpha_0-8(2+\delta)\pm\sqrt{\alpha_0}\sqrt{\alpha_0-16+4\delta^2}],
\end{equation}
where $\alpha_0=4(1+1/\beta_{12})(1+\beta_{12}+\delta^2 \beta_{13})$ in which $\beta_{1j}=T_j/T_1$.

Therefore, the restricted characteristic function for total entropy production $\Delta S^{12}_{tot}$ can be written as
\begin{align}
Z(\lambda,v,\tau|v_0)&=\exp\bigg[-\dfrac{\lambda}{2 \sigma_v^2}[v^2(\tau)-v^2(0)]\bigg]Z_\mathcal{Q}(\lambda,v,\tau|v_0)\nonumber\\
&=e^{(\tau/\tau_\gamma) \mu(\lambda)}\sqrt{\dfrac{m\nu(\lambda)}{2 \pi \Theta}}\exp \bigg[-\dfrac{m v^2}{4\Theta}[2+\delta+2\lambda\delta+\nu(\lambda)]\bigg]\nonumber\\
&\hspace{2.2cm}\times\exp \bigg[-\dfrac{m v_0^2}{4\Theta}[-2-\delta-2\lambda\delta+\nu(\lambda)]\bigg]+\dots
\end{align}
The characteristic function for the partial entropy production is given by
\begin{align}
Z(\lambda)=\int dv\ \int dv_0\ P_{ss}(v_0)\ Z(\lambda,v,\tau|v_0)=e^{(\tau/\tau_\gamma) \mu(\lambda)} g(\lambda)+\dots,
\end{align}
where the prefactor is 
\begin{equation}
g(\lambda)=\dfrac{2\sqrt{(2+\delta)\nu(\lambda)}}{\sqrt{2+\delta+2\lambda\delta+\nu(\lambda)}\sqrt{2+\delta-2\lambda\delta+\nu(\lambda)}}.
\label{g-lambda}
\end{equation}
Here, first term is the denominator comes from  integrating the restricted characteristic function of partial entropy production over the final variable $v$ while the second term in the denominator arises from integrating the restricted characteristic function over the initial steady state ensemble $P_{ss}(v_0)$.

In $g(\lambda)$, both of the denominators have one branch point each for particular choice of $\beta_{12}$, $\beta_{13}$ and $\delta$. Corresponding to first denominator, $\lambda=\lambda_a$ is the zero of $[2+\delta+2\lambda\delta+\nu(\lambda)]$, and $\lambda_a\in(\lambda^\delta_-,\lambda^\delta_+)$ when $[2+\delta+2\delta \lambda^\delta_-]<0$ (condition I). On the other hand, $\lambda=\lambda_b$ is the zero of second denominator having $[2+\delta-2\lambda\delta+\nu(\lambda)]$, and $\lambda_b\in(\lambda^\delta_-,\lambda^\delta_+)$ only when $[2+\delta-2\delta \lambda_+]<0$ (condition II). Condition II is simply given by $\delta\geq 2$.

\begin{figure}
\begin{center}
\includegraphics[width=0.5\linewidth]{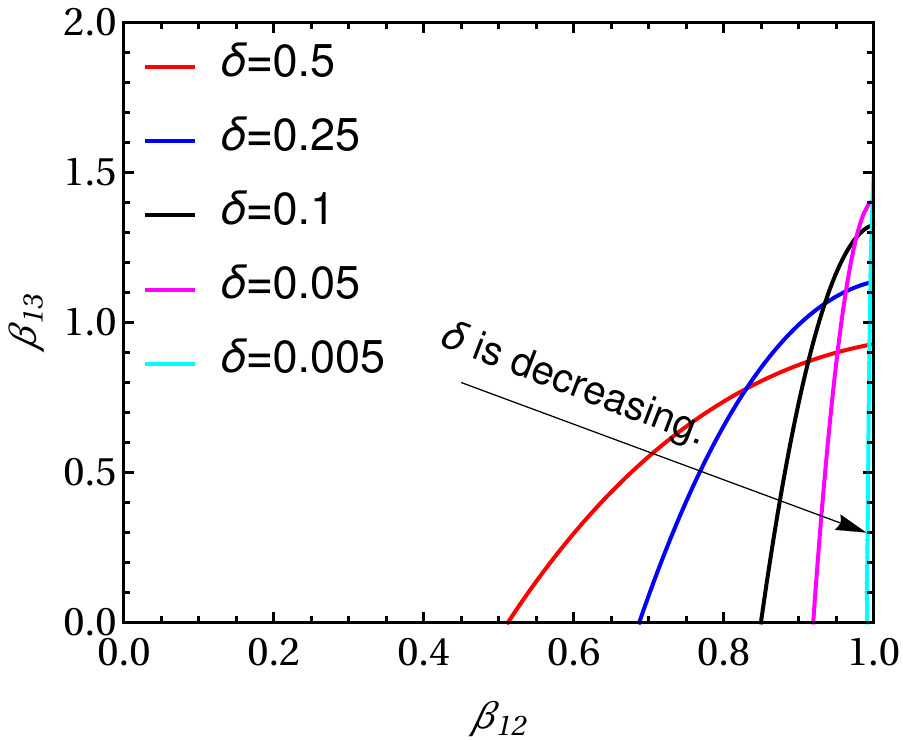}
\caption{The prefactor $g(\lambda)$ is analytic
to the left side of the contours given by condition I for given $\delta$. The arrow indicates the direction of decreasing coupling parameter $\delta$.}
\label{phase-dig}
\end{center}
\end{figure}

$\lambda_{a,b}$ are given as
\begin{align}
\lambda_a&=\dfrac{(\alpha_0-16-4\delta(\delta+4))}{(\alpha_0-16+4 \delta^2)},\\
\lambda_b&=1.
\end{align}
Since we are interested in the weak coupling limit ($\delta\to 0$), using condition I, we plot the phase diagram as shown in \fref{phase-dig} in $(\beta_{12},\beta_{13})$ plane for various values of $\delta$. In the phase diagram, condition I does not hold to the left side of contours for respective $\delta$. From the \fref{phase-dig}, it is clear that there is no singularity present in the prefactor $g(\lambda)$ in the weak coupling limit, \emph{i.e.,} $\delta \to 0$. Hence, $g(\lambda)$ is analytic function of $\lambda$ in the weak coupling limit. 
Therefore, in the limit $\delta\to0$, the probability density function $p(s)$ is given by
\begin{equation}
p(s)\sim e^{(\tau/\tau_\gamma)I(s)},
\end{equation}
where $I(s):=\mu(\lambda^*)+\lambda^* s$ is the large deviation function \cite{Touchette}. The saddle point $\lambda^*(s)$ is the solution of $\mu^\prime(\lambda^*)+s=0$ which gives
\begin{equation}
\lambda^*(s)=\dfrac{1}{2(\alpha_0-16)}\bigg[\alpha_0-8(2+\delta)-2 s\sqrt{\dfrac{\alpha_0(\alpha_0-16+4\delta^2)}{\alpha_0-16+4s^2}}\bigg],
\label{saddle-point-3}
\end{equation}
Therefore,
\begin{equation}
I(s)=1+\dfrac{\delta}{2}+\dfrac{s[\alpha_0-8(2+\delta)]}{2(\alpha_0-16)}-\dfrac{1}{4}\sqrt{\dfrac{\alpha_0(\alpha_0-16+4\delta^2)}{\alpha_0-16+4s^2}}\bigg(1+\dfrac{4 s^2}{\alpha_0-16}\bigg).
\label{ldf}
\end{equation}
The asymmetry function $f(s)$ in this case is
\begin{equation}
f(s)=I(s)-I(-s)=\dfrac{s[\alpha_0-16-8 \delta]}{\alpha_0-16}.
\label{ft-sol}
\end{equation}
Thus, in the weak coupling limit ($\delta \to 0$), the fluctuation theorem is satisfied ($f(s)=s$).

\vskip 2cm

\bibliographystyle{unsrt}
\bibliography{Heattransport.bib}
\end{document}